\documentclass{llncs}
\usepackage[utf8]{inputenc}
\usepackage[T1]{fontenc}
\usepackage{graphicx}        
\usepackage{makeidx}  

\usepackage{bsymb,b2latex}

\usepackage{oz}

\usepackage{breqn}

\usepackage[bookmarks=false]{hyperref}
\usepackage[linewidth=1pt]{mdframed}

\begin{document}
\title{Formal Representation of SysML/KAOS Domain Model \\(Complete Version)}
\titlerunning{Formal Representation of SysML/KAOS Domain Model \\(Complete Version)}  
%
\author{Steve Jeffrey Tueno Fotso\inst{1, 3} \and Amel Mammar\inst{2} \and Régine Laleau\inst{1} \and Marc Frappier\inst{3}}
\authorrunning{Steve Tueno et al.} 
%
\tocauthor{Steve Tueno, Amel Mammar, Régine Laleau, Marc Frappier}
\institute{Université Paris-Est Créteil, LACL, Créteil, France,\\
\email{steve.tuenofotso@univ-paris-est.fr},
\email{laleau@u-pec.fr}
\and
Télécom SudParis, SAMOVAR-CNRS, Evry, France, \\
\email{amel.mammar@telecom-sudparis.eu}
\and
Université de Sherbrooke, GRIL, Québec, Canada, \\
\email{Marc.Frappier@usherbrooke.ca}}

\maketitle              

\paragraph{\today}

\begin{abstract}
Nowadays, the usefulness of a formal language for ensuring the consistency of requirements is well established. 
The work presented here is part of the definition of a 
formally-grounded, model-based requirements engineering  method for critical and complex systems.  Requirements are captured through the \textit{SysML/KAOS} method and the targeted formal specification is written using the\textit{ Event-B} method.
Firstly, an \textit{Event-B} skeleton is produced from the goal hierarchy provided by the SysML/KAOS goal model.
 This skeleton is then completed in a second step by the Event-B specification obtained from system application domain properties that gives rise to the  system structure.
 Considering that the domain is represented using ontologies through the \textit{SysML/KAOS Domain Model} method, is it possible to automatically produce the structural part of  system  Event-B models ?
This paper  proposes a set of generic rules that translate SysML/KAOS domain ontologies into an Event-B specification. They are illustrated  through a case study dealing with a landing gear system.
 Our proposition 
makes it possible to automatically obtain, from a representation of the system application domain in the form of ontologies, the structural part of the Event-B specification which will be used to formally validate the consistency of  system requirements.

\end{abstract}

\section{Introduction}
This article focuses on the development of systems in critical areas such as railway or aeronautics. The implementation of such systems, in view of their complexity, requires several validation steps, more or less formal\footnote{through formal methods}, with regard to the current regulations. Our work is part of the \textit{FORMOSE} project \cite{anr_FORMOSE_reference_link} which integrates  industrial partners involved in the implementation of critical systems for which the regulation imposes  formal validations. 
The contribution presented in this paper represents a straight continuation of our research work on the formal specification of systems whose requirements are captured with \textit{SysML/KAOS} goal models. 
The \textit{Event-B}   method \cite{DBLP:books/daglib/0024570}  has been choosen for the formal validation steps because  it 
involves simple mathematical concepts and 
has  a powerful  refinement logic  facilitating the separation of concerns. Furthermore, it is supported by many industrial tools. 
In \cite{DBLP:conf/iceccs/MatoussiGL11}, we have defined translation rules to produce an Event-B  specification from \textit{SysML/KAOS} goal models. 
Nevertheless, the generated Event-B specification does not contain the system state. This is why in \cite{DBLP:conf/isola/MammarL16}, we have presented the use of ontologies and \textit{UML} class and object diagrams  for domain properties representation and have also introduced a first attempt to complete the Event-B model with specifications obtained from the translation of these domain representations. 
Unfortunately, the proposed approach raised several concerns such as the use of several modeling formalisms for the representation of domain knowledge or the disregard of variable entities. In addition, the proposed translation rules did not take into account several elements of the domain model such as data sets or predicates. 
We have therefore proposed in \cite{sysml_kaos_domain_modeling} a formalism for domain  knowledge representation  through ontologies.
This paper is specifically concerned with establishing  correspondence links between this new formalism called \textit{SysML/KAOS Domain Modeling} and Event-B.  
The proposed approach allows a high-level modeling of domain properties by encapsulating the difficulties inherent in the manipulation of formal specifications. This facilitates system constraining  and enables the expression of more precise and complete properties. The approach  also allows  further reuse and separation of concerns.

The remainder of this paper is structured as follows: Section 2 briefly describes 
our abstraction of the \textit{Event-B} specification language,  the SysML/KAOS requirements engineering method,     the formalization in  Event-B of SysML/KAOS goal models and  the SysML/KAOS domain modeling formalism.
 Follows a presentation, in Section 3,  of the relevant state of the art on the formalization of domain knowledge representations. In Section 4, we describe and illustrate our  matching rules between domain models and Event-B specifications. Finally, Section 5 reports our conclusions and  discusses our future work.

\section{Formalism Overviews}

\subsection{Event-B} \label{event_b_description_section}
\textit{Event-B } is an industrial-strength formal method defined by \textit{J. R. Abrial} in 2010 
for \emph{system modeling} \cite{DBLP:books/daglib/0024570}. It is used to prove the preservation of safety invariants about a system. 
\textit{Event-B} is mostly used for the modeling of closed systems: the modeling of the system is accompanied by that of its environment and of all  interactions likely to occur between them.

\begin{figure}[!h]
\begin{center}
\includegraphics[width=1.0\textwidth]{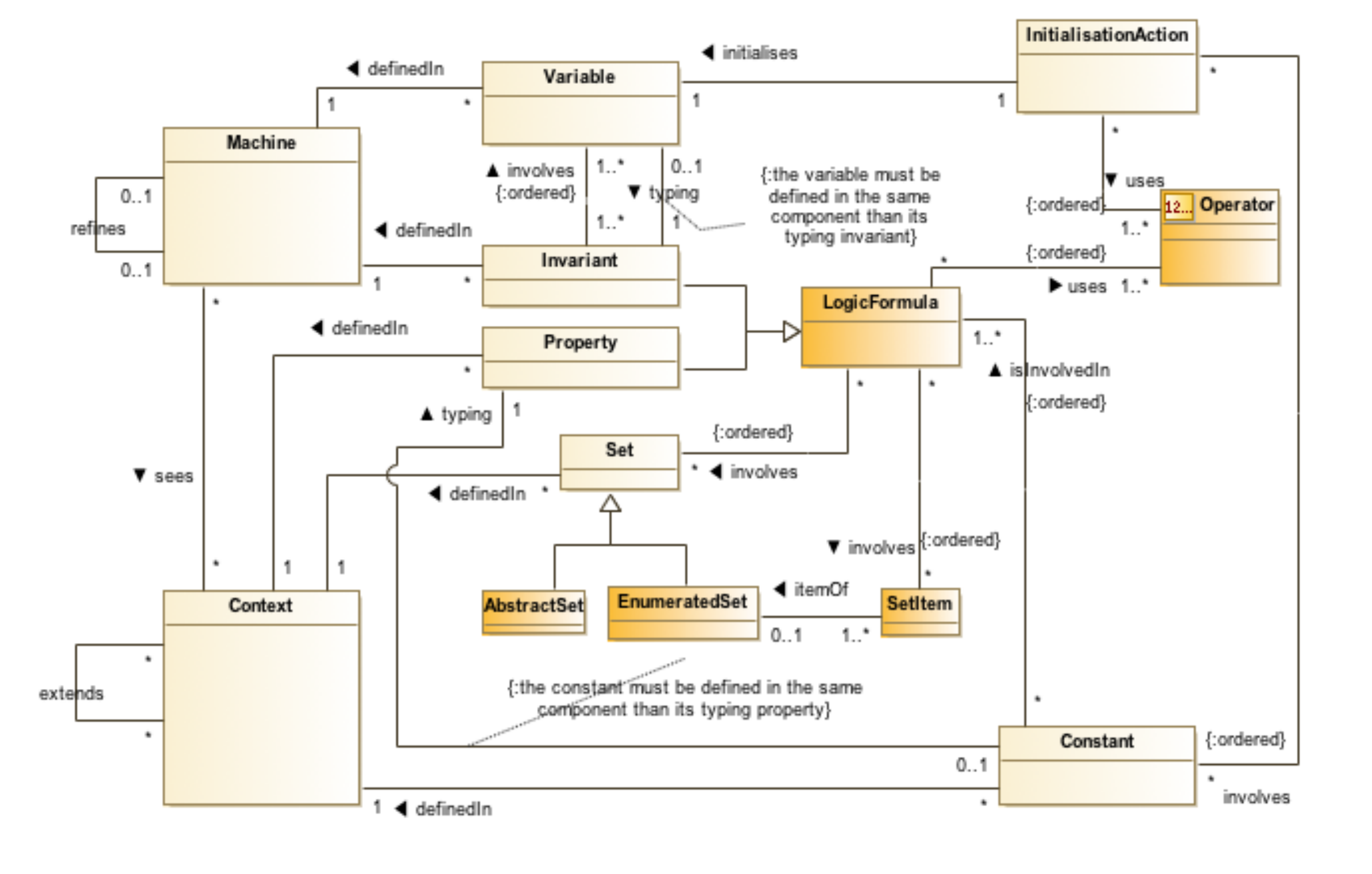}
\end{center}
\caption{\label{eventb_metamodel} Our abstraction of the Event-B specification language}
\end{figure}

 Figure \ref{eventb_metamodel} is an excerpt from our abstraction of the \textit{Event-B} specification language restricted and adjusted to fulfill the expression of our formalization rules. 
We have represented in orange some categories that do not appear explicitly in  Event-B specifications, but which will be useful to better describe our formalization rules.
An \textit{Event-B} model includes a static part called \textbf{\textit{Context}} and a dynamic part called \textbf{\textit{Machine}}.
The \textbf{\textit{context}} contains
 the definitions of abstract and enumerated sets, 
  constants and 
   properties. An enumerated set is constructed by specifying its items which are instances of \textsf{SetItem}.
 The  system state is represented in the   \textbf{\textit{machine}} using variables constrained through invariants and initialised through initialisation actions. Moreover, a machine can see contexts. Properties and invariants can be categorised as instances of \textsf{LogicFormula}. 
An instance of  \textsf{LogicFormula} consists of a certain number of operators applied, according to their order of appearance, on the operands that may be variables, constants,  sets or set items, following their associated order of appearance. 
   An instance of \textsf{InitialisationAction} references the operator and the operands of the assignment.
We describe here some operators and their actions : 
\begin{itemize}
\item[•] \textbf{\textit{Inclusion_OP}}  is used to assert that the first operand is a subset of the second operand  : \\$(Inclusion\_OP, [op_1, op_2]) \Leftrightarrow op_1 \subset op_2$.
\item[•] \textbf{\textit{Belonging_OP}}  is used to assert that the first operand is an element of the second operand  : \\$(Belonging\_OP, [op_1, op_2]) \Leftrightarrow op_1 \in op_2$. 
\item[•] \textbf{\textit{RelationSet_OP}}  is used to construct the set of relations between two operands : \\$(RelationSet\_OP, [op_1, op_2, op_3]) \Leftrightarrow op_1 =  op_2 \leftrightarrow op_3$.
\item[•] \textbf{\textit{FunctionSet_OP}}  is used to construct the set of functional relations between two operands : \\ $(FunctionSet\_OP, [op_1, op_2, op_3]) \Leftrightarrow op_1 =  op_2 \longrightarrow op_3$.
\item[•] \textbf{\textit{Maplet_OP}}  is used to construct a maplet having the operands as antecedent and image :  \\$(Maplet\_OP, [op_1, op_2, op_3]) \Leftrightarrow op_1 =  op_2 \mapsto  op_3$.
\item[•] \textbf{\textit{RelationComposition_OP}}  is used to assert that the first operand  is the result of the composition of the second operand by the third operand :  $(RelationComposition\_OP, [op_1, op_2, op_3]) \Leftrightarrow op_1 =  op_2 \circ op_3$.
\item[•] \textbf{\textit{Equal2SetOf_OP}}  is used to define the elements constituting a set :  \\$(Equal2SetOf\_OP, [op_1, op_2, ...,  op_n]) \Leftrightarrow op_1 =  \{op_2, ..., op_n\}$.
\item[•] \textbf{\textit{Inversion_OP}}  is used to assert that the first operand is the inverse of the second operand  : \\$(Inversion\_OP, [op_1, op_2]) \Leftrightarrow op_1 = op_2^{-1}$.
\item[•] \textbf{\textit{Equality_OP}}  is used to assert that the first operand is equal to the second operand  : \\$(Equality\_OP, [op_1, op_2]) \Leftrightarrow op_1 = op_2$.
\item[•] \textbf{\textit{BecomeEqual2SetOf_OP}}  is used to initialize a variable as a set of elements :  \\$(BecomeEqual2SetOf\_OP,  [va, op_2, ...,  op_n]) \Leftrightarrow va :=  \{op_2, ..., op_n\}$.
\item[•] \textbf{\textit{BecomeEqual2EmptySet_OP}}  is used to initialize a variable as an empty set :  \\$(BecomeEqual2EmptySet\_OP, [va]) \Leftrightarrow va :=  \emptyset$.
\end{itemize}
 
 The system specification can be constructed using stepwise refinement.   A machine can refine another one,  adding new events,    reducing nondeterminacy of existing events,  introducing new state variables, or   replacing abstract variables by more concrete variables.
Furthermore, a context can extend another one in order  to access the elements defined in it and to reuse them for new constructions.

\begin{figure}[!h]
\begin{center}
\includegraphics[width=0.5\textwidth]{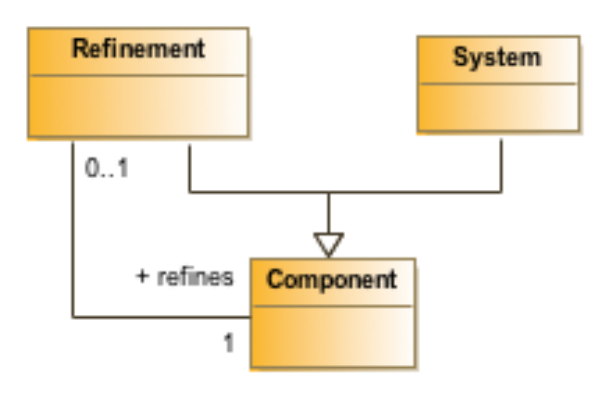}
\end{center}
\caption{\label{BSystem-EventB-min} B System Components}
\end{figure}

 In the rest of this paper, we will illustrate our formal models using \textit{B System}, an  \textit{Event-B} variant   proposed by \textit{ClearSy}, an industrial partner in the \textit{FORMOSE} project,  in its integrated development environment \textit{Atelier B} \cite{clearsy_b_system_link}. A \textit{B System} specification considers the notion of \textsf{Component} to specify machines and contexts, knowing that a component can be a system or a refinement  (figure \ref{BSystem-EventB-min}). 
 Although it is advisable  to always isolate the static and dynamic parts of the \textit{B System} formal model, it is possible to define the two parts within the same component, for simplification purposes. In the following sections, our \textit{B System} models will be presented using this facility.
  
\subsection{SysML/KAOS Requirements Engineering Method}
Requirements engineering focuses on defining and handling requirements. These and all related activities, in order to be carried out, require the choice of an adequate means for  requirements representation. The \textit{KAOS} method \cite{DBLP:books/daglib/0025377,DBLP:conf/isola/MammarL16},
   proposes to represent the requirements in the form of goals, which can be \textit{functional} or \textit{non-functional}, through five sub-models of which the two main ones are : 
 \textbf{the object model} which uses the \textit{UML} class diagram for the representation of domain  vocabulary and \textbf{the goal model} for the determination of  requirements to be satisfied by the system and of expectations with regard to the environment through a goals hierarchy.
 \textit{KAOS} proposes a structured approach to obtaining the requirements based on  expectations formulated by  stakeholders. Unfortunately, it  offers no mechanism to maintain a strong traceability between   those requirements and  deliverables associated with system  design and implementation, making it difficult to validate them against the needs formulated.
 
The \textit{SysML UML profile} has been  specially designed by the Object Management Group (OMG)  for the analysis and specification of complex systems and allows for the capturing of requirements and the maintaining of   traceability links between those requirements and  design diagrams resulting from the system design phase.
Unfortunately,  OMG has not defined a formal semantics and an unambiguous syntax for requirements specification. \textit{SysML/KAOS} \cite{DBLP:conf/inforsid/GnahoS10} therefore proposes to extend the \textit{SysML} metamodel with a set of concepts allowing to represent  
  requirements in \textit{SysML} models as \textit{KAOS}  goals.
  
\begin{figure}[!h]
\begin{center}
\includegraphics[width=0.7\textwidth]{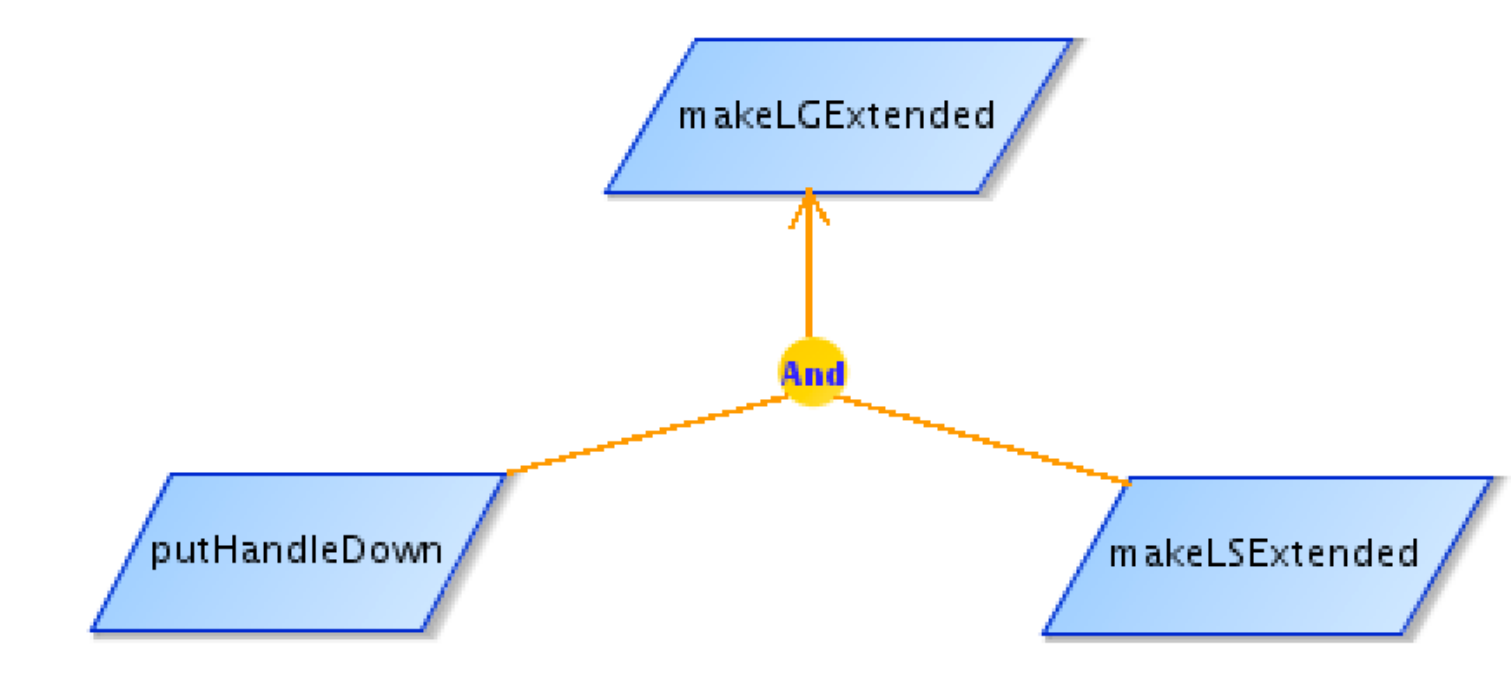}
\end{center}
\caption{\label{lgsystem_goal_model_makelgextended} Excerpt from the landing gear system goal diagram}
\end{figure}

  Figure \ref{lgsystem_goal_model_makelgextended} is an excerpt from the landing gear system \cite{Boniol2014} goal diagram focused on the purpose of  landing gear expansion.  We assume that each aircraft has one landing gear system which is equipped with three landing sets which can be each extended or retracted. We also assume that in the initial state, there is one landing gear named \textit{LG1} which is extended and is associated to one handle named \textit{HD1} which is down and to landing  sets \textit{LS1}, \textit{LS2} and \textit{LS3} which are all extended.

To achieve the root goal, which is the extension of the landing gear  (\textbf{makeLGExtended}), the handle must be put down  (\textbf{putHandleDown}) and  landing gear sets must be extended  (\textbf{makeLSExtended}).

\subsection{From SysML/KAOS Goal Model to Event-B}
The matching between \textit{SysML/KAOS} modeling and \textit{Event-B} specifications is the focus of the work done by \cite{DBLP:conf/iceccs/MatoussiGL11}.
Each layer of abstraction of the goal diagram gives rise to an \textit{Event-B} machine, each goal of the layer giving rise to an event.
The refinement links are materialized within the\textit{ Event-B} specification through a set of proof obligations and refinement links between machines and between events. Figure \ref{lgsystem_event_b_model_without_context_refinment_0} 
 represents
  the \textit{B System} specifications associated with the most abstract layer of the \textit{SysML/KAOS} goal diagram of the Landing Gear System illustrated  through
   Figure \ref{lgsystem_goal_model_makelgextended}. 
   
   \begin{figure}[!h]
   \it SYSTEM

\hspace*{0.20in}\it LandingGearSystem

\bf SETS

\bf CONSTANTS

\bf PROPERTIES

\bf VARIABLES


\bf INVARIANT


\bf INITIALISATION


\bf \it EVENTS

\hspace*{0.20in}\it makeLGExtended\rm =

\hspace*{0.20in}\textbf{BEGIN} 
/* extension of the landing gear */

\hspace*{0.20in}\bf END

\bf END
\caption{\label{lgsystem_event_b_model_without_context_refinment_0} Formalization of the root level of the Landing Gear System goal model}
\end{figure}

As we can see, the state of the system  and the body of  events must be manually completed.  The state of a system is composed of variables, constrained by an invariant, and constants, constrained by properties. 
The objective of our study is 
    to automatically derive this state in
     the \textit{Event-B}  model 
     starting from \textit{SysML/KAOS} domain models.

\subsection{SysML/KAOS Domain Modeling}
\begin{figure*}[!h]
\begin{center}
\includegraphics[width=1.12\textwidth]{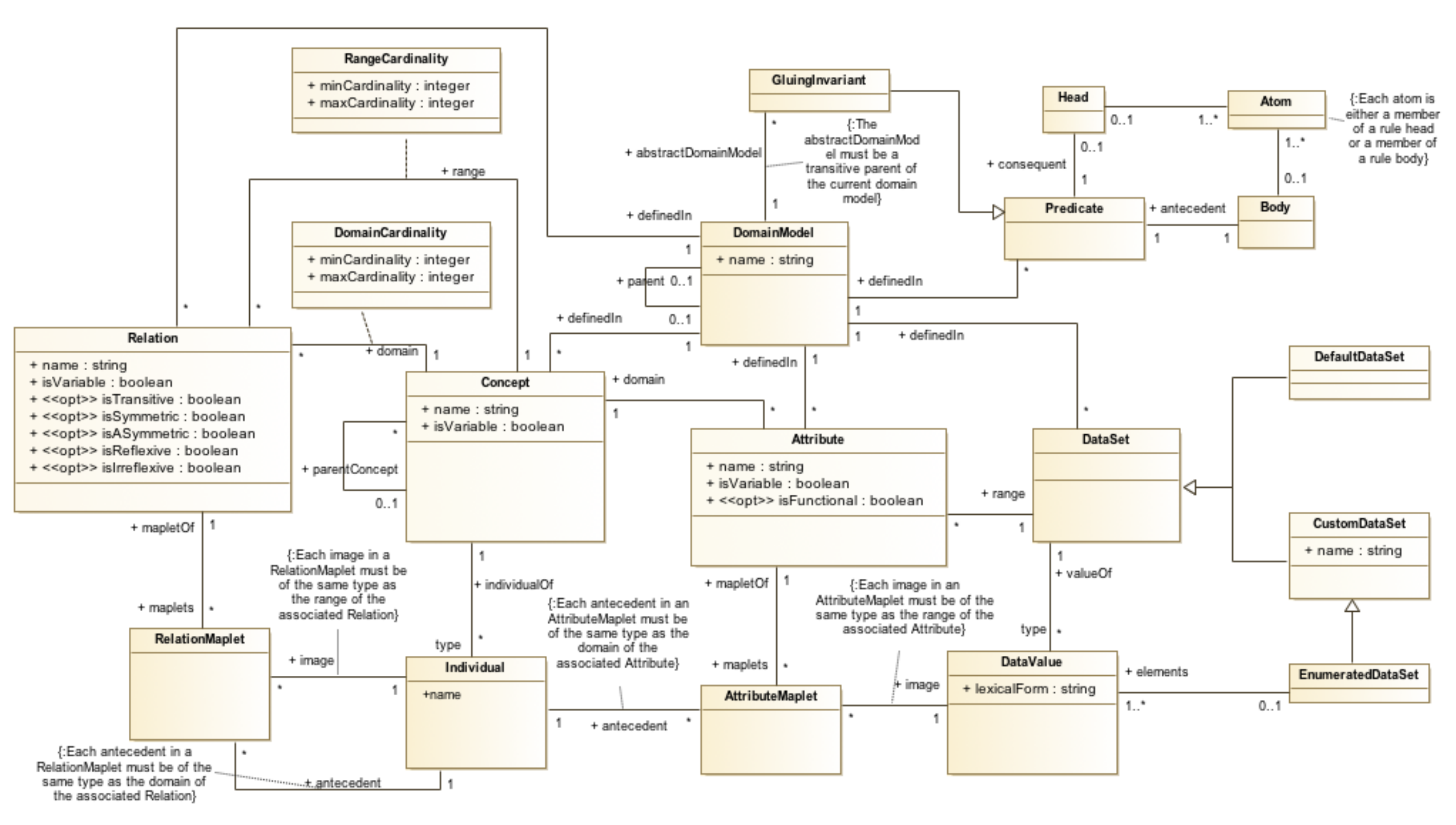}
\end{center}
\caption{\label{our_businessdomain_metamodel} Metamodel associated with SysML/KAOS domain modeling}
\end{figure*}

\begin{figure*}[!h]
\begin{center}
\includegraphics[width=1.1\textwidth]{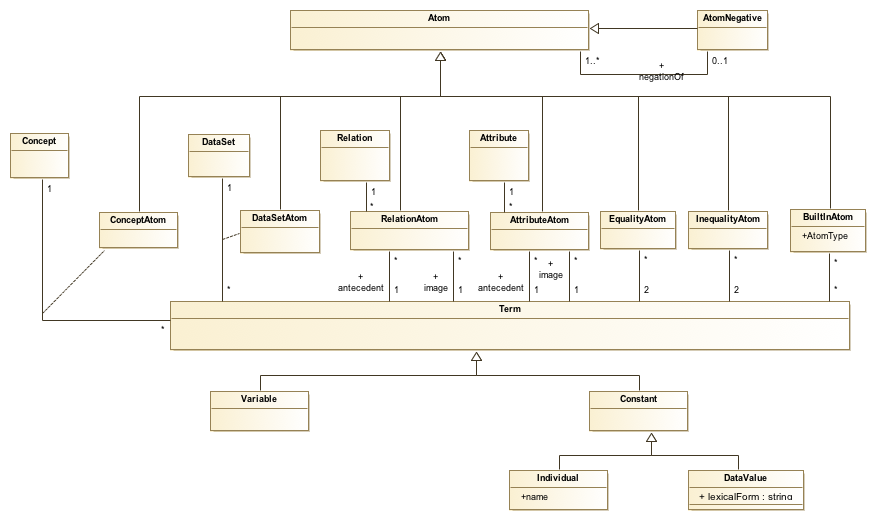}
\end{center}
\caption{\label{our_businessdomain_metamodel_bis} Extension of the metamodel associated with SysML/KAOS domain modeling for atom specification}
\end{figure*}

We present, through Figures \ref{our_businessdomain_metamodel} 
and \ref{our_businessdomain_metamodel_bis}
 the metamodel associated with the \textit{SysML/KAOS}  domain modeling approach \cite{sysml_kaos_domain_modeling} which is  an ontology modeling formalism for the modeling of domain knowledge in the framework of the \textit{SysML/KAOS} requirements engineering method.
 
\begin{figure}[!h]
\begin{center}
\includegraphics[width=0.7\textwidth]{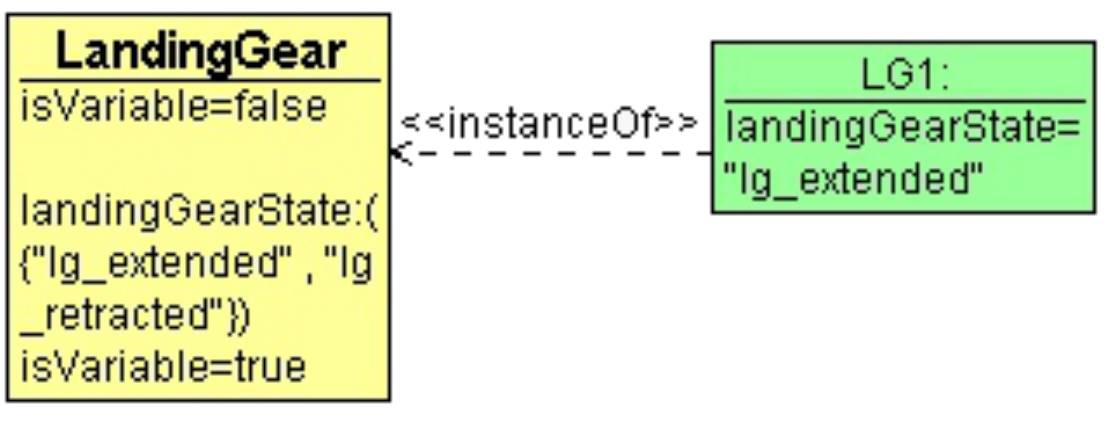}
\end{center}
\caption{\label{lgsystem_refinment_0_ontology} \textit{\textbf{lg\_system\_ref\_0}}: ontology associated to the root level of the landing gear goal model}
\end{figure}

\begin{figure}[!h]
\begin{center}
\includegraphics[width=1\textwidth]{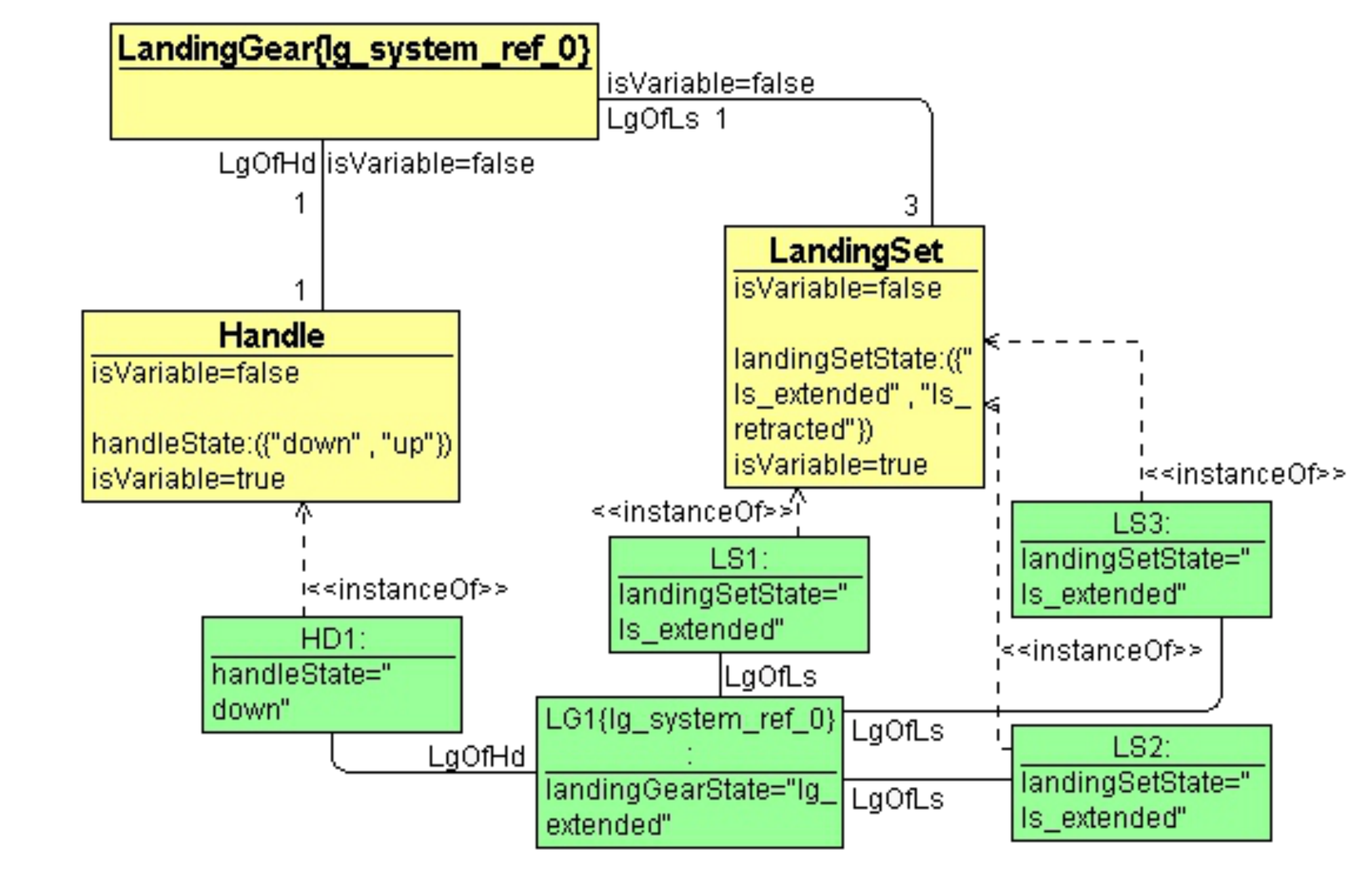}
\end{center}
\caption{\label{lgsystem_refinment_1_ontology}
\textit{\textbf{lg\_system\_ref\_1}}: ontology associated to the first level of refinement of the landing gear goal model}
\end{figure}

 Figure \ref{lgsystem_refinment_0_ontology}    represents  the \textit{SysML/KAOS} domain model associated to the root level of the landing gear system goal model of Figure \ref{lgsystem_goal_model_makelgextended},  and Figure \ref{lgsystem_refinment_1_ontology} represents   the first refinement level. They are illustrated using the syntax proposed by   \textit{OWLGred} \cite{owlgred_reference_link} and,  for readability purposes, we have decided to remove optional characteristics representation. It should be noted that the \textit{individualOf} association is illustrated by \textit{OWLGred} within the figures as a stereotyped link with the tag \textit{<<instanceOf>>}. The domain model associated to the goal diagram root level is named \textbf{\textit{lg\_system\_ref\_0}} and the one associated to the first refinement level is named \textbf{\textit{lg\_system\_ref\_1}}.

Each domain model is associated with a level of refinement of the \textit{SysML/KAOS} goal diagram and is likely to have as its parent, through the \textit{parent} association, another domain model. This allows the child domain model to access and extend some elements defined in the parent domain model.	For example, in  \textit{lg\_system\_ref\_1} (Fig. \ref{lgsystem_refinment_1_ontology}), elements defined in  \textit{lg\_system\_ref\_0} (Fig. \ref{lgsystem_refinment_0_ontology}) are imported and reused. 

A \textit{concept} (instance of metaclass \textit{Concept} of Figure \ref{our_businessdomain_metamodel})   represents a group of individuals sharing common characteristics. It can be declared \textit{variable} (\textsf{isVariable=true}) when the set of its individuals is likely to be updated through  addition or deletion of individuals. Otherwise, it is considered to be \textit{constant} (\textsf{isVariable=false}).
A concept may be associated with another, known as its parent concept, through the \textsf{parentConcept} association, from which it inherits   properties. For example, in \textit{lg\_system\_ref\_0} (Fig. \ref{lgsystem_refinment_0_ontology}),  a \textit{landing gear} is modeled as an instance of Concept named \textit{"LandingGear"}. Since it is impossible to dynamically add or remove a landing gear, the attribute \textsf{isVariable} of  \textbf{\textit{LandingGear}} is set to \textit{false}. \textit{LG1} is modeled as an instance of \textsf{Individual} (Fig. \ref{our_businessdomain_metamodel}) named \textit{"LG1"} \textsf{individual of} \textbf{\textit{LandingGear}}. 

Instances of \textsf{Relation} are used to capture  links between concepts, and instances of \textsf{Attribute} capture  links between concepts and data sets, knowing that data sets (instances of \textsf{DataSet}) are used to group data values (instances of \textsf{DataValue}) having  the same type.
The most basic way to build an instance of \textsf{DataSet} is by listing its elements. This can be done through the \textsf{DataSet} specialization   called \textsf{EnumeratedDataSet}.
A relation  or an attribute  can be declared \textit{variable} if the list of maplets related to it  is likely to change over time. Otherwise, it is considered to be \textit{constant}. Each instance of \textsf{DomainCardinality} (respectively \textsf{RangeCardinality}) makes it possible to define, for an instance of \textsf{Relation} \textit{re}, the minimum and maximum limits of the number of instances of \textsf{Individual}, having the domain (respectively range) of \textit{re} as \textit{type}, that can be put in relation with one instance of \textsf{Individual}, having the range (respectively domain) of \textit{re} as \textit{type}.
 The following constraint is associated with these limits : $(minCardinality \geq 0) \wedge (maxCardinality = * \vee maxCardinality \geq minCardinality)$, knowing that if $maxCardinality=*$, then the maximum limit is \textit{infinity}.
Instances of \textsf{RelationMaplet} are used to define associations between instances of \textsf{Individual} through  instances of \textsf{Relation}. In an identical manner, instances of \textsf{AttributeMaplet} are used to define associations between instances of \textsf{Individual}  and instances of \textsf{DataValue} through  instances of \textsf{Attribute}. 
Optional characteristics can be specified for a relation  : \textit{transitive} (\textsf{isTransitive}, default \textit{false}), \textit{symmetrical} (\textsf{isSymmetric}, default \textit{false}), \textit{asymmetrical} (\textsf{isASymmetric}, default \textit{false}), \textit{reflexive} (\textsf{isReflexive}, default \textit{false}) or \textit{irreflexive} (\textsf{isIrreflexive}, default \textit{false}). Moreover, an attribute can be \textit{functional} (\textsf{isFunctional}, default \textit{true}). For example, in \textit{lg\_system\_ref\_0} (Fig. \ref{lgsystem_refinment_0_ontology}), the possible states of a landing gear is modeled as an instance of \textsf{Attribute} named \textit{"landingGearState"}, having \textbf{\textit{LandingGear}} as domain and as range an instance of \textsf{EnumeratedDataSet} containing two instances of  \textsf{DataValue} of type \textsf{STRING}: \textbf{\textit{"lg\_extended"}} for the extended state and \textbf{\textit{"lg\_retracted"}} for the retracted state. Since it is possible to dynamically change a landing gear state, its \textsf{isVariable} attribute is set to \textit{true}.

The notion of \textsf{Predicate}   is used to represent constraints between different elements of the domain model in the form of \textit{Horn clauses}: each predicate has a body which represents its \textit{antecedent} and a head which represents its \textit{consequent}, body and head designating conjunctions of atoms.

\textsf{GluingInvariant}, specialization of \textsf{Predicate}, is used to represent links between variables and constants defined within a domain model and those appearing in more abstract domain models,  transitively linked to it through the \textit{parent} association. Gluing invariants are  extremely important because they capture relationships between abstract and concrete data during refinement which are used to discharge  proof obligations. 
The following gluing invariant is associated with our case study: if there is at least one landing set having the retracted state, then the state of LG1 is retracted.

\section{Existing Approaches for the Formalization  of Domain Models }

In \cite{DBLP:conf/birthday/BjornerE10}, domain models consist of entities and operations which can be atomic or composite.
Atomic entities correspond to states of the formal model. Composite entities correspond to sets, groups, lists or associations of entities. Furthermore, operations are translated into state-changing actions, composite operations corresponding to composition of actions.
In \cite{DBLP:conf/icfem/WangDS10}, an approach is proposed for the automatic extraction of  domain knowledge, as \textit{OWL} ontologies, from \textit{Z/Object-Z (OZ)} models \cite{DBLP:journals/stt/Doberkat01} :  OZ types and classes are transformed into OWL classes. Relations and functions are transformed into OWL properties, with the \textit{cardinality} restricted to \textit{1} for total functions and the \textit{maxCardinality} restricted to \textit{1} for partial functions. OZ constants are translated into OWL individuals. Rules are also proposed for subsets and  state schemas. Unfortunately, the approach is only interested  in static domain knowledge and it does not propose any rule regarding predicates. Furthermore,  refinement links between models are not handled.
 A similar approach is proposed in \cite{DBLP:conf/icfem/DongSW02}, for the extraction of \textit{DAML}  ontologies \cite{van2001reference} from \textit{Z} models.

An approach for generating  an \textit{Event-B} specification from an \textit{OWL}   ontology \cite{DBLP:reference/snam/SenguptaH14}   is provided in \cite{h.Alkhammash}. The proposed mapping requires the generation of an \textit{ACE (Attempto Controlled English)} version of the \textit{OWL} ontology which serves as the basis for the development of the \textit{Event-B} specification. This is done through a step called \textit{OWL verbalization}. 
The verbalization method,  proposed by \cite{h.Alkhammash}, transforms  \textit{OWL} instances into capitalized proper names, classes into common names, and properties into active and passive verbs. Once the verbalization process has been completed, \cite{h.Alkhammash} proposes a set of rules for obtaining the Event-B specification: classes are translated as \textit{Event-B} sets, properties are translated as relations, etc. In addition, \cite{h.Alkhammash} proposes rules for the \textit{Event-B} representation of property characteristics and  associations between classes or  properties. Unfortunately, the proposal makes no distinction between constant and variable : It does not specify when it is necessary to use constants or variables,  when  it is necessary to express an ontology rule as an invariant or as an axiom. Moreover, the proposal imposes a two-step sequence for the transition from an OWL ontology to an Event-B model, the first step requiring the ontology to be constructed in English. Finally, the approach does not propose anything regarding the referencing from an ontology into another one.

 In \cite{DBLP:conf/ifip2/PoernomoU09}, domain is modeled by defining agents, business entities and relations between them. The paper proposes rules for mapping domain models so designed in \textit{Event-B} specifications : agents are transformed into machines, business entities are transformed into sets, and relations are transformed into \textit{Event-B} variable relations. These rules are certainly sufficient for domain models of  interest for \cite{DBLP:conf/ifip2/PoernomoU09}, but they are very far from covering the extent of \textit{SysML/KAOS} domain modeling formalism.

In \cite{DBLP:journals/scp/AlkhammashBFC15}, domain properties are described through  data-oriented requirements for concepts, attributes and associations and through  constraint-oriented requirements for axioms. Possible states of a \textit{variable} element are represented using UML state machines.
Concepts, attributes and associations arising from  data-oriented requirements are modeled as UML class diagrams and translated to Event-B using  \textit{UML-B} \cite{Snook:2006:UFM:1125808.1125811} : 
nouns and attributes are represented as UML classes and relationships between nouns are represented as UML associations. \textit{UML-B} is also used for the translation of state machines to Event-B variables, invariants and events.
Unfortunately, constraints arising from   constraint-oriented requirements
are modeled using a semi-formal language called \textit{ Structured English}, following a method similar to the \textit{Verbalization} approach described in \cite{h.Alkhammash} and manually translated to Event-B. 
Moreover, it is impossible to rely solely on the representation of an element of the class diagram to know if its state is likely to change dynamically.
The consequence being that in an Event-B model, the same element can appear as a set, a constant or a variable
and its properties are likely to appear both in the \textit{PROPERTIES} and in the \textit{INVARIANT} clauses.

Some rules for passing from an \textit{OWL} ontology representing a domain model to \textit{Event-B} specifications are proposed through a case study in \cite{DBLP:conf/isola/MammarL16}. This case study reveals that each ontology class, having no instance, is modeled as an \textit{Event-B} abstract set. 
The others are 
modeled as an enumerated set. Finally, each object property between two classes is modeled as a constant defines  as a relation. These rules allow the generation of a first version of an \textit{Event-B} specification from  a domain model ontology. Unfortunately, the case study does not address several concerns. For example, object properties are always modeled as constants, despite the fact that they may be variable. 
Moreover, the case study  does not provide  any rule for some domain model elements such as datasets or predicates.   
In the remainder of this paper, we propose to enrich this proposal for a complete mapping of \textit{SysML/KAOS} domain models with \textit{Event-B} specifications.


\newpage
\section{SysML/KAOS Domain Model Formalization}
\begin{figure}[!h]
\begin{center}
\includegraphics[width=1\textwidth]{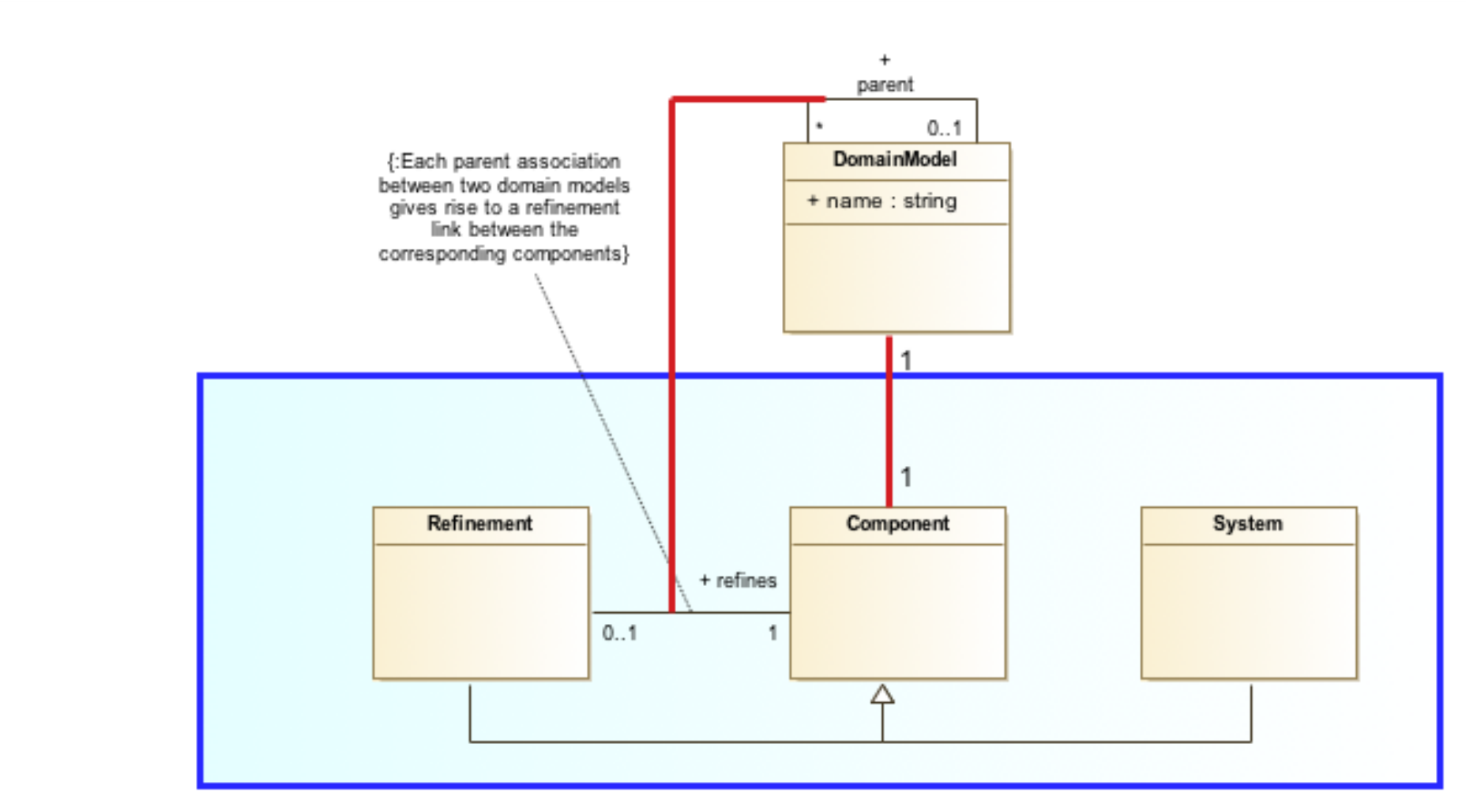}
\end{center}
\caption{\label{OurBusinesDomainModel_min_version_2_EventB_Machine_Context} Correspondence to B System Components}
\end{figure}

\begin{figure}[!h]
\begin{center}
\includegraphics[width=1.1\textwidth]{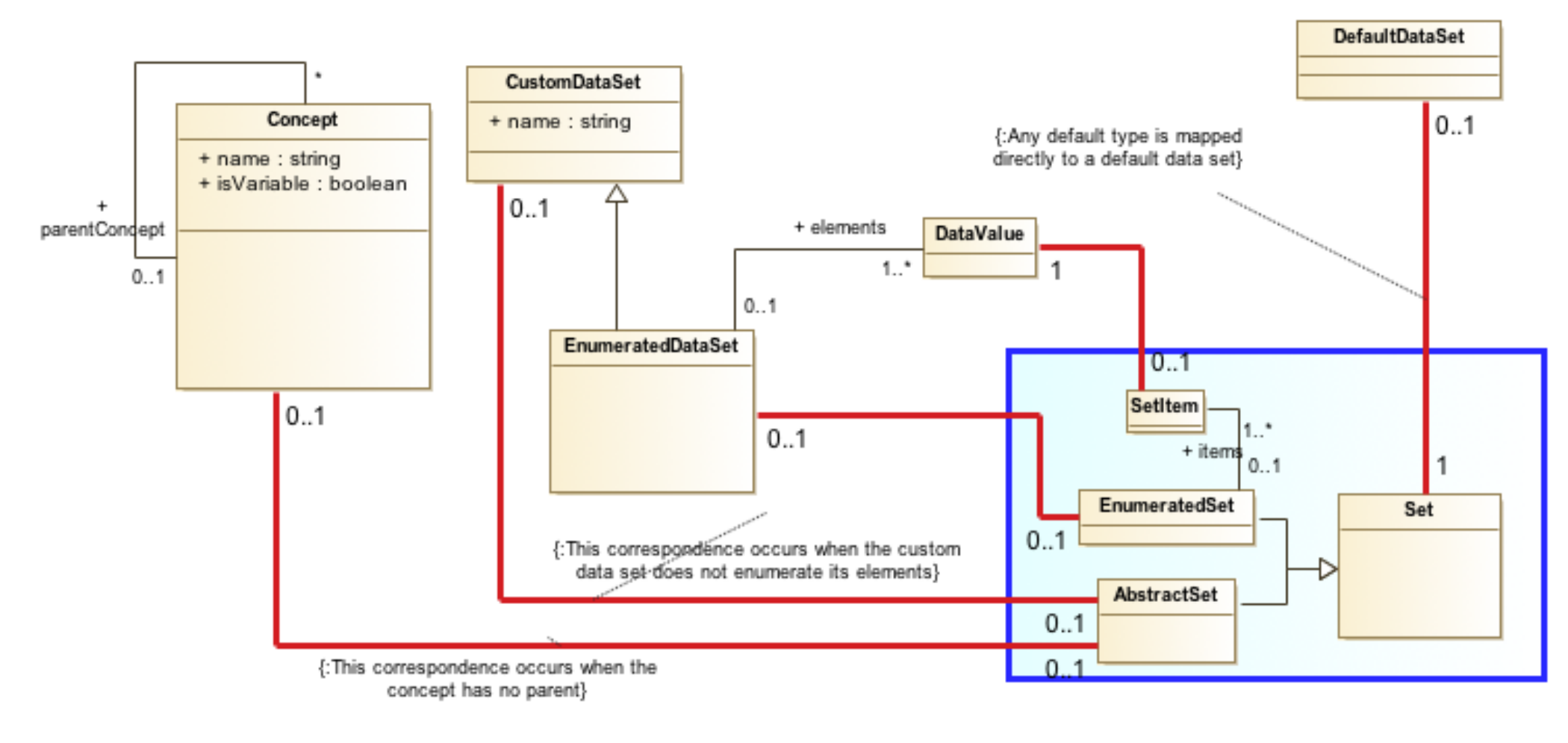}
\end{center}
\caption{\label{OurBusinesDomainModel_min_version_2_EventB_Set} Correspondence to Sets}
\end{figure}

\begin{figure}[!h]
\begin{center}
\includegraphics[width=1.1\textwidth]{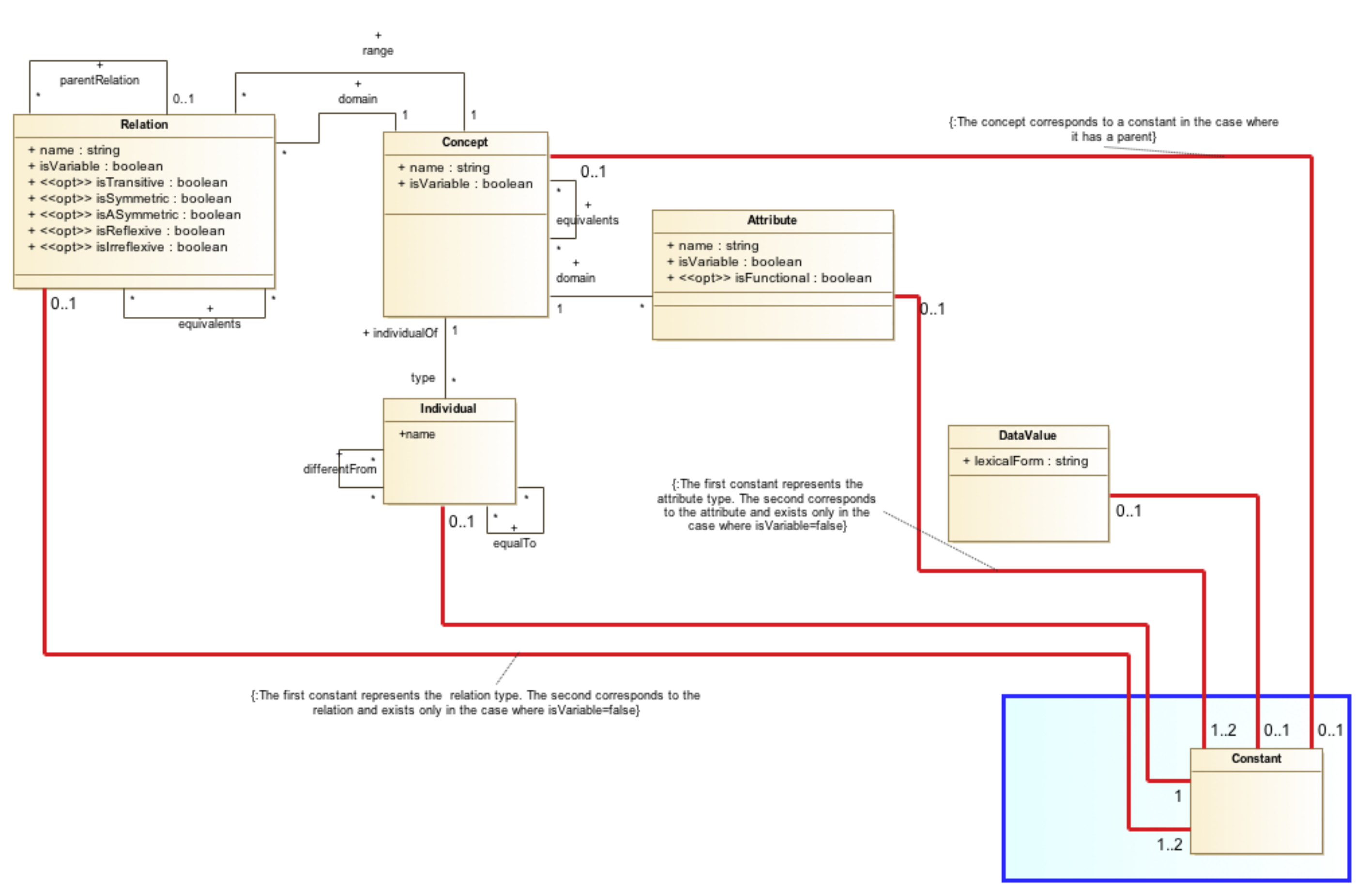}
\end{center}
\caption{\label{OurBusinesDomainModel_min_version_2_EventB_Constant} Correspondence to Constants}
\end{figure}

\begin{figure}[!h]
\begin{center}
\includegraphics[width=0.8\textwidth]{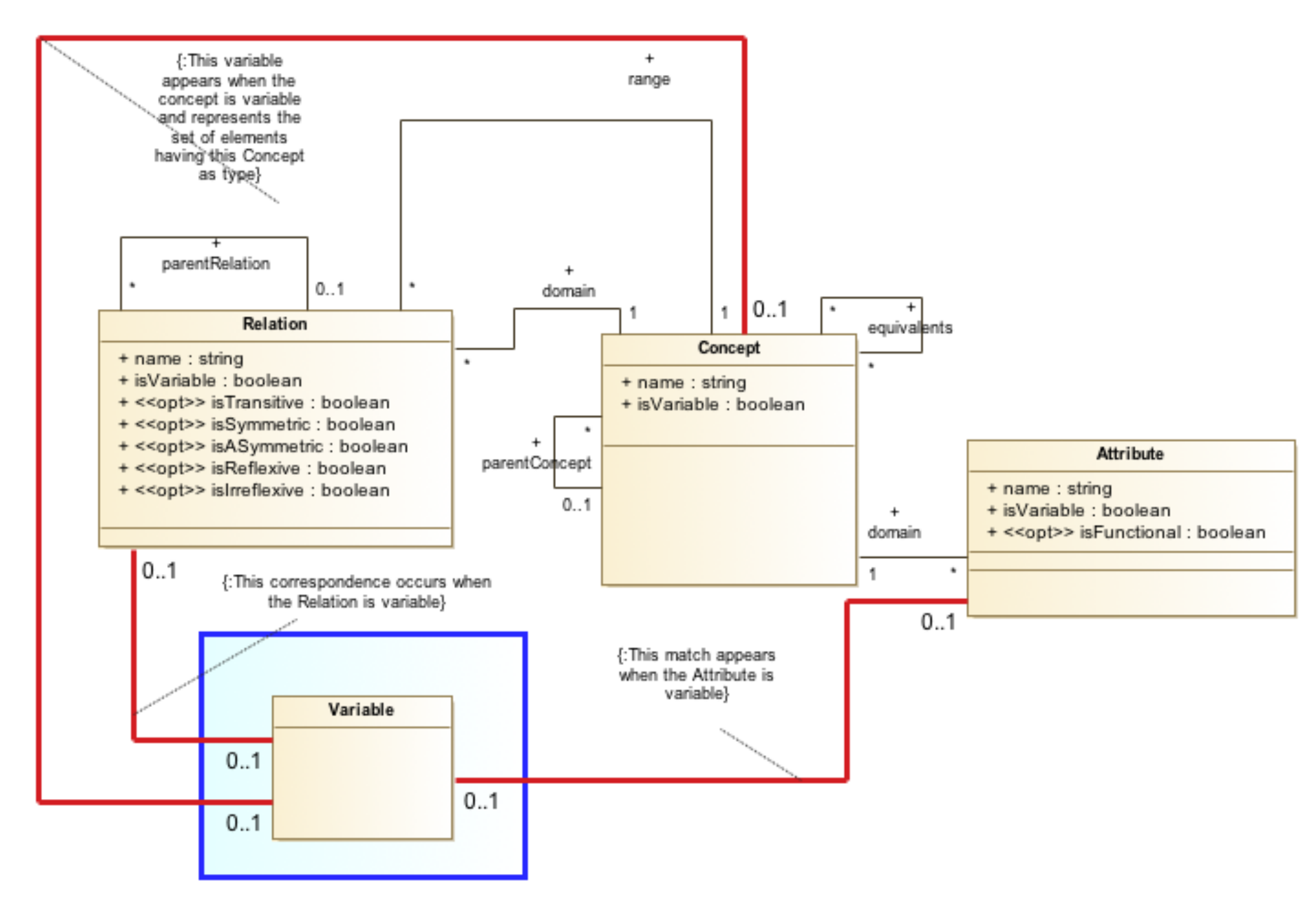}
\end{center}
\caption{\label{OurBusinesDomainModel_min_version_2_EventB_variable} Correspondence to Variables}
\end{figure}

Figures \ref{OurBusinesDomainModel_min_version_2_EventB_Machine_Context},
\ref{OurBusinesDomainModel_min_version_2_EventB_Set} , \ref{OurBusinesDomainModel_min_version_2_EventB_Constant} and \ref{OurBusinesDomainModel_min_version_2_EventB_variable}   are schematizations of  correspondence links between domain models and  B System formal models. Red links represent correspondence links, the part inside the blue rectangle representing the portion of the B System metamodel under consideration.

In the following, we describe a set of rules that allow to obtain  \textit{B System} specification from domain models associated with refinement levels of a \textit{SysML/KAOS} goal model. 
They are illustrated and have been validated using  \textbf{Event-B} : 
\begin{itemize}
\item Regarding the representation of metamodels, we have followed the rules proposed in  \cite{DBLP:conf/kbse/LaleauM00,Snook:2006:UFM:1125808.1125811} for the translation of \textit{UML} class diagrams to \textit{B} specifications:
for example, classes which are not subclasses give rise to abstract sets, each class gives rise to a variable typed as a subset and containing its instances and each association or property gives rise to a variable typed as a relation.
 For example,  \textsf{DomainModel}, \textsf{Concept}, \textsf{Relation}, \textsf{Attribute} and \textsf{DataSet} of the SysML/KAOS domain metamodel (\hyperlink{domain_metamodel_context}{ \textbf{\textit{Domain_Metamodel_Context}}}) and \textsf{Component}, \textsf{Set}, \textsf{LogicFormula} and \textsf{Variable} of the B System metamodel (\hyperlink{eventb_metamodel_context}{ \textbf{\textit{BSystem_Metamodel_Context}}}) give rise to abstract sets representing all their possible instances. Variables appear to capture, for each class, all the currently defined instances. Variables are also used to represent attributes and associations such as  \textsf{ParentConcept}, \textsf{Relation_isVariable}, \textsf{Attribute_isFunctional} of the SysML/KAOS domain metamodel and \textsf{Refines} of the B System metamodel (\hyperlink{eventb_specs_from_ontologies_machine}{\textit{\textbf{Ontologies\_BSystem\_specs\_translation}}} and \hyperlink{eventb_specs_from_ontologies_ref_1_machine}{\textit{\textbf{Ontologies\_BSystem\_specs\_\-translation\_ref\_1}}}). In case of ambiguity as to the nomenclature of an element, its name is prefixed and sufixed by that of the class to which it is attached.

\item Correspondence links between   classes are represented through variables typed as partial injections having 
the \textbf{\textit{Event-B}} representation of the first class as domain and the \textbf{\textit{Event-B}} representation of the second class as range. For example, correspondence links between instances of \textsf{Concept} and instances of \textsf{AbstractSet}  illustrated through figure \ref{OurBusinesDomainModel_min_version_2_EventB_Set}, 
are captured through a variable typed as a partial injective function between \textsf{Concept} and \textsf{AbstractSet} : $Concept\_corresp\_AbstractSet \in Concept \pinj AbstractSet$ (\hyperlink{eventb_specs_from_ontologies_ref_1_machine}{\textit{\textbf{Ontologies\_BSystem\_specs\_\-translation\_ref\_1}}}). 

\item Each rule is represented as an event by following the correspondence links.

\item Whereas no additional precision is given, we consider that all \textit{Event-B} content associated with a refinement level  is defined within a single component (SYSTEM/REFINEMENT) : it is always possible to separate it into two parts: the context for the static part (SETS, CONSTANTS and PROPERTIES) and the machine for the dynamic part (VARIABLES, INVARIANT, INITIALIZATION and EVENTS).

\end{itemize}


 Figures \ref{lgsystem_event_b_model_refinment_0} and \ref{lgsystem_event_b_model_refinment_1}  represents  respectively
  the \textit{B System} specifications associated with the 
  root level of the landing gear system domain model illustrated  through Figure \ref{lgsystem_refinment_0_ontology} and that associated with the first refinement level domain model illustrated through Figure \ref{lgsystem_refinment_1_ontology}.

\begin{figure}[!h]
    \it SYSTEM
\hspace*{0.20in}\it lg\_system\_ref\_0

\textbf{SETS}
\hspace*{0.20in} LandingGear; DataSet\_1= \{lg\_extended, lg\_retracted\}

\textbf{CONSTANTS}
\hspace*{0.20in} T\_landingGearState, LG1

 \textbf{PROPERTIES}

\textsf{(0.1)} \hspace*{0.1in} LG1 $\in$ LandingGear 

\textsf{(0.2)} \hspace*{0.10in} $\wedge$  LandingGear=\{LG1\} 

\textsf{(0.3)} \hspace*{0.10in} $\wedge$  T\_landingGearState = LandingGear $\longrightarrow$ DataSet\_1

 \textbf{VARIABLES}
\hspace*{0.20in} landingGearState


 \textbf{INVARIANT}

\textsf{(0.4)} \hspace*{0.10in} \hspace*{0.20in}  landingGearState $\in$ T\_landingGearState


 \textbf{INITIALISATION}

\textsf{(0.5)} \hspace*{0.10in} \hspace*{0.20in} landingGearState := \{LG1 $\mapsto$ lg\_extended \}


  \textbf{EVENTS}

\hspace*{0.20in} •••

END
\caption{\label{lgsystem_event_b_model_refinment_0} Formalization of the Root Level of the Landing Gear System Domain Model}
\end{figure}

\begin{figure}[!h]
    \it REFINEMENT
\hspace*{0.20in} lg\_system\_ref\_1

 \textbf{REFINES}
\hspace*{0.20in} lg\_system\_ref\_0

\textbf{SETS}
\hspace*{0.20in} Handle; LandingSet; DataSet\_2=\{ls\_extended, ls\_retracted\}; DataSet\_3=\{down, up\}

\textbf{CONSTANTS}
\hspace*{0.20in} T\_LgOfHd,  LgOfHd,  T\_LgOfLs, LgOfLs,  T\_landingSetState, T\_handleState, HD1, LS1, LS2, LS3

 \textbf{PROPERTIES}

\textsf{(1.1)} \hspace*{0.10in} HD1 $\in$ Handle 

\textsf{(1.2)} \hspace*{0.10in} $\wedge$  Handle=\{HD1\} 

\textsf{(1.3)} \hspace*{0.10in} $\wedge$  LS1 $\in$ LandingSet 

\textsf{(1.4)} \hspace*{0.10in} $\wedge$  LS2 $\in$ LandingSet 

\textsf{(1.5)} \hspace*{0.10in} $\wedge$  LS3 $\in$ LandingSet 

\textsf{(1.6)} \hspace*{0.10in} $\wedge$  LandingSet=\{LS1, LS2, LS3\} 

\textsf{(1.7)} \hspace*{0.10in} $\wedge$  T\_LgOfHd = Handle $\leftrightarrow$ LandingGear 

\textsf{(1.8)} \hspace*{0.10in} $\wedge$  LgOfHd $\in$ T\_LgOfHd 

\textsf{(1.9)} \hspace*{0.10in} $\wedge$  $\forall$ xx.(xx $\in$ Handle $\Rightarrow$ card(LgOfHd[\{xx\}])=1) 

\textsf{(1.10)} \hspace*{0.10in} $\wedge$  $\forall$xx.(xx $\in$ LandingGear $\Rightarrow$ card(LgOfHd$^{-1}$[\{xx\}])=1) 

\textsf{(1.11)} \hspace*{0.10in} $\wedge$  LgOfHd = \{HD1 $\mapsto$ LG1 \} 

\textsf{(1.12)} \hspace*{0.10in} $\wedge$  T\_LgOfLs = LandingSet $\leftrightarrow$ LandingGear 

\textsf{(1.13)} \hspace*{0.10in} $\wedge$  LgOfLs $\in$ T\_LgOfLs 

\textsf{(1.14)} \hspace*{0.10in} $\wedge$  $\forall$xx.(xx $\in$ LandingSet $\Rightarrow$ card(LgOfLs[\{xx\}])=1) 

\textsf{(1.15)} \hspace*{0.10in} $\wedge$  $\forall$xx.(xx $\in$ LandingGear $\Rightarrow$ card(LgOfLs$^{-1}$[\{xx\}])=3) 

\textsf{(1.16)} \hspace*{0.10in} $\wedge$  LgOfLs = \{LS1 $\mapsto$ LG1, LS2 $\mapsto$ LG1, LS3 $\mapsto$ LG1 \} 

\textsf{(1.17)} \hspace*{0.10in} $\wedge$ T\_landingSetState = LandingSet $\longrightarrow$ DataSet\_2 

\textsf{(1.18)} \hspace*{0.10in} $\wedge$  T\_handleState = Handle $\longrightarrow$ DataSet\_3 

 \textbf{VARIABLES}
\hspace*{0.20in} landingSetState, handleState


 \textbf{INVARIANT}

\textsf{(1.19)} \hspace*{0.10in}  landingSetState $\in$ T\_landingSetState 

\textsf{(1.20)} \hspace*{0.10in} $\wedge$   handleState $\in$ T\_handleState

\textsf{(1.21)} \hspace*{0.10in} $\wedge$ $ \forall ls. (ls \in LandingSet \wedge landingSetState(ls, ls\_extended)\Rightarrow \\ landingGearState(LG1, lg\_extended))$


 \textbf{INITIALISATION}

\textsf{(1.22)} \hspace*{0.10in} landingSetState := \{LS1 $\mapsto$ ls\_extended, LS2 $\mapsto$ ls\_extended, LS3 $\mapsto$ ls\_extended \} 

\textsf{(1.23)} \hspace*{0.10in} || \hspace*{0.20in} handleState := \{HD1 $\mapsto$ down \}


  \textbf{EVENTS}

\hspace*{0.20in} •••

END
\caption{\label{lgsystem_event_b_model_refinment_1} Formalization of the First Refinement Level of the Landing Gear System Domain Model}
\end{figure}

\newpage

\subsection{Formalization  of SysML/KAOS Domain Modeling and BSystem Formalisms in Event-B}

\hyperlink{eventb_metamodel_context}{ \textbf{\textit{BSystem_Metamodel_Context}}}
  and 
\hyperlink{domain_metamodel_context}{ \textbf{\textit{Domain_Metamodel_Context}}}  
represent respectively the context associated to 
our abstraction of the \textit{BSystem} specification language and that associated to  the SysML/KAOS Domain Metamodel.
\hyperlink{eventb_specs_from_ontologies_machine}{\textit{\textbf{Ontologies\_BSystem\_specs\_translation}}}
  and \hyperlink{eventb_specs_from_ontologies_ref_1_machine}{\textit{\textbf{Ontologies\_BSystem\_specs\_translation\_ref\_1}}}  represent the corresponding variables and the associated invariants.

\MultiHeader{CONTEXT}{Domain\_Metamodel\_Context}
\hypertarget{domain_metamodel_context}{}
\CONTEXT{Domain\_Metamodel\_Context}{}{}
\SETS{
	\Set{DomainModel\_Set}{}
	\Set{Relation\_Set}{}
	\Set{Concept\_Set}{}
	\Set{Relation\_Maplet\_Set}{}
	\Set{Individual\_Set}{}
	\Set{Attribute\_Maplet\_Set}{}
	\Set{Attribute\_Set}{}
	\Set{DataValue\_Set}{}
	\Set{DataSet\_Set}{}
	\Set{RelationCharacteristics\_Set}{}
}
\CONSTANTS{
	\Constant{\_NATURAL}{}
	\Constant{\_INTEGER}{}
	\Constant{\_FLOAT}{}
	\Constant{\_BOOL}{}
	\Constant{\_STRING}{}
	\Constant{isTransitive}{}
	\Constant{isSymmetric}{}
}
\AXIOMS{
	\Axiom{axiom1}{false}{$finite(DataValue\_Set)$}{}
	\Axiom{axiom2}{false}{$\{\_NATURAL, \_INTEGER, \_FLOAT, \_BOOL, \_STRING\} \subseteq{}  DataSet\_Set$}{}
	\Axiom{axiom3}{false}{$partition(\{\_NATURAL, \_INTEGER, \_FLOAT, \\\_BOOL, \_STRING\}, \{\_NATURAL\}, \{\_INTEGER\}, \{\_FLOAT\}, \{\_BOOL\}, \{\_STRING\})$}{}
	\Axiom{axiom4}{false}{$partition(RelationCharacteristics\_Set,\{isTransitive\}, \{isSymmetric\})$}{}
}
\END
\MultiHeader{CONTEXT}{BSystem\_Metamodel\_Context}
\hypertarget{eventb_metamodel_context}{}
\CONTEXT{BSystem\_Metamodel\_Context}{}{}
\SETS{
	\Set{Component\_Set}{}
	\Set{Variable\_Set}{}
	\Set{Constant\_Set}{}
	\Set{Set\_Set}{}
	\Set{SetItem\_Set}{}
	\Set{LogicFormula\_Set}{\\the subset of logical formulas that can directly be expressed within the specification,\\~       without the need for an explicit constructor, will not be contained in this set.\\~       This is for example the case of equality between elements.\\~        }
	\Set{Operator}{}
	\Set{InitialisationAction\_Set}{}
}
\CONSTANTS{
	\Constant{B\_NATURAL}{}
	\Constant{B\_INTEGER}{}
	\Constant{B\_FLOAT}{}
	\Constant{B\_BOOL}{}
	\Constant{B\_STRING}{}
	\Constant{Inclusion\_OP}{}
	\Constant{Belonging\_OP}{}
	\Constant{BecomeEqual2SetOf\_OP}{}
	\Constant{RelationSet\_OP}{}
	\Constant{FunctionSet\_OP}{}
	\Constant{Maplet\_OP}{}
	\Constant{Equal2SetOf\_OP}{}
	\Constant{BecomeEqual2EmptySet\_OP}{}
	\Constant{RelationComposition\_OP}{}
	\Constant{Inversion\_OP}{}
	\Constant{Equality\_OP}{}
}
\AXIOMS{
	\Axiom{axiom1}{false}{$finite(SetItem\_Set)$}{}
	\Axiom{axiom2}{false}{$\{B\_NATURAL, B\_INTEGER, B\_FLOAT, B\_BOOL, B\_STRING\} \subseteq{}  Set\_Set$}{}
	\Axiom{axiom3}{false}{$partition(\{B\_NATURAL, B\_INTEGER, B\_FLOAT, B\_BOOL, B\_STRING\}, \{B\_NATURAL\}, \\\{B\_INTEGER\}, \{B\_FLOAT\}, \{B\_BOOL\}, \{B\_STRING\})$}{}
	\Axiom{axiom4}{false}{$partition(Operator, \{Inclusion\_OP\}, \{Belonging\_OP\}, \{BecomeEqual2SetOf\_OP\}, \{RelationSet\_OP\}, \\\{Maplet\_OP\}, \{Equal2SetOf\_OP\}, \{BecomeEqual2EmptySet\_OP\}, \{FunctionSet\_OP\}, \{RelationComposition\_OP\}, \\\{Inversion\_OP\}, \{Equality\_OP\})$}{}
}
\END
\MultiHeader{MACHINE}{Ontologies\_BSystem\_specs\_translation}
\hypertarget{eventb_specs_from_ontologies_machine}{}
\MACHINE{Ontologies\_BSystem\_specs\_translation}{}{BSystem\_Metamodel\_Context,Domain\_Metamodel\_Context}{}
\VARIABLES{
	\Variable{Component}{}
	\Variable{System}{}
	\Variable{Refinement}{Event-B associations}
	\Variable{Refinement\_refines\_Component}{Domain Model sets}
	\Variable{DomainModel}{Domain Model associations}
	\Variable{DomainModel\_parent\_DomainModel}{correspondences}
	\Variable{DomainModel\_corresp\_Component}{}
}
\INVARIANTS{
	\Invariant{inv0\_1}{false}{$Component\subseteq{}Component\_Set$}{}
	\Invariant{inv0\_2}{false}{$partition(Component, System, Refinement)$}{\\Domain Model}
	\Invariant{inv0\_3}{false}{$DomainModel \subseteq{} DomainModel\_Set$}{}
	\Invariant{inv0\_4}{false}{$DomainModel\_parent\_DomainModel \in{} DomainModel \pinj{} DomainModel$}{}
	\Invariant{inv0\_5}{false}{$DomainModel\_corresp\_Component \in{} DomainModel \pinj{} Component$}{}
	\Invariant{inv0\_6}{false}{$Refinement\_refines\_Component  \in{} Refinement \tinj{} Component$}{}
	\Invariant{inv0\_7}{false}{$\\\forall{}xx\qdot{}(\\~  				\forall{}px\qdot{}(\\~  					(\\~  						xx\in{}dom(DomainModel\_parent\_DomainModel)\\~  						\land{} px=DomainModel\_parent\_DomainModel(xx)\\~  						\land{} px \in{} dom(DomainModel\_corresp\_Component)\\~  						\land{} xx \notin{} dom(DomainModel\_corresp\_Component)\\~  					)\\~  					\limp{}DomainModel\_corresp\_Component(px) \notin{} ran(Refinement\_refines\_Component)\\~  				)\\~  			)$}{}
}
•••
\END
\MultiHeader{MACHINE}{Ontologies\_BSystem\_specs\_translation\_ref\_1}
\MACHINE{Ontologies\_BSystem\_specs\_translation\_ref\_1}{Ontologies\_BSystem\_specs\_translation}{BSystem\_Metamodel\_Context,Domain\_Metamodel\_Context}{}
\VARIABLES{
	\Variable{DomainModel}{}
	\Variable{DomainModel\_parent\_DomainModel}{}
	\Variable{Variable}{}
	\Variable{Constant}{}
	\Variable{Set}{}
	\Variable{SetItem}{}
	\Variable{AbstractSet}{}
	\Variable{EnumeratedSet}{}
	\Variable{Invariant}{}
	\Variable{Property}{}
	\Variable{LogicFormula}{}
	\Variable{InitialisationAction}{\\Event-B associations}
	\Variable{Variable\_definedIn\_Component}{}
	\Variable{Constant\_definedIn\_Component}{}
	\Variable{Set\_definedIn\_Component}{}
	\Variable{LogicFormula\_definedIn\_Component}{}
	\Variable{Invariant\_involves\_Variables}{}
	\Variable{Constant\_isInvolvedIn\_LogicFormulas}{}
	\Variable{LogicFormula\_involves\_Sets}{}
	\Variable{LogicFormula\_involves\_SetItems}{}
	\Variable{LogicFormula\_uses\_Operators}{}
	\Variable{Variable\_typing\_Invariant}{}
	\Variable{Constant\_typing\_Property}{}
	\Variable{SetItem\_itemOf\_EnumeratedSet}{}
	\Variable{InitialisationAction\_uses\_Operators}{}
	\Variable{Variable\_init\_InitialisationAction}{}
	\Variable{InitialisationAction\_involves\_Constants}{\\Domain Model sets}
	\Variable{Concept}{}
	\Variable{Individual}{}
	\Variable{DataValue}{}
	\Variable{DataSet}{}
	\Variable{DefaultDataSet}{}
	\Variable{CustomDataSet}{}
	\Variable{EnumeratedDataSet}{\\*************relations/attributes**************}
	\Variable{Relation}{}
	\Variable{RelationMaplet}{}
	\Variable{AttributeMaplet}{}
	\Variable{Attribute}{Domain Model attributes}
	\Variable{Concept\_isVariable}{\\*************relations/attributes****************}
	\Variable{Relation\_isVariable}{}
	\Variable{Relation\_isTransitive}{}
	\Variable{Relation\_isSymmetric}{}
	\Variable{relation\_isASymmetric}{}
	\Variable{Relation\_isReflexive}{}
	\Variable{Relation\_isIrreflexive}{}
	\Variable{Attribute\_isVariable}{}
	\Variable{Attribute\_isFunctional}{\\Domain Model associations}
	\Variable{Concept\_definedIn\_DomainModel}{}
	\Variable{DataSet\_definedIn\_DomainModel}{}
	\Variable{Concept\_parentConcept\_Concept}{}
	\Variable{Individual\_individualOf\_Concept}{}
	\Variable{DataValue\_valueOf\_DataSet}{}
	\Variable{DataValue\_elements\_EnumeratedDataSet}{}
	\Variable{Relation\_definedIn\_DomainModel}{}
	\Variable{Attribute\_definedIn\_DomainModel}{\\*************relations/attributes***********************}
	\Variable{Relation\_domain\_Concept}{}
	\Variable{Relation\_range\_Concept}{}
	\Variable{Relation\_DomainCardinality\_minCardinality}{}
	\Variable{Relation\_DomainCardinality\_maxCardinality}{}
	\Variable{Relation\_RangeCardinality\_minCardinality}{}
	\Variable{Relation\_RangeCardinality\_maxCardinality}{}
	\Variable{RelationMaplet\_mapletOf\_Relation}{}
	\Variable{RelationMaplet\_antecedent\_Individual}{}
	\Variable{RelationMaplet\_image\_Individual}{}
	\Variable{Attribute\_domain\_Concept}{}
	\Variable{Attribute\_range\_DataSet}{}
	\Variable{AttributeMaplet\_mapletOf\_Attribute}{}
	\Variable{AttributeMaplet\_antecedent\_Individual}{}
	\Variable{AttributeMaplet\_image\_DataValue}{\\correspondences}
	\Variable{Concept\_corresp\_AbstractSet}{}
	\Variable{DomainModel\_corresp\_Component}{}
	\Variable{EnumeratedDataSet\_corresp\_EnumeratedSet}{}
	\Variable{DataValue\_corresp\_SetItem}{}
	\Variable{CustomDataSet\_corresp\_AbstractSet}{}
	\Variable{DefaultDataSet\_corresp\_AbstractSet}{}
	\Variable{Concept\_corresp\_Constant}{}
	\Variable{Individual\_corresp\_Constant}{}
	\Variable{DataValue\_corresp\_Constant}{}
	\Variable{Concept\_corresp\_Variable}{\\*************relations/attributes****************************}
	\Variable{Relation\_Type}{}
	\Variable{Relation\_corresp\_Constant}{}
	\Variable{Relation\_corresp\_Variable}{}
	\Variable{Attribute\_Type}{}
	\Variable{Attribute\_corresp\_Constant}{}
	\Variable{Attribute\_corresp\_Variable}{}
	\Variable{RelationCharacteristic\_corresp\_LogicFormula}{}
	\Variable{RelationMaplet\_corresp\_Constant}{}
	\Variable{DataSet\_corresp\_Set}{}
	\Variable{AttributeMaplet\_corresp\_Constant}{}
}
\INVARIANTS{
	\Invariant{inv1\_1}{false}{$Variable\subseteq{}Variable\_Set$}{}
	\Invariant{inv1\_2}{false}{$Constant\subseteq{}Constant\_Set$}{}
	\Invariant{inv1\_3}{false}{$Set\subseteq{}Set\_Set$}{}
	\Invariant{inv1\_4}{false}{$partition(Set, AbstractSet, EnumeratedSet)$}{}
	\Invariant{inv1\_5}{false}{$SetItem\subseteq{}SetItem\_Set$}{}
	\Invariant{inv1\_6}{false}{$Variable\_definedIn\_Component \in{}  Variable \tfun{} Component$}{}
	\Invariant{inv1\_7}{false}{$Constant\_definedIn\_Component \in{}  Constant \tfun{} Component$}{}
	\Invariant{inv1\_8}{false}{$Set\_definedIn\_Component \in{}  Set \tfun{} Component$}{}
	\Invariant{inv1\_9}{false}{$SetItem\_itemOf\_EnumeratedSet \in{} SetItem \psur{} EnumeratedSet$}{\\Domain Model}
	\Invariant{inv1\_10}{false}{$Concept \subseteq{} Concept\_Set$}{}
	\Invariant{inv1\_11}{false}{$Individual \subseteq{} Individual\_Set$}{}
	\Invariant{inv1\_12}{false}{$DataValue \subseteq{} DataValue\_Set$}{}
	\Invariant{inv1\_13}{false}{$DataSet \subseteq{} DataSet\_Set$}{}
	\Invariant{inv1\_14}{false}{$partition(DataSet, DefaultDataSet, CustomDataSet)$}{}
	\Invariant{inv1\_15}{false}{$EnumeratedDataSet \subseteq{} CustomDataSet$}{}
	\Invariant{inv1\_16}{false}{$Concept\_isVariable  \in{} Concept \tfun{} BOOL$}{}
	\Invariant{inv1\_17}{false}{$Concept\_definedIn\_DomainModel \in{}  Concept \tfun{} DomainModel$}{}
	\Invariant{inv1\_18}{false}{$DataSet\_definedIn\_DomainModel \in{}  DataSet \tfun{} DomainModel$}{}
	\Invariant{inv1\_19}{false}{$Concept\_parentConcept\_Concept \in{} Concept \pfun{} Concept$}{}
	\Invariant{inv1\_20}{false}{$Individual\_individualOf\_Concept \in{} Individual \tfun{} Concept$}{}
	\Invariant{inv1\_21}{false}{$DataValue\_valueOf\_DataSet \in{} DataValue \tfun{}  DataSet$}{}
	\Invariant{inv1\_22}{false}{$DataValue\_elements\_EnumeratedDataSet \in{} DataValue \psur{}  EnumeratedDataSet$}{}
	\Invariant{inv1\_23}{false}{$Concept\_corresp\_AbstractSet \in{} Concept \pinj{} AbstractSet$}{}
	\Invariant{inv1\_24}{false}{$EnumeratedDataSet\_corresp\_EnumeratedSet \in{} EnumeratedDataSet \pinj{} EnumeratedSet$}{}
	\Invariant{inv1\_25}{false}{$DataValue\_corresp\_SetItem \in{} DataValue \pinj{} SetItem$}{}
	\Invariant{inv1\_26}{false}{$\forall{}xx\qdot{}(xx\in{}EnumeratedDataSet \land{} xx \notin{} dom(EnumeratedDataSet\_corresp\_EnumeratedSet)\limp{}DataValue\_elements\_EnumeratedDataSet\converse{}[\{xx\}]\binter{}dom(DataValue\_corresp\_SetItem)=\emptyset{})$}{}
	\Invariant{inv1\_27}{false}{$CustomDataSet\_corresp\_AbstractSet \in{} CustomDataSet \pinj{} AbstractSet$}{}
	\Invariant{inv1\_28}{false}{$\{\_NATURAL, \_INTEGER, \_FLOAT, \_BOOL, \_STRING\} \binter{}  CustomDataSet = \emptyset{}$}{}
	\Invariant{inv1\_29}{false}{$DefaultDataSet\_corresp\_AbstractSet \in{} DefaultDataSet \pinj{} AbstractSet$}{}
	\Invariant{inv1\_30}{false}{$\{B\_NATURAL, B\_INTEGER, B\_FLOAT, B\_BOOL, B\_STRING\} \binter{}  EnumeratedSet = \emptyset{}$}{}
	\Invariant{inv1\_31}{false}{$Concept\_corresp\_Constant \in{} Concept \pinj{} Constant$}{}
	\Invariant{inv1\_33}{false}{$LogicFormula \subseteq{} LogicFormula\_Set$}{}
	\Invariant{inv1\_34}{false}{$Property \subseteq{} LogicFormula$}{}
	\Invariant{inv1\_35}{false}{$Invariant \subseteq{} LogicFormula$}{}
	\Invariant{inv1\_36}{false}{$LogicFormula\_definedIn\_Component \in{}  LogicFormula \tfun{} Component$}{}
	\Invariant{inv1\_37}{false}{$Invariant\_involves\_Variables \in{}  Invariant \tfun{} (\natn{}\pfun{}Variable)$}{\\logic formula operands can be variables, constants, sets or set items, indexed by their appearance order number. The first operand is indexed by 1, no matter it's type.}
	\Invariant{inv1\_38}{false}{$ran(union(ran(Invariant\_involves\_Variables)))=Variable$}{}
	\Invariant{inv1\_39}{false}{$Constant\_isInvolvedIn\_LogicFormulas \in{}  Constant \tfun{} \pown{}(\natn{}\cprod{}LogicFormula)$}{\\When appearance order does not matter, we may index all constants using the same number.}
	\Invariant{inv1\_40}{false}{$\forall{}cons\qdot{}(cons\in{}Constant\limp{}ran(Constant\_isInvolvedIn\_LogicFormulas(cons))\binter{}Property\neq{}\emptyset{})$}{}
	\Invariant{inv1\_41}{false}{$LogicFormula\_involves\_Sets \in{}  LogicFormula \tfun{} (\natn{}\pfun{}Set)$}{}
	\Invariant{inv1\_42}{false}{$LogicFormula\_uses\_Operators \in{}  LogicFormula \tfun{} (\natn{}\pfun{}Operator)$}{}
	\Invariant{inv1\_44}{false}{$Individual\_corresp\_Constant \in{} Individual \pinj{} Constant$}{}
	\Invariant{inv1\_45}{false}{$DataValue\_corresp\_Constant \in{} DataValue \pinj{} Constant$}{}
	\Invariant{inv1\_46}{false}{$Concept\_corresp\_Variable \in{} Concept \pinj{} Variable$}{}
	\Invariant{inv1\_47}{false}{$InitialisationAction\subseteq{} InitialisationAction\_Set$}{}
	\Invariant{inv1\_49}{false}{$InitialisationAction\_uses\_Operators \in{}  InitialisationAction \tfun{} (\natn{}\pfun{}Operator)$}{}
	\Invariant{inv1\_50}{false}{$Variable\_init\_InitialisationAction \in{}  Variable \tbij{} InitialisationAction$}{\\for initialisation actions, the assigned operand is the involved variable.}
	\Invariant{inv1\_52}{false}{$InitialisationAction\_involves\_Constants \in{}  InitialisationAction \tfun{} (\natn{}\pfun{}Constant)$}{\\*************relations/attributes**********************************************************************}
	\Invariant{inv1\_53}{false}{$Relation \subseteq{} Relation\_Set$}{}
	\Invariant{inv1\_56}{false}{$RelationMaplet \subseteq{} Relation\_Maplet\_Set$}{}
	\Invariant{inv1\_57}{false}{$AttributeMaplet \subseteq{} Attribute\_Maplet\_Set$}{}
	\Invariant{inv1\_58}{false}{$Attribute \subseteq{} Attribute\_Set$}{}
	\Invariant{inv1\_59}{false}{$Relation\_isVariable  \in{} Relation \tfun{} BOOL$}{}
	\Invariant{inv1\_60}{false}{$Relation\_isTransitive  \in{} Relation \pfun{} BOOL$}{}
	\Invariant{inv1\_61}{false}{$Relation\_isSymmetric  \in{} Relation \pfun{} BOOL$}{}
	\Invariant{inv1\_62}{false}{$relation\_isASymmetric  \in{} Relation \pfun{} BOOL$}{}
	\Invariant{inv1\_63}{false}{$Relation\_isReflexive  \in{} Relation \pfun{} BOOL$}{}
	\Invariant{inv1\_64}{false}{$Relation\_isIrreflexive  \in{} Relation \pfun{} BOOL$}{}
	\Invariant{inv1\_65}{false}{$Relation\_DomainCardinality\_minCardinality  \in{} Relation \pfun{} \nat{}$}{}
	\Invariant{inv1\_66}{false}{$Relation\_DomainCardinality\_maxCardinality  \in{} Relation \pfun{} (\nat{} \bunion{} \{-1\})$}{}
	\Invariant{inv1\_67}{false}{$Relation\_RangeCardinality\_minCardinality  \in{} Relation \pfun{} \nat{}$}{}
	\Invariant{inv1\_68}{false}{$Relation\_RangeCardinality\_maxCardinality  \in{} Relation \pfun{} (\nat{} \bunion{} \{-1\})$}{}
	\Invariant{inv1\_69}{false}{$Attribute\_isVariable  \in{} Attribute \tfun{} BOOL$}{}
	\Invariant{inv1\_70}{false}{$Attribute\_isFunctional  \in{} Attribute \pfun{} BOOL$}{}
	\Invariant{inv1\_71}{false}{$Relation\_definedIn\_DomainModel \in{}  Relation \tfun{} DomainModel$}{}
	\Invariant{inv1\_72}{false}{$Attribute\_definedIn\_DomainModel \in{}  Attribute \tfun{} DomainModel$}{}
	\Invariant{inv1\_73}{false}{$Relation\_domain\_Concept \in{}  Relation \tfun{} Concept$}{}
	\Invariant{inv1\_74}{false}{$Relation\_range\_Concept \in{}  Relation \tfun{} Concept$}{}
	\Invariant{inv1\_77}{false}{$RelationMaplet\_mapletOf\_Relation \in{}  RelationMaplet \tfun{} Relation$}{}
	\Invariant{inv1\_78}{false}{$RelationMaplet\_antecedent\_Individual \in{}  RelationMaplet \tfun{} Individual$}{}
	\Invariant{inv1\_79}{false}{$RelationMaplet\_image\_Individual \in{}  RelationMaplet \tfun{} Individual$}{}
	\Invariant{inv1\_80}{false}{$Attribute\_domain\_Concept \in{}  Attribute \tfun{} Concept$}{}
	\Invariant{inv1\_81}{false}{$Attribute\_range\_DataSet \in{}  Attribute \tfun{} DataSet$}{}
	\Invariant{inv1\_82}{false}{$AttributeMaplet\_mapletOf\_Attribute \in{}  AttributeMaplet \tfun{} Attribute$}{}
	\Invariant{inv1\_83}{false}{$AttributeMaplet\_antecedent\_Individual \in{}  AttributeMaplet \tfun{} Individual$}{}
	\Invariant{inv1\_84}{false}{$AttributeMaplet\_image\_DataValue \in{}  AttributeMaplet \tfun{} DataValue$}{}
	\Invariant{inv1\_85}{false}{$\forall{}rm\qdot{}(rm\in{}RelationMaplet\limp{}Individual\_individualOf\_Concept(RelationMaplet\_antecedent\_Individual(rm))=Relation\_domain\_Concept(RelationMaplet\_mapletOf\_Relation(rm)))$}{}
	\Invariant{inv1\_86}{false}{$\forall{}rm\qdot{}(rm\in{}RelationMaplet\limp{}Individual\_individualOf\_Concept(RelationMaplet\_image\_Individual(rm))=Relation\_range\_Concept(RelationMaplet\_mapletOf\_Relation(rm)))$}{}
	\Invariant{inv1\_87}{false}{$\forall{}am\qdot{}(am\in{}AttributeMaplet\limp{}Individual\_individualOf\_Concept(AttributeMaplet\_antecedent\_Individual(am))=Attribute\_domain\_Concept(AttributeMaplet\_mapletOf\_Attribute(am)))$}{}
	\Invariant{inv1\_88}{false}{$\forall{}am\qdot{}(am\in{}AttributeMaplet\limp{}DataValue\_valueOf\_DataSet(AttributeMaplet\_image\_DataValue(am))=Attribute\_range\_DataSet(AttributeMaplet\_mapletOf\_Attribute(am)))$}{}
	\Invariant{inv1\_89}{false}{$Relation\_Type \in{} Relation \pinj{} Constant$}{}
	\Invariant{inv1\_90}{false}{$Relation\_corresp\_Constant \in{} Relation \pinj{} Constant$}{}
	\Invariant{inv1\_91}{false}{$Relation\_corresp\_Variable \in{} Relation \pinj{} Variable$}{}
	\Invariant{inv1\_92}{false}{$\forall{} re\qdot{}(re\in{}dom(Relation\_Type)\leqv{}(re\in{}dom(Relation\_corresp\_Constant) \lor{} (re\in{}dom(Relation\_corresp\_Variable))))$}{}
	\Invariant{inv1\_93}{false}{$Attribute\_Type \in{} Attribute \pinj{} Constant$}{}
	\Invariant{inv1\_94}{false}{$Attribute\_corresp\_Constant \in{} Attribute \pinj{} Constant$}{}
	\Invariant{inv1\_95}{false}{$Attribute\_corresp\_Variable \in{} Attribute \pinj{} Variable$}{}
	\Invariant{inv1\_96}{false}{$\forall{} re\qdot{}(re\in{}dom(Attribute\_Type)\leqv{}(re\in{}dom(Attribute\_corresp\_Constant) \lor{} (re\in{}dom(Attribute\_corresp\_Variable))))$}{}
	\Invariant{inv1\_97}{false}{$Variable\_typing\_Invariant \in{} Variable \tinj{} Invariant$}{}
	\Invariant{inv1\_98}{false}{$Constant\_typing\_Property \in{} Constant \tinj{} Property$}{}
	\Invariant{inv1\_99}{false}{$RelationCharacteristic\_corresp\_LogicFormula \in{}  (Relation\pfun{}RelationCharacteristics\_Set)\pinj{}LogicFormula$}{}
	\Invariant{inv1\_100}{false}{$RelationMaplet\_corresp\_Constant \in{} RelationMaplet\pinj{} Constant$}{}
	\Invariant{inv1\_101}{false}{$DataSet\_corresp\_Set \in{} DataSet\pinj{}Set$}{}
	\Invariant{inv1\_102}{false}{$AttributeMaplet\_corresp\_Constant \in{} AttributeMaplet\pinj{} Constant$}{}
	\Invariant{inv1\_103}{false}{$LogicFormula\_involves\_SetItems \in{}  LogicFormula \pfun{} (\natn{}\pfun{}SetItem)$}{}
	\Invariant{inv1\_104}{false}{$EnumeratedDataSet\_corresp\_EnumeratedSet \subseteq{}  DataSet\_corresp\_Set$}{}
	\Invariant{inv1\_105}{false}{$CustomDataSet\_corresp\_AbstractSet \subseteq{}  DataSet\_corresp\_Set$}{}
}
\hypertarget{initialize_default_datasets_evt}{}
\EVENTS{
\EVT{initialize\_default\_datasets}{false}{ordinary}{}{}{
	\ANY{
		\Param{DM}{true}{}
		\Param{o\_DM}{true}{}
	}
	\GUARDS{true}{
		\Guard{grd0}{false}{$dom(DomainModel\_corresp\_Component)\setminus{}dom(DomainModel\_parent\_DomainModel) \neq{}\emptyset{}$}{true}{}
		\Guard{grd1}{false}{$DefaultDataSet = \emptyset{}$}{true}{}
		\Guard{grd2}{false}{$DM \in{}  dom(DomainModel\_corresp\_Component)$}{true}{}
		\Guard{grd3}{false}{$DM \notin{} dom(DomainModel\_parent\_DomainModel)$}{true}{}
		\Guard{grd4}{false}{$AbstractSet \binter{} \{B\_NATURAL, B\_INTEGER, B\_FLOAT, B\_BOOL, B\_STRING\} = \emptyset{}$}{true}{}
		\Guard{grd5}{false}{$o\_DM = DomainModel\_corresp\_Component(DM)$}{true}{}
	}
	\ACTIONS{true}{
		\Action{act1}{$DefaultDataSet \bcmeq{} \{\_NATURAL, \_INTEGER, \_FLOAT, \_BOOL, \_STRING\}$}{true}{}
		\Action{act2}{$DataSet \bcmeq{} DataSet \bunion{} \{\_NATURAL, \_INTEGER, \_FLOAT, \_BOOL, \_STRING\}$}{true}{}
		\Action{act3}{$DataSet\_definedIn\_DomainModel \bcmeq{}  DataSet\_definedIn\_DomainModel \bunion{} \{(xx\mapsto{}yy)|xx\in{}\{\_NATURAL, \\\_INTEGER, \_FLOAT, \_BOOL, \_STRING\}\land{}yy=DM\}$}{true}{}
		\Action{act4}{$AbstractSet \bcmeq{}  AbstractSet \bunion{} \{B\_NATURAL, B\_INTEGER, B\_FLOAT, B\_BOOL, B\_STRING\}$}{true}{}
		\Action{act5}{$Set \bcmeq{}  Set \bunion{} \{B\_NATURAL, B\_INTEGER, B\_FLOAT, B\_BOOL, B\_STRING\}$}{true}{}
		\Action{act6}{$DefaultDataSet\_corresp\_AbstractSet \bcmeq{} \{\_NATURAL\mapsto{}B\_NATURAL, \_INTEGER\mapsto{}B\_INTEGER,\\ \_FLOAT\mapsto{}B\_FLOAT, \_BOOL\mapsto{}B\_BOOL, \_STRING\mapsto{}B\_STRING\}$}{true}{}
		\Action{act7}{$Set\_definedIn\_Component \bcmeq{}  Set\_definedIn\_Component \bunion{} \{(xx\mapsto{}yy)|xx\in{}\{B\_NATURAL, \\B\_INTEGER, B\_FLOAT, B\_BOOL, B\_STRING\}\land{}yy=o\_DM\}$}{true}{}
		\Action{act8}{$DataSet\_corresp\_Set \bcmeq{}  DataSet\_corresp\_Set \ovl{} \{\_NATURAL\mapsto{}B\_NATURAL, \_INTEGER\mapsto{}B\_INTEGER, \_FLOAT\mapsto{}B\_FLOAT, \_BOOL\mapsto{}B\_BOOL, \_STRING\mapsto{}B\_STRING\}$}{true}{}
	}
}
}
•••
\END

\subsection{From Domain Models to B System Specifications}

\subsubsection{B System Components}

\paragraph{Rule 1: Domain model without parent}

\MACHINE{Ontologies\_BSystem\_specs\_translation}{}{BSystem\_Metamodel\_Context,Domain\_Metamodel\_Context}{}

\EVT{rule\_1}{false}{ordinary}{}{correspondence of a domain model not associated to a parent domain model}{
	\ANY{
		\Param{DM}{true}{}
		\Param{o\_DM}{true}{}
	}
	\GUARDS{true}{
		\Guard{grd0}{false}{$DomainModel \setminus{} (dom(DomainModel\_corresp\_Component) \bunion{} dom(DomainModel\_parent\_DomainModel)) \neq{}\emptyset{}$}{true}{}
		\Guard{grd1}{false}{$DM \in{} DomainModel$}{true}{}
		\Guard{grd2}{false}{$DM \notin{} dom(DomainModel\_corresp\_Component)$}{true}{}
		\Guard{grd3}{false}{$DM \notin{} dom(DomainModel\_parent\_DomainModel)$}{true}{}
		\Guard{grd4}{false}{$Component\_Set \setminus{} Component \neq{}\emptyset{}$}{true}{}
		\Guard{grd5}{false}{$o\_DM \in{} Component\_Set \setminus{} Component$}{true}{}
	}
	\ACTIONS{true}{
		\Action{act1}{$System \bcmeq{}  System \bunion{}  \{o\_DM\}$}{true}{}
		\Action{act2}{$Component \bcmeq{}  Component \bunion{}  \{o\_DM\}$}{true}{}
		\Action{act3}{$DomainModel\_corresp\_Component(DM)\bcmeq{}o\_DM$}{true}{}
	}
}

\END

Any domain model that is not associated with another domain model (Fig. \ref{OurBusinesDomainModel_min_version_2_EventB_Machine_Context}), through the \textit{parent} association, gives rise to a system component. 
\textbf{\underline{Example : }} in Figure \ref{lgsystem_event_b_model_refinment_0}, the root level domain model is translated into a  system component named   \textbf{\textit{lg\_system\_ref\_0}}. 

\paragraph{Rule 2: Domain model with parent}
\MACHINE{Ontologies\_BSystem\_specs\_translation}{}{BSystem\_Metamodel\_Context,Domain\_Metamodel\_Context}{}

\EVT{rule\_2}{false}{ordinary}{}{\\correspondence of a domain model associated to a parent domain model}{
	\ANY{
		\Param{DM}{true}{}
		\Param{PDM}{true}{}
		\Param{o\_DM}{true}{}
	}
	\GUARDS{true}{
		\Guard{grd0}{false}{$dom(DomainModel\_parent\_DomainModel) \setminus{}  dom(DomainModel\_corresp\_Component) \neq{} \emptyset{}$}{true}{}
		\Guard{grd1}{false}{$DM \in{} dom(DomainModel\_parent\_DomainModel)$}{true}{}
		\Guard{grd2}{false}{$DM \notin{} dom(DomainModel\_corresp\_Component)$}{true}{}
		\Guard{grd3}{false}{$dom(DomainModel\_corresp\_Component) \neq{} \emptyset{}$}{true}{}
		\Guard{grd4}{false}{$PDM \in{} dom(DomainModel\_corresp\_Component)$}{true}{}
		\Guard{grd5}{false}{$DomainModel\_parent\_DomainModel(DM)=PDM$}{true}{}
		\Guard{grd6}{false}{$Component\_Set \setminus{} Component \neq{}\emptyset{}$}{true}{}
		\Guard{grd7}{false}{$o\_DM \in{} Component\_Set \setminus{} Component $}{true}{}
	}
	\ACTIONS{true}{
		\Action{act1}{$Refinement \bcmeq{}  Refinement \bunion{}  \{o\_DM\}$}{true}{}
		\Action{act2}{$Component \bcmeq{}  Component \bunion{}  \{o\_DM\}$}{true}{}
		\Action{act3}{$Refinement\_refines\_Component(o\_DM) \bcmeq{} DomainModel\_corresp\_Component(PDM)$}{true}{}
		\Action{act4}{$DomainModel\_corresp\_Component(DM)\bcmeq{}o\_DM$}{true}{}
	}
}

\END

A domain model associated with another one representing its parent (Fig. \ref{OurBusinesDomainModel_min_version_2_EventB_Machine_Context})  gives rise to a refinement component. The  refinement component must refine the component corresponding  to the parent domain model. \textbf{\underline{Example : }} in Figure \ref{lgsystem_event_b_model_refinment_1}, the first refinement level domain model is translated into a  refinement component named   \textbf{\textit{lg\_system\_ref\_1}} refining \textbf{\textit{lg\_system\_ref\_0}}.

\newpage

\subsubsection{B System Sets}

\paragraph{Rule 3: Concept without parent}
\MACHINE{Ontologies\_BSystem\_specs\_translation\_ref\_1}{Ontologies\_BSystem\_specs\_translation}{BSystem\_Metamodel\_Context,Domain\_Metamodel\_Context}{}

\EVT{rule\_3}{false}{ordinary}{}{\\correspondence of a concept not associated to a parent concept}{
	\ANY{
		\Param{CO}{true}{}
		\Param{o\_CO}{true}{}
	}
	\GUARDS{true}{
		\Guard{grd0}{false}{$Concept \setminus{} (dom(Concept\_parentConcept\_Concept) \bunion{} dom(Concept\_corresp\_AbstractSet)) \neq{}\emptyset{}$}{true}{}
		\Guard{grd1}{false}{$CO \in{} Concept$}{true}{}
		\Guard{grd2}{false}{$CO \notin{} dom(Concept\_parentConcept\_Concept)$}{true}{}
		\Guard{grd3}{false}{$CO \notin{} dom(Concept\_corresp\_AbstractSet)$}{true}{}
		\Guard{grd4}{false}{$Concept\_definedIn\_DomainModel(CO) \in{} dom(DomainModel\_corresp\_Component)$}{true}{}
		\Guard{grd5}{false}{$Set\_Set \setminus{} Set \neq{}\emptyset{}$}{true}{}
		\Guard{grd6}{false}{$o\_CO \in{} Set\_Set \setminus{} Set $}{true}{}
	}
	\ACTIONS{true}{
		\Action{act1}{$AbstractSet \bcmeq{}  AbstractSet \bunion{}  \{o\_CO\}$}{true}{}
		\Action{act2}{$Set \bcmeq{}  Set \bunion{}  \{o\_CO\}$}{true}{}
		\Action{act3}{$Concept\_corresp\_AbstractSet(CO)\bcmeq{}o\_CO$}{true}{}
		\Action{act4}{$Set\_definedIn\_Component(o\_CO) \bcmeq{} DomainModel\_corresp\_Component(\\Concept\_definedIn\_DomainModel(CO))$}{true}{}
	}
}

\END

Any  concept  that is not associated with another one known as its parent concept (Fig. \ref{OurBusinesDomainModel_min_version_2_EventB_Set}), through the \textsf{parentConcept} association, gives rise to a \textit{B System} abstract set. \textbf{\underline{Example : }} in Figure \ref{lgsystem_event_b_model_refinment_0}, the abstract set   \textit{LandingGear} appears because of \textsf{Concept} instance \textbf{\textit{LandingGear}}.

\paragraph{Rule 4: Enumerated data set}
\MACHINE{Ontologies\_BSystem\_specs\_translation\_ref\_1}{Ontologies\_BSystem\_specs\_translation}{BSystem\_Metamodel\_Context,Domain\_Metamodel\_Context}{}

\EVT{rule\_4}{false}{ordinary}{}{\\correspondence of an instance of EnumeratedDataSet}{
	\ANY{
		\Param{EDS}{true}{}
		\Param{o\_EDS}{true}{}
		\Param{elements}{true}{}
		\Param{o\_elements}{true}{}
		\Param{mapping\_elements\_o\_elements}{true}{}
	}
	\GUARDS{true}{
		\Guard{grd0}{false}{$EnumeratedDataSet \setminus{} dom(DataSet\_corresp\_Set) \neq{}\emptyset{}$}{true}{}
		\Guard{grd1}{false}{$EDS \in{} EnumeratedDataSet$}{true}{}
		\Guard{grd2}{false}{$EDS \notin{} dom(DataSet\_corresp\_Set)$}{true}{}
		\Guard{grd4}{false}{$DataSet\_definedIn\_DomainModel(EDS) \in{} dom(DomainModel\_corresp\_Component)$}{true}{}
		\Guard{grd5}{false}{$Set\_Set \setminus{} Set \neq{}\emptyset{}$}{true}{}
		\Guard{grd6}{false}{$o\_EDS \in{} Set\_Set \setminus{} Set$}{true}{}
		\Guard{grd8}{false}{$o\_EDS \notin{}  \{B\_NATURAL, B\_INTEGER, B\_FLOAT, B\_BOOL, B\_STRING\}$}{true}{\\elements}
		\Guard{grd9}{false}{$o\_elements \subseteq{}  SetItem\_Set \setminus{} SetItem$}{true}{}
		\Guard{grd11}{false}{$elements = DataValue\_elements\_EnumeratedDataSet\converse{}[\{EDS\}]$}{true}{}
		\Guard{grd12}{false}{$card(o\_elements) = card(elements)$}{true}{}
		\Guard{grd13}{false}{$mapping\_elements\_o\_elements \in{} elements \tbij{} o\_elements$}{true}{}
	}
	\ACTIONS{true}{
		\Action{act1}{$EnumeratedSet \bcmeq{}  EnumeratedSet \bunion{}  \{o\_EDS\}$}{true}{}
		\Action{act2}{$Set \bcmeq{}  Set \bunion{}  \{o\_EDS\}$}{true}{}
		\Action{act3}{$EnumeratedDataSet\_corresp\_EnumeratedSet(EDS)\bcmeq{}o\_EDS$}{true}{}
		\Action{act4}{$Set\_definedIn\_Component(o\_EDS) \bcmeq{} DomainModel\_corresp\_Component(\\DataSet\_definedIn\_DomainModel(EDS))$}{true}{\\elements}
		\Action{act5}{$SetItem \bcmeq{} SetItem \bunion{}  o\_elements$}{true}{}
		\Action{act6}{$SetItem\_itemOf\_EnumeratedSet \bcmeq{}  SetItem\_itemOf\_EnumeratedSet \bunion{} (o\_elements \cprod{} \{o\_EDS\})$}{true}{}
		\Action{act7}{$DataValue\_corresp\_SetItem \bcmeq{}  DataValue\_corresp\_SetItem \bunion{} mapping\_elements\_o\_elements$}{true}{}
		\Action{act8}{$DataSet\_corresp\_Set \bcmeq{}  DataSet\_corresp\_Set \ovl{} \{EDS\mapsto{}o\_EDS\}$}{true}{}
	}
}

\END

\paragraph{Rule 5 : Custom data set not defined through an enumeration}
\MACHINE{Ontologies\_BSystem\_specs\_translation\_ref\_1}{Ontologies\_BSystem\_specs\_translation}{BSystem\_Metamodel\_Context,Domain\_Metamodel\_Context}{}

\EVT{rule\_5}{false}{ordinary}{}{\\correspondence of an instance of CustomDataSet which is not an instance of EnumeratedDataSet}{
	\ANY{
		\Param{CS}{true}{}
		\Param{o\_CS}{true}{}
	}
	\GUARDS{true}{
		\Guard{grd0}{false}{$CustomDataSet \setminus{} (EnumeratedDataSet \bunion{} dom(DataSet\_corresp\_Set)) \neq{}\emptyset{}$}{true}{}
		\Guard{grd1}{false}{$CS \in{} CustomDataSet$}{true}{}
		\Guard{grd2}{false}{$CS \notin{} EnumeratedDataSet$}{true}{}
		\Guard{grd3}{false}{$CS \notin{} dom(DataSet\_corresp\_Set)$}{true}{}
		\Guard{grd4}{false}{$DataSet\_definedIn\_DomainModel(CS) \in{} dom(DomainModel\_corresp\_Component)$}{true}{}
		\Guard{grd5}{false}{$Set\_Set \setminus{} Set \neq{}\emptyset{}$}{true}{}
		\Guard{grd6}{false}{$o\_CS \in{} Set\_Set \setminus{} Set$}{true}{}
	}
	\ACTIONS{true}{
		\Action{act1}{$AbstractSet \bcmeq{}  AbstractSet \bunion{}  \{o\_CS\}$}{true}{}
		\Action{act2}{$Set \bcmeq{}  Set \bunion{}  \{o\_CS\}$}{true}{}
		\Action{act3}{$CustomDataSet\_corresp\_AbstractSet(CS)\bcmeq{}o\_CS$}{true}{}
		\Action{act4}{$Set\_definedIn\_Component(o\_CS) \bcmeq{} DomainModel\_corresp\_Component(\\DataSet\_definedIn\_DomainModel(CS))$}{true}{}
		\Action{act5}{$DataSet\_corresp\_Set \bcmeq{}  DataSet\_corresp\_Set \ovl{} \{CS\mapsto{}o\_CS\}$}{true}{}
	}
}

\END

 Any instance of \textsf{CustomDataSet}, defined through an enumeration, gives rise to a \textit{B System} enumerated set.
  \textbf{\underline{Example : }} in Figure \ref{lgsystem_event_b_model_refinment_0},
   the data set \textbf{\textit{\{"lg\_extended", "lg\_retracted"\}}}, defined in domain model represented in Figure  (Fig. \ref{lgsystem_refinment_0_ontology}),  gives rise to the enumerated set \textit{DataSet\_1=\{lg\_extended, lg\_retracted\}}.

 Any instance of \textsf{DefaultDataSet} is mapped directly to a \textit{B System} default data set (\textsf{NATURAL}, \textsf{INTEGER}, \textsf{FLOAT}, \textsf{STRING} or \textsf{BOOL})  following the \hyperlink{initialize_default_datasets_evt}{\textit{\textbf{initialize\_default\_datasets }}} event.

\subsubsection{B System Constants}

\paragraph{Rule 6 : Concept with parent}

\MACHINE{Ontologies\_BSystem\_specs\_translation\_ref\_1}{Ontologies\_BSystem\_specs\_translation}{BSystem\_Metamodel\_Context,Domain\_Metamodel\_Context}{}

\EVT{rule\_6\_1}{false}{ordinary}{}{\\correspondence of a concept associated to a parent concept (where the parent concept corresponds to an abstract set)}{
	\ANY{
		\Param{CO}{true}{}
		\Param{o\_CO}{true}{}
		\Param{PCO}{true}{}
		\Param{o\_lg}{true}{}
		\Param{o\_PCO}{true}{}
	}
	\GUARDS{true}{
		\Guard{grd0}{false}{$dom(Concept\_parentConcept\_Concept) \setminus{} dom(Concept\_corresp\_Constant) \neq{}\emptyset{}$}{true}{}
		\Guard{grd1}{false}{$CO \in{} dom(Concept\_parentConcept\_Concept) \setminus{} dom(Concept\_corresp\_Constant)$}{true}{}
		\Guard{grd2}{false}{$dom(Concept\_corresp\_AbstractSet) \neq{} \emptyset{}$}{true}{}
		\Guard{grd3}{false}{$PCO \in{} dom(Concept\_corresp\_AbstractSet)$}{true}{}
		\Guard{grd4}{false}{$Concept\_parentConcept\_Concept(CO)=PCO$}{true}{}
		\Guard{grd5}{false}{$Concept\_definedIn\_DomainModel(CO) \in{} dom(DomainModel\_corresp\_Component)$}{true}{}
		\Guard{grd6}{false}{$Constant\_Set \setminus{} Constant \neq{}\emptyset{}$}{true}{}
		\Guard{grd7}{false}{$o\_CO \in{} Constant\_Set \setminus{} Constant$}{true}{}
		\Guard{grd8}{false}{$LogicFormula\_Set \setminus{} LogicFormula \neq{} \emptyset{}$}{true}{}
		\Guard{grd9}{false}{$o\_lg \in{} LogicFormula\_Set \setminus{} LogicFormula$}{true}{}
		\Guard{grd10}{false}{$o\_PCO \in{} AbstractSet$}{true}{}
		\Guard{grd11}{false}{$o\_PCO = Concept\_corresp\_AbstractSet(PCO)$}{true}{}
	}
	\ACTIONS{true}{
		\Action{act1}{$Constant \bcmeq{}  Constant \bunion{}  \{o\_CO\}$}{true}{}
		\Action{act2}{$Concept\_corresp\_Constant(CO)\bcmeq{}o\_CO$}{true}{}
		\Action{act3}{$Constant\_definedIn\_Component(o\_CO) \bcmeq{} DomainModel\_corresp\_Component(\\Concept\_definedIn\_DomainModel(CO))$}{true}{}
		\Action{act4}{$Property \bcmeq{} Property \bunion{} \{o\_lg\}$}{true}{}
		\Action{act5}{$LogicFormula \bcmeq{} LogicFormula \bunion{} \{o\_lg\}$}{true}{}
		\Action{act6}{$LogicFormula\_uses\_Operators(o\_lg) \bcmeq{} \{1\mapsto{}Inclusion\_OP\}$}{true}{}
		\Action{act7}{$Constant\_isInvolvedIn\_LogicFormulas(o\_CO) \bcmeq{} \{1\mapsto{}o\_lg\}$}{true}{}
		\Action{act8}{$LogicFormula\_involves\_Sets(o\_lg) \bcmeq{} \{2\mapsto{}o\_PCO\}$}{true}{}
		\Action{act9}{$LogicFormula\_definedIn\_Component(o\_lg) \bcmeq{} DomainModel\_corresp\_Component(\\Concept\_definedIn\_DomainModel(CO))$}{true}{}
		\Action{act10}{$Constant\_typing\_Property(o\_CO) \bcmeq{}  o\_lg$}{true}{}
	}
}
\EVT{rule\_6\_2}{false}{ordinary}{}{\\correspondence of a concept associated to a parent concept (where the parent concept corresponds to a constant)}{
	\ANY{
		\Param{CO}{true}{}
		\Param{o\_CO}{true}{}
		\Param{PCO}{true}{}
		\Param{o\_lg}{true}{}
		\Param{o\_PCO}{true}{}
	}
	\GUARDS{true}{
		\Guard{grd0}{false}{$dom(Concept\_parentConcept\_Concept) \setminus{} dom(Concept\_corresp\_Constant) \neq{}\emptyset{}$}{true}{}
		\Guard{grd1}{false}{$CO \in{} dom(Concept\_parentConcept\_Concept) \setminus{} dom(Concept\_corresp\_Constant)$}{true}{}
		\Guard{grd2}{false}{$dom(Concept\_corresp\_Constant) \neq{} \emptyset{}$}{true}{}
		\Guard{grd3}{false}{$PCO \in{} dom(Concept\_corresp\_Constant)$}{true}{}
		\Guard{grd4}{false}{$Concept\_parentConcept\_Concept(CO)=PCO$}{true}{}
		\Guard{grd5}{false}{$Concept\_definedIn\_DomainModel(CO) \in{} dom(DomainModel\_corresp\_Component)$}{true}{}
		\Guard{grd6}{false}{$Constant\_Set \setminus{} Constant \neq{}\emptyset{}$}{true}{}
		\Guard{grd7}{false}{$o\_CO \in{} Constant\_Set \setminus{} Constant$}{true}{}
		\Guard{grd8}{false}{$LogicFormula\_Set \setminus{} LogicFormula \neq{} \emptyset{}$}{true}{}
		\Guard{grd9}{false}{$o\_lg \in{} LogicFormula\_Set \setminus{} LogicFormula$}{true}{}
		\Guard{grd10}{false}{$o\_PCO \in{} Constant$}{true}{}
		\Guard{grd11}{false}{$o\_PCO = Concept\_corresp\_Constant(PCO)$}{true}{}
	}
	\ACTIONS{true}{
		\Action{act1}{$Constant \bcmeq{}  Constant \bunion{}  \{o\_CO\}$}{true}{}
		\Action{act2}{$Concept\_corresp\_Constant(CO)\bcmeq{}o\_CO$}{true}{}
		\Action{act3}{$Constant\_definedIn\_Component(o\_CO) \bcmeq{} DomainModel\_corresp\_Component(\\Concept\_definedIn\_DomainModel(CO))$}{true}{}
		\Action{act4}{$Property \bcmeq{} Property \bunion{} \{o\_lg\}$}{true}{}
		\Action{act5}{$LogicFormula \bcmeq{} LogicFormula \bunion{} \{o\_lg\}$}{true}{}
		\Action{act6}{$LogicFormula\_uses\_Operators(o\_lg) \bcmeq{} \{1\mapsto{}Inclusion\_OP\}$}{true}{}
		\Action{act7}{$Constant\_isInvolvedIn\_LogicFormulas\bcmeq{}Constant\_isInvolvedIn\_LogicFormulas\ovl{}\{(o\_CO\mapsto{}\{1\mapsto{}o\_lg\}), o\_PCO\mapsto{}Constant\_isInvolvedIn\_LogicFormulas(o\_PCO)\bunion{}\{2\mapsto{}o\_lg\}\}$}{true}{}
		\Action{act8}{$LogicFormula\_involves\_Sets(o\_lg) \bcmeq{} \emptyset{}$}{true}{}
		\Action{act9}{$LogicFormula\_definedIn\_Component(o\_lg) \bcmeq{} DomainModel\_corresp\_Component(\\Concept\_definedIn\_DomainModel(CO))$}{true}{}
		\Action{act10}{$Constant\_typing\_Property(o\_CO) \bcmeq{}  o\_lg$}{true}{}
	}
}

\END

Any  concept  associated with another one known as its parent concept (Fig. \ref{OurBusinesDomainModel_min_version_2_EventB_Machine_Context}), through the \textsf{parentConcept} association, gives rise to a constant typed as a subset of the \textit{B System} element corresponding to the parent concept. 

Each individual (or data value) gives rise to a constant having its name (or with his \textit{lexicalForm} typed as value) and each instance of \textsf{CustomDataSet}, not defined through an enumeration of its elements, unlike \textbf{\textit{DataSet\_1}} of Figure \ref{lgsystem_event_b_model_refinment_0}, gives rise to a constant having its name. \textbf{\underline{Example : }} in Figure \ref{lgsystem_event_b_model_refinment_1}, the constant named \textit{HD1} is the correspondent of the individual \textbf{\textit{HD1}}.

\paragraph{Rule 7 : Individual}
\MACHINE{Ontologies\_BSystem\_specs\_translation\_ref\_1}{Ontologies\_BSystem\_specs\_translation}{BSystem\_Metamodel\_Context,Domain\_Metamodel\_Context}{}

\EVT{rule\_7\_1}{false}{ordinary}{}{\\correspondence of an instance of Individual (where the  concept corresponds to an abstract set)}{
	\ANY{
		\Param{ind}{true}{}
		\Param{o\_ind}{true}{}
		\Param{CO}{true}{}
		\Param{o\_lg}{true}{}
		\Param{o\_CO}{true}{}
	}
	\GUARDS{true}{
		\Guard{grd0}{false}{$dom(Individual\_individualOf\_Concept) \setminus{} dom(Individual\_corresp\_Constant) \neq{}\emptyset{}$}{true}{}
		\Guard{grd1}{false}{$ind \in{} dom(Individual\_individualOf\_Concept) \setminus{} dom(Individual\_corresp\_Constant)$}{true}{}
		\Guard{grd2}{false}{$dom(Concept\_corresp\_AbstractSet) \neq{} \emptyset{}$}{true}{}
		\Guard{grd3}{false}{$CO \in{} dom(Concept\_corresp\_AbstractSet)$}{true}{}
		\Guard{grd4}{false}{$Individual\_individualOf\_Concept(ind)=CO$}{true}{}
		\Guard{grd5}{false}{$Concept\_definedIn\_DomainModel(CO) \in{} dom(DomainModel\_corresp\_Component)$}{true}{}
		\Guard{grd6}{false}{$Constant\_Set \setminus{} Constant \neq{}\emptyset{}$}{true}{}
		\Guard{grd7}{false}{$o\_ind \in{} Constant\_Set \setminus{} Constant$}{true}{}
		\Guard{grd8}{false}{$LogicFormula\_Set \setminus{} LogicFormula \neq{} \emptyset{}$}{true}{}
		\Guard{grd9}{false}{$o\_lg \in{} LogicFormula\_Set \setminus{} LogicFormula$}{true}{}
		\Guard{grd10}{false}{$o\_CO \in{} AbstractSet$}{true}{}
		\Guard{grd11}{false}{$o\_CO = Concept\_corresp\_AbstractSet(CO)$}{true}{}
	}
	\ACTIONS{true}{
		\Action{act1}{$Constant \bcmeq{}  Constant \bunion{}  \{o\_ind\}$}{true}{}
		\Action{act2}{$Individual\_corresp\_Constant(ind)\bcmeq{}o\_ind$}{true}{}
		\Action{act3}{$Constant\_definedIn\_Component(o\_ind) \bcmeq{} DomainModel\_corresp\_Component(\\Concept\_definedIn\_DomainModel(CO))$}{true}{}
		\Action{act4}{$Property \bcmeq{} Property \bunion{} \{o\_lg\}$}{true}{}
		\Action{act5}{$LogicFormula \bcmeq{} LogicFormula \bunion{} \{o\_lg\}$}{true}{}
		\Action{act6}{$LogicFormula\_uses\_Operators(o\_lg) \bcmeq{} \{1\mapsto{}Belonging\_OP\}$}{true}{}
		\Action{act7}{$Constant\_isInvolvedIn\_LogicFormulas(o\_ind) \bcmeq{} \{1\mapsto{}o\_lg\}$}{true}{}
		\Action{act8}{$LogicFormula\_involves\_Sets(o\_lg) \bcmeq{} \{2\mapsto{}o\_CO\}$}{true}{}
		\Action{act9}{$LogicFormula\_definedIn\_Component(o\_lg) \bcmeq{} DomainModel\_corresp\_Component(\\Concept\_definedIn\_DomainModel(CO))$}{true}{}
		\Action{act10}{$Constant\_typing\_Property(o\_ind) \bcmeq{}  o\_lg$}{true}{}
	}
}
\EVT{rule\_7\_2}{false}{ordinary}{}{\\correspondence of an instance of Individual (where the  concept corresponds to a constant)}{
	\ANY{
		\Param{ind}{true}{}
		\Param{o\_ind}{true}{}
		\Param{CO}{true}{}
		\Param{o\_lg}{true}{}
		\Param{o\_CO}{true}{}
	}
	\GUARDS{true}{
		\Guard{grd0}{false}{$dom(Individual\_individualOf\_Concept) \setminus{} dom(Individual\_corresp\_Constant) \neq{}\emptyset{}$}{true}{}
		\Guard{grd1}{false}{$ind \in{} dom(Individual\_individualOf\_Concept) \setminus{} dom(Individual\_corresp\_Constant)$}{true}{}
		\Guard{grd2}{false}{$dom(Concept\_corresp\_Constant) \neq{} \emptyset{}$}{true}{}
		\Guard{grd3}{false}{$CO \in{} dom(Concept\_corresp\_Constant)$}{true}{}
		\Guard{grd4}{false}{$Individual\_individualOf\_Concept(ind)=CO$}{true}{}
		\Guard{grd5}{false}{$Concept\_definedIn\_DomainModel(CO) \in{} dom(DomainModel\_corresp\_Component)$}{true}{}
		\Guard{grd6}{false}{$Constant\_Set \setminus{} Constant \neq{}\emptyset{}$}{true}{}
		\Guard{grd7}{false}{$o\_ind \in{} Constant\_Set \setminus{} Constant$}{true}{}
		\Guard{grd8}{false}{$LogicFormula\_Set \setminus{} LogicFormula \neq{} \emptyset{}$}{true}{}
		\Guard{grd9}{false}{$o\_lg \in{} LogicFormula\_Set \setminus{} LogicFormula$}{true}{}
		\Guard{grd10}{false}{$o\_CO \in{} Constant$}{true}{}
		\Guard{grd11}{false}{$o\_CO = Concept\_corresp\_Constant(CO)$}{true}{}
	}
	\ACTIONS{true}{
		\Action{act1}{$Constant \bcmeq{}  Constant \bunion{}  \{o\_ind\}$}{true}{}
		\Action{act2}{$Individual\_corresp\_Constant(ind)\bcmeq{}o\_ind$}{true}{}
		\Action{act3}{$Constant\_definedIn\_Component(o\_ind) \bcmeq{} DomainModel\_corresp\_Component(\\Concept\_definedIn\_DomainModel(CO))$}{true}{}
		\Action{act4}{$Property \bcmeq{} Property \bunion{} \{o\_lg\}$}{true}{}
		\Action{act5}{$LogicFormula \bcmeq{} LogicFormula \bunion{} \{o\_lg\}$}{true}{}
		\Action{act6}{$LogicFormula\_uses\_Operators(o\_lg) \bcmeq{} \{1\mapsto{}Belonging\_OP\}$}{true}{}
		\Action{act7}{$Constant\_isInvolvedIn\_LogicFormulas\bcmeq{}Constant\_isInvolvedIn\_LogicFormulas\ovl{}\{(o\_ind\mapsto{}\{1\mapsto{}o\_lg\}), o\_CO\mapsto{}Constant\_isInvolvedIn\_LogicFormulas(o\_CO)\bunion{}\{2\mapsto{}o\_lg\}\}$}{true}{}
		\Action{act8}{$LogicFormula\_involves\_Sets(o\_lg) \bcmeq{} \emptyset{}$}{true}{}
		\Action{act9}{$LogicFormula\_definedIn\_Component(o\_lg) \bcmeq{} DomainModel\_corresp\_Component(\\Concept\_definedIn\_DomainModel(CO))$}{true}{}
		\Action{act10}{$Constant\_typing\_Property(o\_ind) \bcmeq{}  o\_lg$}{true}{}
	}
}

\END

\paragraph{Rule 8 : Data value}

\MACHINE{Ontologies\_BSystem\_specs\_translation\_ref\_1}{Ontologies\_BSystem\_specs\_translation}{BSystem\_Metamodel\_Context,Domain\_Metamodel\_Context}{}

\EVT{rule\_8}{false}{ordinary}{}{\\correspondence of an instance of DataValue (When the data set is an instance of CustomDataSet not instance of EnumeratedDataSet\\~    (for this last case, the rule for instances of EnumeratedDataSet also handles data values) ) }{
	\ANY{
		\Param{dva}{true}{}
		\Param{o\_dva}{true}{}
		\Param{DS}{true}{}
		\Param{o\_lg}{true}{}
		\Param{o\_DS}{true}{}
	}
	\GUARDS{true}{
		\Guard{grd0}{false}{$dom(DataValue\_valueOf\_DataSet) \setminus{} dom(DataValue\_corresp\_Constant) \neq{}\emptyset{}$}{true}{}
		\Guard{grd1}{false}{$dva \in{} dom(DataValue\_valueOf\_DataSet) \setminus{} dom(DataValue\_corresp\_Constant)$}{true}{}
		\Guard{grd2}{false}{$dom(CustomDataSet\_corresp\_AbstractSet) \neq{} \emptyset{}$}{true}{}
		\Guard{grd3}{false}{$DS \in{} dom(CustomDataSet\_corresp\_AbstractSet)$}{true}{}
		\Guard{grd4}{false}{$DataValue\_valueOf\_DataSet(dva)=DS$}{true}{}
		\Guard{grd5}{false}{$DataSet\_definedIn\_DomainModel(DS) \in{} dom(DomainModel\_corresp\_Component)$}{true}{}
		\Guard{grd6}{false}{$Constant\_Set \setminus{} Constant \neq{}\emptyset{}$}{true}{}
		\Guard{grd7}{false}{$o\_dva \in{} Constant\_Set \setminus{} Constant$}{true}{}
		\Guard{grd8}{false}{$LogicFormula\_Set \setminus{} LogicFormula \neq{} \emptyset{}$}{true}{}
		\Guard{grd9}{false}{$o\_lg \in{} LogicFormula\_Set \setminus{} LogicFormula$}{true}{}
		\Guard{grd10}{false}{$o\_DS \in{} AbstractSet$}{true}{}
		\Guard{grd11}{false}{$o\_DS = CustomDataSet\_corresp\_AbstractSet(DS)$}{true}{}
	}
	\ACTIONS{true}{
		\Action{act1}{$Constant \bcmeq{}  Constant \bunion{}  \{o\_dva\}$}{true}{}
		\Action{act2}{$DataValue\_corresp\_Constant(dva)\bcmeq{}o\_dva$}{true}{}
		\Action{act3}{$Constant\_definedIn\_Component(o\_dva) \bcmeq{} DomainModel\_corresp\_Component(\\DataSet\_definedIn\_DomainModel(DS))$}{true}{}
		\Action{act4}{$Property \bcmeq{} Property \bunion{} \{o\_lg\}$}{true}{}
		\Action{act5}{$LogicFormula \bcmeq{} LogicFormula \bunion{} \{o\_lg\}$}{true}{}
		\Action{act6}{$LogicFormula\_uses\_Operators(o\_lg) \bcmeq{} \{1\mapsto{}Belonging\_OP\}$}{true}{}
		\Action{act7}{$Constant\_isInvolvedIn\_LogicFormulas(o\_dva) \bcmeq{} \{1\mapsto{}o\_lg\}$}{true}{}
		\Action{act8}{$LogicFormula\_involves\_Sets(o\_lg) \bcmeq{} \{2\mapsto{}o\_DS\}$}{true}{}
		\Action{act9}{$LogicFormula\_definedIn\_Component(o\_lg) \bcmeq{} DomainModel\_corresp\_Component(\\DataSet\_definedIn\_DomainModel(DS))$}{true}{}
		\Action{act10}{$Constant\_typing\_Property(o\_dva) \bcmeq{}  o\_lg$}{true}{}
	}
}

\END

\paragraph{Rule 10 : Constant relation}

\MACHINE{Ontologies\_BSystem\_specs\_translation\_ref\_1}{Ontologies\_BSystem\_specs\_translation}{BSystem\_Metamodel\_Context,Domain\_Metamodel\_Context}{}

\EVT{rule\_10\_1}{false}{ordinary}{}{\\correspondence of an instance of Relation having its isVariable property set to false (case where domain and range correspond to abstract sets)}{
	\ANY{
		\Param{RE}{true}{}
		\Param{T\_RE}{true}{}
		\Param{o\_RE}{true}{}
		\Param{CO1}{true}{}
		\Param{o\_CO1}{true}{}
		\Param{CO2}{true}{}
		\Param{o\_CO2}{true}{}
		\Param{o\_lg1}{true}{}
		\Param{o\_lg2}{true}{}
		\Param{DM}{true}{}
	}
	\GUARDS{true}{
		\Guard{grd0}{false}{$Relation\_isVariable\converse{}[\{FALSE\}] \setminus{} dom(Relation\_Type) \neq{}\emptyset{}$}{true}{}
		\Guard{grd1}{false}{$RE \in{} Relation\_isVariable\converse{}[\{FALSE\}] \setminus{} dom(Relation\_Type)$}{true}{}
		\Guard{grd2}{false}{$dom(Concept\_corresp\_AbstractSet) \neq{} \emptyset{}$}{true}{}
		\Guard{grd3}{false}{$CO1 = Relation\_domain\_Concept(RE)$}{true}{}
		\Guard{grd4}{false}{$CO2 = Relation\_range\_Concept(RE)$}{true}{}
		\Guard{grd5}{false}{$\{CO1, CO2\} \subseteq{} dom(Concept\_corresp\_AbstractSet)$}{true}{}
		\Guard{grd6}{false}{$Relation\_definedIn\_DomainModel(RE) \in{} dom(DomainModel\_corresp\_Component)$}{true}{}
		\Guard{grd7}{false}{$Constant\_Set \setminus{} Constant \neq{}\emptyset{}$}{true}{}
		\Guard{grd8}{false}{$\{T\_RE, o\_RE\} \subseteq{} Constant\_Set \setminus{} Constant$}{true}{}
		\Guard{grd9}{false}{$LogicFormula\_Set \setminus{} LogicFormula \neq{} \emptyset{}$}{true}{}
		\Guard{grd10}{false}{$\{o\_lg1,o\_lg2\}  \subseteq{} LogicFormula\_Set \setminus{} LogicFormula$}{true}{}
		\Guard{grd11}{false}{$o\_CO1 = Concept\_corresp\_AbstractSet(CO1)$}{true}{}
		\Guard{grd12}{false}{$o\_CO2 = Concept\_corresp\_AbstractSet(CO2)$}{true}{}
		\Guard{grd13}{false}{$DM = Relation\_definedIn\_DomainModel(RE)$}{true}{}
		\Guard{grd14}{false}{$T\_RE \neq{}  o\_RE$}{true}{}
		\Guard{grd15}{false}{$o\_lg1 \neq{}  o\_lg2$}{true}{}
	}
	\ACTIONS{true}{
		\Action{act1}{$Constant \bcmeq{}  Constant \bunion{}  \{T\_RE, o\_RE\}$}{true}{}
		\Action{act2}{$Relation\_Type(RE)\bcmeq{}T\_RE$}{true}{}
		\Action{act3}{$Relation\_corresp\_Constant(RE)\bcmeq{}o\_RE$}{true}{}
		\Action{act4}{$Constant\_definedIn\_Component\bcmeq{}Constant\_definedIn\_Component\bunion{}\\\{o\_RE\mapsto{} DomainModel\_corresp\_Component(DM), T\_RE\mapsto{} DomainModel\_corresp\_Component(DM)\}$}{true}{}
		\Action{act5}{$Property \bcmeq{} Property \bunion{} \{o\_lg1,o\_lg2\}$}{true}{}
		\Action{act6}{$LogicFormula \bcmeq{} LogicFormula \bunion{} \{o\_lg1,o\_lg2\}$}{true}{}
		\Action{act7}{$Constant\_typing\_Property\bcmeq{}  Constant\_typing\_Property \bunion{} \{T\_RE\mapsto{}o\_lg1, o\_RE\mapsto{}o\_lg2\}$}{true}{}
		\Action{act8}{$Constant\_isInvolvedIn\_LogicFormulas\bcmeq{}  Constant\_isInvolvedIn\_LogicFormulas \bunion{} \{T\_RE\mapsto{}\{1\mapsto{}o\_lg1, 2\mapsto{}o\_lg2\}, o\_RE\mapsto{}\{1\mapsto{}o\_lg2\}\}$}{true}{}
		\Action{act9}{$LogicFormula\_uses\_Operators\bcmeq{}  LogicFormula\_uses\_Operators \bunion{} \{o\_lg1\mapsto{}\{1\mapsto{}RelationSet\_OP\}, o\_lg2\mapsto{}\{1\mapsto{}Belonging\_OP\}\}$}{true}{}
		\Action{act10}{$LogicFormula\_involves\_Sets\bcmeq{}  LogicFormula\_involves\_Sets \bunion{} \{o\_lg1\mapsto{}\{2\mapsto{}o\_CO1, 3\mapsto{}o\_CO2\}, o\_lg2\mapsto{}\emptyset{}\}$}{true}{}
		\Action{act11}{$LogicFormula\_definedIn\_Component\bcmeq{}LogicFormula\_definedIn\_Component\bunion{}\{\\o\_lg1\mapsto{} DomainModel\_corresp\_Component(DM), o\_lg2\mapsto{} DomainModel\_corresp\_Component(DM)\}$}{true}{}
	}
}
\EVT{rule\_10\_2}{false}{ordinary}{}{\\correspondence of an instance of Relation having its isVariable property set to false (case where domain  corresponds to an abstract set and range corresponds to a constant)}{
	\ANY{
		\Param{RE}{true}{}
		\Param{T\_RE}{true}{}
		\Param{o\_RE}{true}{}
		\Param{CO1}{true}{}
		\Param{o\_CO1}{true}{}
		\Param{CO2}{true}{}
		\Param{o\_CO2}{true}{}
		\Param{o\_lg1}{true}{}
		\Param{o\_lg2}{true}{}
		\Param{DM}{true}{}
	}
	\GUARDS{true}{
		\Guard{grd0}{false}{$Relation\_isVariable\converse{}[\{FALSE\}] \setminus{} dom(Relation\_Type) \neq{}\emptyset{}$}{true}{}
		\Guard{grd1}{false}{$RE \in{} Relation\_isVariable\converse{}[\{FALSE\}] \setminus{} dom(Relation\_Type)$}{true}{}
		\Guard{grd2}{false}{$dom(Concept\_corresp\_AbstractSet) \neq{} \emptyset{}$}{true}{}
		\Guard{grd3}{false}{$CO1 = Relation\_domain\_Concept(RE)$}{true}{}
		\Guard{grd4}{false}{$CO1 \in{} dom(Concept\_corresp\_AbstractSet)$}{true}{}
		\Guard{grd5}{false}{$dom(Concept\_corresp\_Constant) \neq{} \emptyset{}$}{true}{}
		\Guard{grd6}{false}{$CO2 = Relation\_range\_Concept(RE)$}{true}{}
		\Guard{grd7}{false}{$CO2 \in{} dom(Concept\_corresp\_Constant)$}{true}{}
		\Guard{grd8}{false}{$Relation\_definedIn\_DomainModel(RE) \in{} dom(DomainModel\_corresp\_Component)$}{true}{}
		\Guard{grd9}{false}{$Constant\_Set \setminus{} Constant \neq{}\emptyset{}$}{true}{}
		\Guard{grd10}{false}{$\{T\_RE, o\_RE\} \subseteq{} Constant\_Set \setminus{} Constant$}{true}{}
		\Guard{grd11}{false}{$LogicFormula\_Set \setminus{} LogicFormula \neq{} \emptyset{}$}{true}{}
		\Guard{grd12}{false}{$\{o\_lg1,o\_lg2\}  \subseteq{} LogicFormula\_Set \setminus{} LogicFormula$}{true}{}
		\Guard{grd13}{false}{$o\_CO1 = Concept\_corresp\_AbstractSet(CO1)$}{true}{}
		\Guard{grd14}{false}{$o\_CO2 = Concept\_corresp\_Constant(CO2)$}{true}{}
		\Guard{grd15}{false}{$DM = Relation\_definedIn\_DomainModel(RE)$}{true}{}
		\Guard{grd16}{false}{$T\_RE \neq{}  o\_RE$}{true}{}
		\Guard{grd17}{false}{$o\_lg1 \neq{}  o\_lg2$}{true}{}
	}
	\ACTIONS{true}{
		\Action{act1}{$Constant \bcmeq{}  Constant \bunion{}  \{T\_RE, o\_RE\}$}{true}{}
		\Action{act2}{$Relation\_Type(RE)\bcmeq{}T\_RE$}{true}{}
		\Action{act3}{$Relation\_corresp\_Constant(RE)\bcmeq{}o\_RE$}{true}{}
		\Action{act4}{$Constant\_definedIn\_Component\bcmeq{}Constant\_definedIn\_Component\bunion{}\\\{o\_RE\mapsto{} DomainModel\_corresp\_Component(DM), T\_RE\mapsto{} DomainModel\_corresp\_Component(DM)\}$}{true}{}
		\Action{act5}{$Property \bcmeq{} Property \bunion{} \{o\_lg1,o\_lg2\}$}{true}{}
		\Action{act6}{$LogicFormula \bcmeq{} LogicFormula \bunion{} \{o\_lg1,o\_lg2\}$}{true}{}
		\Action{act7}{$Constant\_typing\_Property\bcmeq{}  Constant\_typing\_Property \bunion{} \{T\_RE\mapsto{}o\_lg1, o\_RE\mapsto{}o\_lg2\}$}{true}{}
		\Action{act8}{$Constant\_isInvolvedIn\_LogicFormulas\bcmeq{}  Constant\_isInvolvedIn\_LogicFormulas \ovl{} \{T\_RE\mapsto{}\{1\mapsto{}o\_lg1, 2\mapsto{}o\_lg2\}, o\_RE\mapsto{}\{1\mapsto{}o\_lg2\}, o\_CO2\mapsto{}\{3\mapsto{}o\_lg1\}\bunion{}Constant\_isInvolvedIn\_LogicFormulas(o\_CO2)\}$}{true}{}
		\Action{act9}{$LogicFormula\_uses\_Operators\bcmeq{}  LogicFormula\_uses\_Operators \bunion{} \{o\_lg1\mapsto{}\{1\mapsto{}RelationSet\_OP\}, o\_lg2\mapsto{}\{2\mapsto{}Belonging\_OP\}\}$}{true}{}
		\Action{act10}{$LogicFormula\_involves\_Sets\bcmeq{}  LogicFormula\_involves\_Sets \bunion{} \{o\_lg1\mapsto{}\{2\mapsto{}o\_CO1\}, o\_lg2\mapsto{}\emptyset{}\}$}{true}{}
		\Action{act11}{$LogicFormula\_definedIn\_Component\bcmeq{}LogicFormula\_definedIn\_Component\bunion{}\{\\o\_lg1\mapsto{} DomainModel\_corresp\_Component(DM), o\_lg2\mapsto{} DomainModel\_corresp\_Component(DM)\}$}{true}{}
	}
}
\EVT{rule\_10\_3}{false}{ordinary}{}{\\correspondence of an instance of Relation having its isVariable property set to false (case where range  corresponds to an abstract set and domain corresponds to a constant)}{
	\ANY{
		\Param{RE}{true}{}
		\Param{T\_RE}{true}{}
		\Param{o\_RE}{true}{}
		\Param{CO1}{true}{}
		\Param{o\_CO1}{true}{}
		\Param{CO2}{true}{}
		\Param{o\_CO2}{true}{}
		\Param{o\_lg1}{true}{}
		\Param{o\_lg2}{true}{}
		\Param{DM}{true}{}
	}
	\GUARDS{true}{
		\Guard{grd0}{false}{$Relation\_isVariable\converse{}[\{FALSE\}] \setminus{} dom(Relation\_Type) \neq{}\emptyset{}$}{true}{}
		\Guard{grd1}{false}{$RE \in{} Relation\_isVariable\converse{}[\{FALSE\}] \setminus{} dom(Relation\_Type)$}{true}{}
		\Guard{grd2}{false}{$dom(Concept\_corresp\_Constant) \neq{} \emptyset{}$}{true}{}
		\Guard{grd3}{false}{$CO1 = Relation\_domain\_Concept(RE)$}{true}{}
		\Guard{grd4}{false}{$CO1 \in{} dom(Concept\_corresp\_Constant)$}{true}{}
		\Guard{grd5}{false}{$dom(Concept\_corresp\_AbstractSet) \neq{} \emptyset{}$}{true}{}
		\Guard{grd6}{false}{$CO2 = Relation\_range\_Concept(RE)$}{true}{}
		\Guard{grd7}{false}{$CO2 \in{} dom(Concept\_corresp\_AbstractSet)$}{true}{}
		\Guard{grd8}{false}{$Relation\_definedIn\_DomainModel(RE) \in{} dom(DomainModel\_corresp\_Component)$}{true}{}
		\Guard{grd9}{false}{$Constant\_Set \setminus{} Constant \neq{}\emptyset{}$}{true}{}
		\Guard{grd10}{false}{$\{T\_RE, o\_RE\} \subseteq{} Constant\_Set \setminus{} Constant$}{true}{}
		\Guard{grd11}{false}{$LogicFormula\_Set \setminus{} LogicFormula \neq{} \emptyset{}$}{true}{}
		\Guard{grd12}{false}{$\{o\_lg1,o\_lg2\}  \subseteq{} LogicFormula\_Set \setminus{} LogicFormula$}{true}{}
		\Guard{grd13}{false}{$o\_CO2 = Concept\_corresp\_AbstractSet(CO2)$}{true}{}
		\Guard{grd14}{false}{$o\_CO1 = Concept\_corresp\_Constant(CO1)$}{true}{}
		\Guard{grd15}{false}{$DM = Relation\_definedIn\_DomainModel(RE)$}{true}{}
		\Guard{grd16}{false}{$T\_RE \neq{}  o\_RE$}{true}{}
		\Guard{grd17}{false}{$o\_lg1 \neq{}  o\_lg2$}{true}{}
	}
	\ACTIONS{true}{
		\Action{act1}{$Constant \bcmeq{}  Constant \bunion{}  \{T\_RE, o\_RE\}$}{true}{}
		\Action{act2}{$Relation\_Type(RE)\bcmeq{}T\_RE$}{true}{}
		\Action{act3}{$Relation\_corresp\_Constant(RE)\bcmeq{}o\_RE$}{true}{}
		\Action{act4}{$Constant\_definedIn\_Component\bcmeq{}Constant\_definedIn\_Component\bunion{}\\\{o\_RE\mapsto{} DomainModel\_corresp\_Component(DM), T\_RE\mapsto{} DomainModel\_corresp\_Component(DM)\}$}{true}{}
		\Action{act5}{$Property \bcmeq{} Property \bunion{} \{o\_lg1,o\_lg2\}$}{true}{}
		\Action{act6}{$LogicFormula \bcmeq{} LogicFormula \bunion{} \{o\_lg1,o\_lg2\}$}{true}{}
		\Action{act7}{$Constant\_typing\_Property\bcmeq{}  Constant\_typing\_Property \bunion{} \{T\_RE\mapsto{}o\_lg1, o\_RE\mapsto{}o\_lg2\}$}{true}{}
		\Action{act8}{$Constant\_isInvolvedIn\_LogicFormulas\bcmeq{}  Constant\_isInvolvedIn\_LogicFormulas \ovl{} \{T\_RE\mapsto{}\{1\mapsto{}o\_lg1, 2\mapsto{}o\_lg2\}, o\_RE\mapsto{}\{1\mapsto{}o\_lg2\}, o\_CO1\mapsto{}\{2\mapsto{}o\_lg1\}\bunion{}Constant\_isInvolvedIn\_LogicFormulas(o\_CO1)\}$}{true}{}
		\Action{act9}{$LogicFormula\_uses\_Operators\bcmeq{}  LogicFormula\_uses\_Operators \bunion{} \{o\_lg1\mapsto{}\{1\mapsto{}RelationSet\_OP\}, o\_lg2\mapsto{}\{1\mapsto{}Belonging\_OP\}\}$}{true}{}
		\Action{act10}{$LogicFormula\_involves\_Sets\bcmeq{}  LogicFormula\_involves\_Sets \bunion{} \{o\_lg1\mapsto{}\{3\mapsto{}o\_CO2\}, o\_lg2\mapsto{}\emptyset{}\}$}{true}{}
		\Action{act11}{$LogicFormula\_definedIn\_Component\bcmeq{}LogicFormula\_definedIn\_Component\bunion{}\{\\o\_lg1\mapsto{} DomainModel\_corresp\_Component(DM), o\_lg2\mapsto{} DomainModel\_corresp\_Component(DM)\}$}{true}{}
	}
}
\EVT{rule\_10\_4}{false}{ordinary}{}{\\correspondence of an instance of Relation having its isVariable property set to false (case where domain and range correspond to constants)}{
	\ANY{
		\Param{RE}{true}{}
		\Param{T\_RE}{true}{}
		\Param{o\_RE}{true}{}
		\Param{CO1}{true}{}
		\Param{o\_CO1}{true}{}
		\Param{CO2}{true}{}
		\Param{o\_CO2}{true}{}
		\Param{o\_lg1}{true}{}
		\Param{o\_lg2}{true}{}
		\Param{DM}{true}{}
	}
	\GUARDS{true}{
		\Guard{grd0}{false}{$Relation\_isVariable\converse{}[\{FALSE\}] \setminus{} dom(Relation\_Type) \neq{}\emptyset{}$}{true}{}
		\Guard{grd1}{false}{$RE \in{} Relation\_isVariable\converse{}[\{FALSE\}] \setminus{} dom(Relation\_Type)$}{true}{}
		\Guard{grd2}{false}{$dom(Concept\_corresp\_Constant) \neq{} \emptyset{}$}{true}{}
		\Guard{grd3}{false}{$CO1 = Relation\_domain\_Concept(RE)$}{true}{}
		\Guard{grd4}{false}{$CO2 = Relation\_range\_Concept(RE)$}{true}{}
		\Guard{grd5}{false}{$\{CO1, CO2\} \subseteq{} dom(Concept\_corresp\_Constant)$}{true}{}
		\Guard{grd6}{false}{$Relation\_definedIn\_DomainModel(RE) \in{} dom(DomainModel\_corresp\_Component)$}{true}{}
		\Guard{grd7}{false}{$Constant\_Set \setminus{} Constant \neq{}\emptyset{}$}{true}{}
		\Guard{grd8}{false}{$\{T\_RE, o\_RE\} \subseteq{} Constant\_Set \setminus{} Constant$}{true}{}
		\Guard{grd9}{false}{$LogicFormula\_Set \setminus{} LogicFormula \neq{} \emptyset{}$}{true}{}
		\Guard{grd10}{false}{$\{o\_lg1,o\_lg2\}  \subseteq{} LogicFormula\_Set \setminus{} LogicFormula$}{true}{}
		\Guard{grd11}{false}{$o\_CO1 = Concept\_corresp\_Constant(CO1)$}{true}{}
		\Guard{grd12}{false}{$o\_CO2 = Concept\_corresp\_Constant(CO2)$}{true}{}
		\Guard{grd13}{false}{$DM = Relation\_definedIn\_DomainModel(RE)$}{true}{}
		\Guard{grd14}{false}{$T\_RE \neq{}  o\_RE$}{true}{}
		\Guard{grd15}{false}{$o\_lg1 \neq{}  o\_lg2$}{true}{}
		\Guard{grd16}{false}{$o\_CO1 \neq{}  o\_CO2$}{true}{}
	}
	\ACTIONS{true}{
		\Action{act1}{$Constant \bcmeq{}  Constant \bunion{}  \{T\_RE, o\_RE\}$}{true}{}
		\Action{act2}{$Relation\_Type(RE)\bcmeq{}T\_RE$}{true}{}
		\Action{act3}{$Relation\_corresp\_Constant(RE)\bcmeq{}o\_RE$}{true}{}
		\Action{act4}{$Constant\_definedIn\_Component\bcmeq{}Constant\_definedIn\_Component\bunion{}\\\{o\_RE\mapsto{} DomainModel\_corresp\_Component(DM), T\_RE\mapsto{} DomainModel\_corresp\_Component(DM)\}$}{true}{}
		\Action{act5}{$Property \bcmeq{} Property \bunion{} \{o\_lg1,o\_lg2\}$}{true}{}
		\Action{act6}{$LogicFormula \bcmeq{} LogicFormula \bunion{} \{o\_lg1,o\_lg2\}$}{true}{}
		\Action{act7}{$Constant\_typing\_Property\bcmeq{}  Constant\_typing\_Property \bunion{} \{T\_RE\mapsto{}o\_lg1, o\_RE\mapsto{}o\_lg2\}$}{true}{}
		\Action{act8}{$Constant\_isInvolvedIn\_LogicFormulas\bcmeq{}  Constant\_isInvolvedIn\_LogicFormulas \ovl{} \{T\_RE\mapsto{}\{1\mapsto{}o\_lg1, 2\mapsto{}o\_lg2\}, o\_RE\mapsto{}\{1\mapsto{}o\_lg2\}, o\_CO1\mapsto{}\{2\mapsto{}o\_lg1\}\bunion{}Constant\_isInvolvedIn\_LogicFormulas(o\_CO1),\\ o\_CO2\mapsto{}\{3\mapsto{}o\_lg1\}\bunion{}Constant\_isInvolvedIn\_LogicFormulas(o\_CO2)\}$}{true}{}
		\Action{act9}{$LogicFormula\_uses\_Operators\bcmeq{}  LogicFormula\_uses\_Operators \bunion{} \{o\_lg1\mapsto{}\{1\mapsto{}RelationSet\_OP\}, o\_lg2\mapsto{}\{1\mapsto{}Belonging\_OP\}\}$}{true}{}
		\Action{act10}{$LogicFormula\_involves\_Sets\bcmeq{}  LogicFormula\_involves\_Sets \bunion{} \{o\_lg1\mapsto{}\emptyset{}, o\_lg2\mapsto{}\emptyset{}\}$}{true}{}
		\Action{act11}{$LogicFormula\_definedIn\_Component\bcmeq{}LogicFormula\_definedIn\_Component\bunion{}\{\\o\_lg1\mapsto{} DomainModel\_corresp\_Component(DM), o\_lg2\mapsto{} DomainModel\_corresp\_Component(DM)\}$}{true}{}
	}
}

\END

\paragraph{Rule 11 : relation and attribute maplet}

\MACHINE{Ontologies\_BSystem\_specs\_translation\_ref\_1}{Ontologies\_BSystem\_specs\_translation}{BSystem\_Metamodel\_Context,Domain\_Metamodel\_Context}{}

\EVT{rule\_11\_1}{false}{ordinary}{}{\\correspondence of an instance of RelationMaplet}{
	\ANY{
		\Param{remap}{true}{}
		\Param{o\_remap}{true}{}
		\Param{RE}{true}{}
		\Param{antecedent}{true}{}
		\Param{image}{true}{}
		\Param{o\_lg}{true}{}
		\Param{o\_antecedent}{true}{}
		\Param{o\_image}{true}{}
	}
	\GUARDS{true}{
		\Guard{grd0}{false}{$RelationMaplet \setminus{} dom(RelationMaplet\_corresp\_Constant) \neq{}\emptyset{}$}{true}{}
		\Guard{grd1}{false}{$remap \in{} RelationMaplet \setminus{} dom(RelationMaplet\_corresp\_Constant)$}{true}{}
		\Guard{grd2}{false}{$dom(Relation\_corresp\_Constant)\bunion{}dom(Relation\_corresp\_Variable) \neq{} \emptyset{}$}{true}{}
		\Guard{grd3}{false}{$RelationMaplet\_mapletOf\_Relation(remap)=RE$}{true}{}
		\Guard{grd4}{false}{$RE \in{} dom(Relation\_corresp\_Constant)\bunion{}dom(Relation\_corresp\_Variable)$}{true}{}
		\Guard{grd5}{false}{$Relation\_definedIn\_DomainModel(RE) \in{} dom(DomainModel\_corresp\_Component)$}{true}{}
		\Guard{grd6}{false}{$Constant\_Set \setminus{} Constant \neq{}\emptyset{}$}{true}{}
		\Guard{grd7}{false}{$o\_remap \in{} Constant\_Set \setminus{} Constant$}{true}{}
		\Guard{grd8}{false}{$LogicFormula\_Set \setminus{} LogicFormula \neq{} \emptyset{}$}{true}{}
		\Guard{grd9}{false}{$o\_lg \in{} LogicFormula\_Set \setminus{} LogicFormula$}{true}{}
		\Guard{grd10}{false}{$antecedent = RelationMaplet\_antecedent\_Individual(remap)$}{true}{}
		\Guard{grd11}{false}{$image = RelationMaplet\_image\_Individual(remap)$}{true}{}
		\Guard{grd12}{false}{$\{antecedent, image\} \subseteq{}    dom(Individual\_corresp\_Constant)$}{true}{}
		\Guard{grd13}{false}{$o\_antecedent = Individual\_corresp\_Constant(antecedent)$}{true}{}
		\Guard{grd14}{false}{$o\_image = Individual\_corresp\_Constant(image)$}{true}{}
		\Guard{grd15}{false}{$o\_antecedent \neq{}  o\_image$}{true}{\\then, for each relation already treated for which all the maplets have been processed,\\~        if it is variable, we generate the initialization, otherwise, we generate the closure property,\\~        knowing that the maplets give rise to variables in case of variable relation and constants\\~        in case of constant relationship }
	}
	\ACTIONS{true}{
		\Action{act1}{$Constant \bcmeq{}  Constant \bunion{}  \{o\_remap\}$}{true}{}
		\Action{act2}{$RelationMaplet\_corresp\_Constant(remap)\bcmeq{}o\_remap$}{true}{}
		\Action{act3}{$Constant\_definedIn\_Component(o\_remap) \bcmeq{} DomainModel\_corresp\_Component(\\Relation\_definedIn\_DomainModel(RE))$}{true}{}
		\Action{act4}{$Property \bcmeq{} Property \bunion{} \{o\_lg\}$}{true}{}
		\Action{act5}{$LogicFormula \bcmeq{} LogicFormula \bunion{} \{o\_lg\}$}{true}{}
		\Action{act6}{$LogicFormula\_uses\_Operators(o\_lg) \bcmeq{} \{1\mapsto{}Maplet\_OP\}$}{true}{}
		\Action{act7}{$Constant\_isInvolvedIn\_LogicFormulas\bcmeq{}  Constant\_isInvolvedIn\_LogicFormulas \ovl{} \{o\_remap\mapsto{}\{1\mapsto{}o\_lg\}, o\_antecedent\mapsto{}\{2\mapsto{}o\_lg\}\bunion{}Constant\_isInvolvedIn\_LogicFormulas(o\_antecedent), o\_image\mapsto{}\{3\mapsto{}o\_lg\}\bunion{}Constant\_isInvolvedIn\_LogicFormulas(o\_image)\}$}{true}{}
		\Action{act8}{$LogicFormula\_involves\_Sets(o\_lg) \bcmeq{} \emptyset{}$}{true}{}
		\Action{act9}{$LogicFormula\_definedIn\_Component(o\_lg) \bcmeq{} DomainModel\_corresp\_Component(\\Relation\_definedIn\_DomainModel(RE))$}{true}{}
		\Action{act10}{$Constant\_typing\_Property(o\_remap) \bcmeq{}  o\_lg$}{true}{}
	}
}
\EVT{rule\_11\_2\_1}{false}{ordinary}{}{\\correspondence  of an instance of AttributeMaplet (case where the image (of type DataValue) corresponds to a constant (it can also corresponds to a set item)}{
	\ANY{
		\Param{atmap}{true}{}
		\Param{o\_atmap}{true}{}
		\Param{AT}{true}{}
		\Param{antecedent}{true}{}
		\Param{image}{true}{}
		\Param{o\_lg}{true}{}
		\Param{o\_antecedent}{true}{}
		\Param{o\_image}{true}{}
	}
	\GUARDS{true}{
		\Guard{grd0}{false}{$AttributeMaplet \setminus{} dom(AttributeMaplet\_corresp\_Constant) \neq{}\emptyset{}$}{true}{}
		\Guard{grd1}{false}{$atmap \in{} AttributeMaplet \setminus{} dom(AttributeMaplet\_corresp\_Constant)$}{true}{}
		\Guard{grd2}{false}{$dom(Attribute\_corresp\_Constant)\bunion{}dom(Attribute\_corresp\_Variable) \neq{} \emptyset{}$}{true}{}
		\Guard{grd3}{false}{$AttributeMaplet\_mapletOf\_Attribute(atmap)=AT$}{true}{}
		\Guard{grd4}{false}{$AT \in{} dom(Attribute\_corresp\_Constant)\bunion{}dom(Attribute\_corresp\_Variable)$}{true}{}
		\Guard{grd5}{false}{$Attribute\_definedIn\_DomainModel(AT) \in{} dom(DomainModel\_corresp\_Component)$}{true}{}
		\Guard{grd6}{false}{$Constant\_Set \setminus{} Constant \neq{}\emptyset{}$}{true}{}
		\Guard{grd7}{false}{$o\_atmap \in{} Constant\_Set \setminus{} Constant$}{true}{}
		\Guard{grd8}{false}{$LogicFormula\_Set \setminus{} LogicFormula \neq{} \emptyset{}$}{true}{}
		\Guard{grd9}{false}{$o\_lg \in{} LogicFormula\_Set \setminus{} LogicFormula$}{true}{}
		\Guard{grd10}{false}{$antecedent = AttributeMaplet\_antecedent\_Individual(atmap)$}{true}{}
		\Guard{grd11}{false}{$image = AttributeMaplet\_image\_DataValue(atmap)$}{true}{}
		\Guard{grd12}{false}{$antecedent \in{}    dom(Individual\_corresp\_Constant)$}{true}{}
		\Guard{grd13}{false}{$image \in{}    dom(DataValue\_corresp\_Constant)$}{true}{}
		\Guard{grd14}{false}{$o\_antecedent = Individual\_corresp\_Constant(antecedent)$}{true}{}
		\Guard{grd15}{false}{$o\_image = DataValue\_corresp\_Constant(image)$}{true}{}
		\Guard{grd16}{false}{$o\_antecedent \neq{}  o\_image$}{true}{}
	}
	\ACTIONS{true}{
		\Action{act1}{$Constant \bcmeq{}  Constant \bunion{}  \{o\_atmap\}$}{true}{}
		\Action{act2}{$AttributeMaplet\_corresp\_Constant(atmap)\bcmeq{}o\_atmap$}{true}{}
		\Action{act3}{$Constant\_definedIn\_Component(o\_atmap) \bcmeq{} DomainModel\_corresp\_Component(\\Attribute\_definedIn\_DomainModel(AT))$}{true}{}
		\Action{act4}{$Property \bcmeq{} Property \bunion{} \{o\_lg\}$}{true}{}
		\Action{act5}{$LogicFormula \bcmeq{} LogicFormula \bunion{} \{o\_lg\}$}{true}{}
		\Action{act6}{$LogicFormula\_uses\_Operators(o\_lg) \bcmeq{} \{1\mapsto{}Maplet\_OP\}$}{true}{}
		\Action{act7}{$Constant\_isInvolvedIn\_LogicFormulas\bcmeq{}  Constant\_isInvolvedIn\_LogicFormulas \ovl{} \{o\_atmap\mapsto{}\{1\mapsto{}o\_lg\}, o\_antecedent\mapsto{}\{2\mapsto{}o\_lg\}\bunion{}Constant\_isInvolvedIn\_LogicFormulas(o\_antecedent), o\_image\mapsto{}\{3\mapsto{}o\_lg\}\bunion{}Constant\_isInvolvedIn\_LogicFormulas(o\_image)\}$}{true}{}
		\Action{act8}{$LogicFormula\_involves\_Sets(o\_lg) \bcmeq{} \emptyset{}$}{true}{}
		\Action{act9}{$LogicFormula\_definedIn\_Component(o\_lg) \bcmeq{} DomainModel\_corresp\_Component(\\Attribute\_definedIn\_DomainModel(AT))$}{true}{}
		\Action{act10}{$Constant\_typing\_Property(o\_atmap) \bcmeq{}  o\_lg$}{true}{}
	}
}
\EVT{rule\_11\_2\_2}{false}{ordinary}{}{\\correspondence  of an instance of AttributeMaplet (case where the image (of type DataValue) corresponds to a set item}{
	\ANY{
		\Param{atmap}{true}{}
		\Param{o\_atmap}{true}{}
		\Param{AT}{true}{}
		\Param{antecedent}{true}{}
		\Param{image}{true}{}
		\Param{o\_lg}{true}{}
		\Param{o\_antecedent}{true}{}
		\Param{o\_image}{true}{}
	}
	\GUARDS{true}{
		\Guard{grd0}{false}{$AttributeMaplet \setminus{} dom(AttributeMaplet\_corresp\_Constant) \neq{}\emptyset{}$}{true}{}
		\Guard{grd1}{false}{$atmap \in{} AttributeMaplet \setminus{} dom(AttributeMaplet\_corresp\_Constant)$}{true}{}
		\Guard{grd2}{false}{$dom(Attribute\_corresp\_Constant)\bunion{}dom(Attribute\_corresp\_Variable) \neq{} \emptyset{}$}{true}{}
		\Guard{grd3}{false}{$AttributeMaplet\_mapletOf\_Attribute(atmap)=AT$}{true}{}
		\Guard{grd4}{false}{$AT \in{} dom(Attribute\_corresp\_Constant)\bunion{}dom(Attribute\_corresp\_Variable)$}{true}{}
		\Guard{grd5}{false}{$Attribute\_definedIn\_DomainModel(AT) \in{} dom(DomainModel\_corresp\_Component)$}{true}{}
		\Guard{grd6}{false}{$Constant\_Set \setminus{} Constant \neq{}\emptyset{}$}{true}{}
		\Guard{grd7}{false}{$o\_atmap \in{} Constant\_Set \setminus{} Constant$}{true}{}
		\Guard{grd8}{false}{$LogicFormula\_Set \setminus{} LogicFormula \neq{} \emptyset{}$}{true}{}
		\Guard{grd9}{false}{$o\_lg \in{} LogicFormula\_Set \setminus{} LogicFormula$}{true}{}
		\Guard{grd10}{false}{$antecedent = AttributeMaplet\_antecedent\_Individual(atmap)$}{true}{}
		\Guard{grd11}{false}{$image = AttributeMaplet\_image\_DataValue(atmap)$}{true}{}
		\Guard{grd12}{false}{$antecedent \in{}    dom(Individual\_corresp\_Constant)$}{true}{}
		\Guard{grd13}{false}{$image \in{}    dom(DataValue\_corresp\_SetItem)$}{true}{}
		\Guard{grd14}{false}{$o\_antecedent = Individual\_corresp\_Constant(antecedent)$}{true}{}
		\Guard{grd15}{false}{$o\_image = DataValue\_corresp\_SetItem(image)$}{true}{}
	}
	\ACTIONS{true}{
		\Action{act1}{$Constant \bcmeq{}  Constant \bunion{}  \{o\_atmap\}$}{true}{}
		\Action{act2}{$AttributeMaplet\_corresp\_Constant(atmap)\bcmeq{}o\_atmap$}{true}{}
		\Action{act3}{$Constant\_definedIn\_Component(o\_atmap) \bcmeq{} DomainModel\_corresp\_Component(\\Attribute\_definedIn\_DomainModel(AT))$}{true}{}
		\Action{act4}{$Property \bcmeq{} Property \bunion{} \{o\_lg\}$}{true}{}
		\Action{act5}{$LogicFormula \bcmeq{} LogicFormula \bunion{} \{o\_lg\}$}{true}{}
		\Action{act6}{$LogicFormula\_uses\_Operators(o\_lg) \bcmeq{} \{1\mapsto{}Maplet\_OP\}$}{true}{}
		\Action{act7}{$Constant\_isInvolvedIn\_LogicFormulas\bcmeq{}  Constant\_isInvolvedIn\_LogicFormulas \ovl{} \{o\_atmap\mapsto{}\{1\mapsto{}o\_lg\}, o\_antecedent\mapsto{}\{2\mapsto{}o\_lg\}\bunion{}Constant\_isInvolvedIn\_LogicFormulas(o\_antecedent)\}$}{true}{}
		\Action{act8}{$LogicFormula\_involves\_Sets(o\_lg) \bcmeq{} \emptyset{}$}{true}{}
		\Action{act9}{$LogicFormula\_involves\_SetItems(o\_lg) \bcmeq{} \{3\mapsto{}o\_image\}$}{true}{}
		\Action{act10}{$LogicFormula\_definedIn\_Component(o\_lg) \bcmeq{} DomainModel\_corresp\_Component(\\Attribute\_definedIn\_DomainModel(AT))$}{true}{}
		\Action{act11}{$Constant\_typing\_Property(o\_atmap) \bcmeq{}  o\_lg$}{true}{}
	}
}

\END

Each relation   gives rise to a constant representing the type of its associated \textit{B System} element and defined as the set of relations between the \textit{B System} element corresponding to the relation domain and the one corresponding to the relation range.  Moreover, if the relation has its \textsf{isVariable} attribute  set to \textit{false}, it is translated through a second constant. 
\textbf{\underline{Example : }}  in Figure \ref{lgsystem_event_b_model_refinment_1},  \textbf{\textit{LgOfHd}}, for which \textsf{isVariable} is set to \textit{false}, is translated into  a  constant named   \textit{LgOfHd} and having as type \textbf{\textit{T\_LgOfHd}} defined as the set of relations between \textbf{\textit{Handle}} and \textbf{\textit{LandingGear}} (assertions \textit{\textsf{1.7}} and \textit{\textsf{1.8}}).

\paragraph{Rule 14 : Constant  attribute}
\MACHINE{Ontologies\_BSystem\_specs\_translation\_ref\_1}{Ontologies\_BSystem\_specs\_translation}{BSystem\_Metamodel\_Context,Domain\_Metamodel\_Context}{}

\EVT{rule\_14\_1}{false}{ordinary}{}{\\correspondence of an instance of Attribute having its isVariable property set to false and its isFunctional property set to false (case where the domain  corresponds to an abstract set, knowing that the range always corresponds to a set )}{
	\ANY{
		\Param{AT}{true}{}
		\Param{T\_AT}{true}{}
		\Param{o\_AT}{true}{}
		\Param{CO}{true}{}
		\Param{o\_CO}{true}{}
		\Param{DS}{true}{}
		\Param{o\_DS}{true}{}
		\Param{o\_lg1}{true}{}
		\Param{o\_lg2}{true}{}
		\Param{DM}{true}{}
	}
	\GUARDS{true}{
		\Guard{grd0}{false}{$Attribute\_isVariable\converse{}[\{FALSE\}] \setminus{} dom(Attribute\_Type) \neq{}\emptyset{}$}{true}{}
		\Guard{grd1}{false}{$AT \in{} Attribute\_isVariable\converse{}[\{FALSE\}] \setminus{} dom(Attribute\_Type)$}{true}{}
		\Guard{grd2}{false}{$dom(Concept\_corresp\_AbstractSet) \neq{} \emptyset{}$}{true}{}
		\Guard{grd3}{false}{$CO = Attribute\_domain\_Concept(AT)$}{true}{}
		\Guard{grd4}{false}{$CO \in{} dom(Concept\_corresp\_AbstractSet)$}{true}{}
		\Guard{grd5}{false}{$dom(DataSet\_corresp\_Set) \neq{} \emptyset{}$}{true}{}
		\Guard{grd6}{false}{$DS = Attribute\_range\_DataSet(AT)$}{true}{}
		\Guard{grd7}{false}{$DS \in{} dom(DataSet\_corresp\_Set)$}{true}{}
		\Guard{grd8}{false}{$Attribute\_definedIn\_DomainModel(AT) \in{} dom(DomainModel\_corresp\_Component)$}{true}{}
		\Guard{grd9}{false}{$Constant\_Set \setminus{} Constant \neq{}\emptyset{}$}{true}{}
		\Guard{grd10}{false}{$\{T\_AT, o\_AT\} \subseteq{} Constant\_Set \setminus{} Constant$}{true}{}
		\Guard{grd11}{false}{$LogicFormula\_Set \setminus{} LogicFormula \neq{} \emptyset{}$}{true}{}
		\Guard{grd12}{false}{$\{o\_lg1,o\_lg2\}  \subseteq{} LogicFormula\_Set \setminus{} LogicFormula$}{true}{}
		\Guard{grd13}{false}{$o\_CO = Concept\_corresp\_AbstractSet(CO)$}{true}{}
		\Guard{grd14}{false}{$o\_DS = DataSet\_corresp\_Set(DS)$}{true}{}
		\Guard{grd15}{false}{$DM = Attribute\_definedIn\_DomainModel(AT)$}{true}{}
		\Guard{grd16}{false}{$T\_AT \neq{}  o\_AT$}{true}{}
		\Guard{grd17}{false}{$o\_lg1 \neq{}  o\_lg2$}{true}{}
		\Guard{grd18}{false}{$AT \in{} Attribute\_isFunctional\converse{}[\{FALSE\}]$}{true}{}
	}
	\ACTIONS{true}{
		\Action{act1}{$Constant \bcmeq{}  Constant \bunion{}  \{T\_AT, o\_AT\}$}{true}{}
		\Action{act2}{$Attribute\_Type(AT)\bcmeq{}T\_AT$}{true}{}
		\Action{act3}{$Attribute\_corresp\_Constant(AT)\bcmeq{}o\_AT$}{true}{}
		\Action{act4}{$Constant\_definedIn\_Component\bcmeq{}Constant\_definedIn\_Component\bunion{}\\\{o\_AT\mapsto{} DomainModel\_corresp\_Component(DM), T\_AT\mapsto{} DomainModel\_corresp\_Component(DM)\}$}{true}{}
		\Action{act5}{$Property \bcmeq{} Property \bunion{} \{o\_lg1,o\_lg2\}$}{true}{}
		\Action{act6}{$LogicFormula \bcmeq{} LogicFormula \bunion{} \{o\_lg1,o\_lg2\}$}{true}{}
		\Action{act7}{$Constant\_typing\_Property\bcmeq{}  Constant\_typing\_Property \bunion{} \{T\_AT\mapsto{}o\_lg1, o\_AT\mapsto{}o\_lg2\}$}{true}{}
		\Action{act8}{$Constant\_isInvolvedIn\_LogicFormulas\bcmeq{}  Constant\_isInvolvedIn\_LogicFormulas \bunion{} \{T\_AT\mapsto{}\{1\mapsto{}o\_lg1, 2\mapsto{}o\_lg2\}, o\_AT\mapsto{}\{1\mapsto{}o\_lg2\}\}$}{true}{}
		\Action{act9}{$LogicFormula\_uses\_Operators\bcmeq{}  LogicFormula\_uses\_Operators \bunion{} \{o\_lg1\mapsto{}\{1\mapsto{}RelationSet\_OP\}, o\_lg2\mapsto{}\{1\mapsto{}Belonging\_OP\}\}$}{true}{}
		\Action{act10}{$LogicFormula\_involves\_Sets\bcmeq{}  LogicFormula\_involves\_Sets \bunion{} \{o\_lg1\mapsto{}\{2\mapsto{}o\_CO, 3\mapsto{}o\_DS\}, o\_lg2\mapsto{}\emptyset{}\}$}{true}{}
		\Action{act11}{$LogicFormula\_definedIn\_Component\bcmeq{}LogicFormula\_definedIn\_Component\bunion{}\{\\o\_lg1\mapsto{} DomainModel\_corresp\_Component(DM), o\_lg2\mapsto{} DomainModel\_corresp\_Component(DM)\}$}{true}{}
	}
}
\EVT{rule\_14\_2}{false}{ordinary}{}{\\correspondence of an instance of Attribute having its isVariable property set to false and its isFunctional property set to false (case where the domain  corresponds to a constant, knowing that the range always corresponds to a set )}{
	\ANY{
		\Param{AT}{true}{}
		\Param{T\_AT}{true}{}
		\Param{o\_AT}{true}{}
		\Param{CO}{true}{}
		\Param{o\_CO}{true}{}
		\Param{DS}{true}{}
		\Param{o\_DS}{true}{}
		\Param{o\_lg1}{true}{}
		\Param{o\_lg2}{true}{}
		\Param{DM}{true}{}
	}
	\GUARDS{true}{
		\Guard{grd0}{false}{$Attribute\_isVariable\converse{}[\{FALSE\}] \setminus{} dom(Attribute\_Type) \neq{}\emptyset{}$}{true}{}
		\Guard{grd1}{false}{$AT \in{} Attribute\_isVariable\converse{}[\{FALSE\}] \setminus{} dom(Attribute\_Type)$}{true}{}
		\Guard{grd2}{false}{$dom(Concept\_corresp\_Constant) \neq{} \emptyset{}$}{true}{}
		\Guard{grd3}{false}{$CO = Attribute\_domain\_Concept(AT)$}{true}{}
		\Guard{grd4}{false}{$CO \in{} dom(Concept\_corresp\_Constant)$}{true}{}
		\Guard{grd5}{false}{$dom(DataSet\_corresp\_Set) \neq{} \emptyset{}$}{true}{}
		\Guard{grd6}{false}{$DS = Attribute\_range\_DataSet(AT)$}{true}{}
		\Guard{grd7}{false}{$DS \in{} dom(DataSet\_corresp\_Set)$}{true}{}
		\Guard{grd8}{false}{$Attribute\_definedIn\_DomainModel(AT) \in{} dom(DomainModel\_corresp\_Component)$}{true}{}
		\Guard{grd9}{false}{$Constant\_Set \setminus{} Constant \neq{}\emptyset{}$}{true}{}
		\Guard{grd10}{false}{$\{T\_AT, o\_AT\} \subseteq{} Constant\_Set \setminus{} Constant$}{true}{}
		\Guard{grd11}{false}{$LogicFormula\_Set \setminus{} LogicFormula \neq{} \emptyset{}$}{true}{}
		\Guard{grd12}{false}{$\{o\_lg1,o\_lg2\}  \subseteq{} LogicFormula\_Set \setminus{} LogicFormula$}{true}{}
		\Guard{grd13}{false}{$o\_CO = Concept\_corresp\_Constant(CO)$}{true}{}
		\Guard{grd14}{false}{$o\_DS = DataSet\_corresp\_Set(DS)$}{true}{}
		\Guard{grd15}{false}{$DM = Attribute\_definedIn\_DomainModel(AT)$}{true}{}
		\Guard{grd16}{false}{$T\_AT \neq{}  o\_AT$}{true}{}
		\Guard{grd17}{false}{$o\_lg1 \neq{}  o\_lg2$}{true}{}
		\Guard{grd18}{false}{$AT \in{} Attribute\_isFunctional\converse{}[\{FALSE\}]$}{true}{}
	}
	\ACTIONS{true}{
		\Action{act1}{$Constant \bcmeq{}  Constant \bunion{}  \{T\_AT, o\_AT\}$}{true}{}
		\Action{act2}{$Attribute\_Type(AT)\bcmeq{}T\_AT$}{true}{}
		\Action{act3}{$Attribute\_corresp\_Constant(AT)\bcmeq{}o\_AT$}{true}{}
		\Action{act4}{$Constant\_definedIn\_Component\bcmeq{}Constant\_definedIn\_Component\bunion{}\\\{o\_AT\mapsto{} DomainModel\_corresp\_Component(DM), T\_AT\mapsto{} DomainModel\_corresp\_Component(DM)\}$}{true}{}
		\Action{act5}{$Property \bcmeq{} Property \bunion{} \{o\_lg1,o\_lg2\}$}{true}{}
		\Action{act6}{$LogicFormula \bcmeq{} LogicFormula \bunion{} \{o\_lg1,o\_lg2\}$}{true}{}
		\Action{act7}{$Constant\_typing\_Property\bcmeq{}  Constant\_typing\_Property \bunion{} \{T\_AT\mapsto{}o\_lg1, o\_AT\mapsto{}o\_lg2\}$}{true}{}
		\Action{act8}{$Constant\_isInvolvedIn\_LogicFormulas\bcmeq{}  Constant\_isInvolvedIn\_LogicFormulas \ovl{} \{T\_AT\mapsto{}\{1\mapsto{}o\_lg1, 2\mapsto{}o\_lg2\}, o\_AT\mapsto{}\{1\mapsto{}o\_lg2\}, o\_CO\mapsto{}\{2\mapsto{}o\_lg1\}\bunion{}Constant\_isInvolvedIn\_LogicFormulas(o\_CO)\}$}{true}{}
		\Action{act9}{$LogicFormula\_uses\_Operators\bcmeq{}  LogicFormula\_uses\_Operators \bunion{} \{o\_lg1\mapsto{}\{1\mapsto{}RelationSet\_OP\}, o\_lg2\mapsto{}\{1\mapsto{}Belonging\_OP\}\}$}{true}{}
		\Action{act10}{$LogicFormula\_involves\_Sets\bcmeq{}  LogicFormula\_involves\_Sets \bunion{} \{o\_lg1\mapsto{}\{3\mapsto{}o\_DS\}, o\_lg2\mapsto{}\emptyset{}\}$}{true}{}
		\Action{act11}{$LogicFormula\_definedIn\_Component\bcmeq{}LogicFormula\_definedIn\_Component\bunion{}\{\\o\_lg1\mapsto{} DomainModel\_corresp\_Component(DM), o\_lg2\mapsto{} DomainModel\_corresp\_Component(DM)\}$}{true}{}
	}
}
\EVT{rule\_14\_3}{false}{ordinary}{}{\\correspondence of an instance of Attribute having its isVariable property set to false and its isFunctional property set to true (case where the domain  corresponds to an abstract set, knowing that the range always corresponds to a set )}{
	\ANY{
		\Param{AT}{true}{}
		\Param{T\_AT}{true}{}
		\Param{o\_AT}{true}{}
		\Param{CO}{true}{}
		\Param{o\_CO}{true}{}
		\Param{DS}{true}{}
		\Param{o\_DS}{true}{}
		\Param{o\_lg1}{true}{}
		\Param{o\_lg2}{true}{}
		\Param{DM}{true}{}
	}
	\GUARDS{true}{
		\Guard{grd0}{false}{$Attribute\_isVariable\converse{}[\{FALSE\}] \setminus{} dom(Attribute\_Type) \neq{}\emptyset{}$}{true}{}
		\Guard{grd1}{false}{$AT \in{} Attribute\_isVariable\converse{}[\{FALSE\}] \setminus{} dom(Attribute\_Type)$}{true}{}
		\Guard{grd2}{false}{$dom(Concept\_corresp\_AbstractSet) \neq{} \emptyset{}$}{true}{}
		\Guard{grd3}{false}{$CO = Attribute\_domain\_Concept(AT)$}{true}{}
		\Guard{grd4}{false}{$CO \in{} dom(Concept\_corresp\_AbstractSet)$}{true}{}
		\Guard{grd5}{false}{$dom(DataSet\_corresp\_Set) \neq{} \emptyset{}$}{true}{}
		\Guard{grd6}{false}{$DS = Attribute\_range\_DataSet(AT)$}{true}{}
		\Guard{grd7}{false}{$DS \in{} dom(DataSet\_corresp\_Set)$}{true}{}
		\Guard{grd8}{false}{$Attribute\_definedIn\_DomainModel(AT) \in{} dom(DomainModel\_corresp\_Component)$}{true}{}
		\Guard{grd9}{false}{$Constant\_Set \setminus{} Constant \neq{}\emptyset{}$}{true}{}
		\Guard{grd10}{false}{$\{T\_AT, o\_AT\} \subseteq{} Constant\_Set \setminus{} Constant$}{true}{}
		\Guard{grd11}{false}{$LogicFormula\_Set \setminus{} LogicFormula \neq{} \emptyset{}$}{true}{}
		\Guard{grd12}{false}{$\{o\_lg1,o\_lg2\}  \subseteq{} LogicFormula\_Set \setminus{} LogicFormula$}{true}{}
		\Guard{grd13}{false}{$o\_CO = Concept\_corresp\_AbstractSet(CO)$}{true}{}
		\Guard{grd14}{false}{$o\_DS = DataSet\_corresp\_Set(DS)$}{true}{}
		\Guard{grd15}{false}{$DM = Attribute\_definedIn\_DomainModel(AT)$}{true}{}
		\Guard{grd16}{false}{$T\_AT \neq{}  o\_AT$}{true}{}
		\Guard{grd17}{false}{$o\_lg1 \neq{}  o\_lg2$}{true}{}
		\Guard{grd18}{false}{$AT \in{} Attribute\_isFunctional\converse{}[\{TRUE\}]$}{true}{}
	}
	\ACTIONS{true}{
		\Action{act1}{$Constant \bcmeq{}  Constant \bunion{}  \{T\_AT, o\_AT\}$}{true}{}
		\Action{act2}{$Attribute\_Type(AT)\bcmeq{}T\_AT$}{true}{}
		\Action{act3}{$Attribute\_corresp\_Constant(AT)\bcmeq{}o\_AT$}{true}{}
		\Action{act4}{$Constant\_definedIn\_Component\bcmeq{}Constant\_definedIn\_Component\bunion{}\\\{o\_AT\mapsto{} DomainModel\_corresp\_Component(DM), T\_AT\mapsto{} DomainModel\_corresp\_Component(DM)\}$}{true}{}
		\Action{act5}{$Property \bcmeq{} Property \bunion{} \{o\_lg1,o\_lg2\}$}{true}{}
		\Action{act6}{$LogicFormula \bcmeq{} LogicFormula \bunion{} \{o\_lg1,o\_lg2\}$}{true}{}
		\Action{act7}{$Constant\_typing\_Property\bcmeq{}  Constant\_typing\_Property \bunion{} \{T\_AT\mapsto{}o\_lg1, o\_AT\mapsto{}o\_lg2\}$}{true}{}
		\Action{act8}{$Constant\_isInvolvedIn\_LogicFormulas\bcmeq{}  Constant\_isInvolvedIn\_LogicFormulas \bunion{} \{T\_AT\mapsto{}\{1\mapsto{}o\_lg1, 2\mapsto{}o\_lg2\}, o\_AT\mapsto{}\{1\mapsto{}o\_lg2\}\}$}{true}{}
		\Action{act9}{$LogicFormula\_uses\_Operators\bcmeq{}  LogicFormula\_uses\_Operators \bunion{} \{o\_lg1\mapsto{}\{1\mapsto{}FunctionSet\_OP\}, o\_lg2\mapsto{}\{1\mapsto{}Belonging\_OP\}\}$}{true}{}
		\Action{act10}{$LogicFormula\_involves\_Sets\bcmeq{}  LogicFormula\_involves\_Sets \bunion{} \{o\_lg1\mapsto{}\{2\mapsto{}o\_CO, 3\mapsto{}o\_DS\}, o\_lg2\mapsto{}\emptyset{}\}$}{true}{}
		\Action{act11}{$LogicFormula\_definedIn\_Component\bcmeq{}LogicFormula\_definedIn\_Component\bunion{}\{\\o\_lg1\mapsto{} DomainModel\_corresp\_Component(DM), o\_lg2\mapsto{} DomainModel\_corresp\_Component(DM)\}$}{true}{}
	}
}
\EVT{rule\_14\_4}{false}{ordinary}{}{\\correspondence of an instance of Attribute having its isVariable property set to false and its isFunctional property set to true (case where the domain  corresponds to a constant, knowing that the range always corresponds to a set )}{
	\ANY{
		\Param{AT}{true}{}
		\Param{T\_AT}{true}{}
		\Param{o\_AT}{true}{}
		\Param{CO}{true}{}
		\Param{o\_CO}{true}{}
		\Param{DS}{true}{}
		\Param{o\_DS}{true}{}
		\Param{o\_lg1}{true}{}
		\Param{o\_lg2}{true}{}
		\Param{DM}{true}{}
	}
	\GUARDS{true}{
		\Guard{grd0}{false}{$Attribute\_isVariable\converse{}[\{FALSE\}] \setminus{} dom(Attribute\_Type) \neq{}\emptyset{}$}{true}{}
		\Guard{grd1}{false}{$AT \in{} Attribute\_isVariable\converse{}[\{FALSE\}] \setminus{} dom(Attribute\_Type)$}{true}{}
		\Guard{grd2}{false}{$dom(Concept\_corresp\_Constant) \neq{} \emptyset{}$}{true}{}
		\Guard{grd3}{false}{$CO = Attribute\_domain\_Concept(AT)$}{true}{}
		\Guard{grd4}{false}{$CO \in{} dom(Concept\_corresp\_Constant)$}{true}{}
		\Guard{grd5}{false}{$dom(DataSet\_corresp\_Set) \neq{} \emptyset{}$}{true}{}
		\Guard{grd6}{false}{$DS = Attribute\_range\_DataSet(AT)$}{true}{}
		\Guard{grd7}{false}{$DS \in{} dom(DataSet\_corresp\_Set)$}{true}{}
		\Guard{grd8}{false}{$Attribute\_definedIn\_DomainModel(AT) \in{} dom(DomainModel\_corresp\_Component)$}{true}{}
		\Guard{grd9}{false}{$Constant\_Set \setminus{} Constant \neq{}\emptyset{}$}{true}{}
		\Guard{grd10}{false}{$\{T\_AT, o\_AT\} \subseteq{} Constant\_Set \setminus{} Constant$}{true}{}
		\Guard{grd11}{false}{$LogicFormula\_Set \setminus{} LogicFormula \neq{} \emptyset{}$}{true}{}
		\Guard{grd12}{false}{$\{o\_lg1,o\_lg2\}  \subseteq{} LogicFormula\_Set \setminus{} LogicFormula$}{true}{}
		\Guard{grd13}{false}{$o\_CO = Concept\_corresp\_Constant(CO)$}{true}{}
		\Guard{grd14}{false}{$o\_DS = DataSet\_corresp\_Set(DS)$}{true}{}
		\Guard{grd15}{false}{$DM = Attribute\_definedIn\_DomainModel(AT)$}{true}{}
		\Guard{grd16}{false}{$T\_AT \neq{}  o\_AT$}{true}{}
		\Guard{grd17}{false}{$o\_lg1 \neq{}  o\_lg2$}{true}{}
		\Guard{grd18}{false}{$AT \in{} Attribute\_isFunctional\converse{}[\{TRUE\}]$}{true}{}
	}
	\ACTIONS{true}{
		\Action{act1}{$Constant \bcmeq{}  Constant \bunion{}  \{T\_AT, o\_AT\}$}{true}{}
		\Action{act2}{$Attribute\_Type(AT)\bcmeq{}T\_AT$}{true}{}
		\Action{act3}{$Attribute\_corresp\_Constant(AT)\bcmeq{}o\_AT$}{true}{}
		\Action{act4}{$Constant\_definedIn\_Component\bcmeq{}Constant\_definedIn\_Component\bunion{}\\\{o\_AT\mapsto{} DomainModel\_corresp\_Component(DM), T\_AT\mapsto{} DomainModel\_corresp\_Component(DM)\}$}{true}{}
		\Action{act5}{$Property \bcmeq{} Property \bunion{} \{o\_lg1,o\_lg2\}$}{true}{}
		\Action{act6}{$LogicFormula \bcmeq{} LogicFormula \bunion{} \{o\_lg1,o\_lg2\}$}{true}{}
		\Action{act7}{$Constant\_typing\_Property\bcmeq{}  Constant\_typing\_Property \bunion{} \{T\_AT\mapsto{}o\_lg1, o\_AT\mapsto{}o\_lg2\}$}{true}{}
		\Action{act8}{$Constant\_isInvolvedIn\_LogicFormulas\bcmeq{}  Constant\_isInvolvedIn\_LogicFormulas \ovl{} \{T\_AT\mapsto{}\{1\mapsto{}o\_lg1, 2\mapsto{}o\_lg2\}, o\_AT\mapsto{}\{1\mapsto{}o\_lg2\}, o\_CO\mapsto{}\{2\mapsto{}o\_lg1\}\bunion{}Constant\_isInvolvedIn\_LogicFormulas(o\_CO)\}$}{true}{}
		\Action{act9}{$LogicFormula\_uses\_Operators\bcmeq{}  LogicFormula\_uses\_Operators \bunion{} \{o\_lg1\mapsto{}\{1\mapsto{}FunctionSet\_OP\}, o\_lg2\mapsto{}\{1\mapsto{}Belonging\_OP\}\}$}{true}{}
		\Action{act10}{$LogicFormula\_involves\_Sets\bcmeq{}  LogicFormula\_involves\_Sets \bunion{} \{o\_lg1\mapsto{}\{3\mapsto{}o\_DS\}, o\_lg2\mapsto{}\emptyset{}\}$}{true}{}
		\Action{act11}{$LogicFormula\_definedIn\_Component\bcmeq{}LogicFormula\_definedIn\_Component\bunion{}\{\\o\_lg1\mapsto{} DomainModel\_corresp\_Component(DM), o\_lg2\mapsto{} DomainModel\_corresp\_Component(DM)\}$}{true}{}
	}
}

\END

Similarly to relations, each attribute gives rise to a constant representing the type of its associated \textit{B System} element and, in the case when \textsf{isVariable} is set to \textit{false},  to another constant having its name. However, when the \textsf{isFunctional} attribute is set to \textit{true}, the constant representing the type is defined as the set of functions between the \textit{B System} element corresponding to the attribute domain and the one corresponding to the attribute range. The \textit{B System} element corresponding to the attribute is then typed as a function.
\textbf{\underline{Example : }} in Figure \ref{lgsystem_event_b_model_refinment_0}, \textit{landingGearState} is typed as a function (assertions \textit{\textsf{0.3}} and \textit{\textsf{0.4}}) since its type is the set of functions between \textbf{\textit{LandingGear}} and \textbf{\textit{DataSet\_1}} (\textbf{\textit{DataSet\_1}}=\{lg\_extended, lg\_retracted\}).

\subsubsection{B System Variables}

\paragraph{Rule 9 : Variable concept}
\MACHINE{Ontologies\_BSystem\_specs\_translation\_ref\_1}{Ontologies\_BSystem\_specs\_translation}{BSystem\_Metamodel\_Context,Domain\_Metamodel\_Context}{}

\EVT{rule\_9\_1}{false}{ordinary}{}{\\handling the variability of a concept and initializing the associated variable (when the  concept corresponds to an abstract set)}{
	\ANY{
		\Param{CO}{true}{}
		\Param{x\_CO}{true}{}
		\Param{o\_lg}{true}{}
		\Param{o\_CO}{true}{}
		\Param{o\_ia}{true}{}
		\Param{o\_inds}{true}{}
		\Param{bij\_o\_inds}{true}{}
	}
	\GUARDS{true}{
		\Guard{grd0}{false}{$(dom(Concept\_corresp\_AbstractSet)\binter{}Concept\_isVariable\converse{}[\{TRUE\}]) \setminus{} dom(Concept\_corresp\_Variable) \neq{}\emptyset{}$}{true}{}
		\Guard{grd1}{false}{$CO \in{} (dom(Concept\_corresp\_AbstractSet)\binter{}Concept\_isVariable\converse{}[\{TRUE\}]) \setminus{} dom(Concept\_corresp\_Variable)$}{true}{}
		\Guard{grd2}{false}{$Concept\_definedIn\_DomainModel(CO) \in{} dom(DomainModel\_corresp\_Component)$}{true}{}
		\Guard{grd3}{false}{$Individual\_individualOf\_Concept\converse{}[\{CO\}] \subseteq{}  dom(Individual\_corresp\_Constant)$}{true}{}
		\Guard{grd4}{false}{$Variable\_Set \setminus{} Variable \neq{}\emptyset{}$}{true}{}
		\Guard{grd5}{false}{$x\_CO \in{} Variable\_Set \setminus{} Variable$}{true}{}
		\Guard{grd6}{false}{$LogicFormula\_Set \setminus{} LogicFormula \neq{} \emptyset{}$}{true}{}
		\Guard{grd7}{false}{$o\_lg \in{} LogicFormula\_Set \setminus{} LogicFormula$}{true}{}
		\Guard{grd8}{false}{$o\_CO \in{} AbstractSet$}{true}{}
		\Guard{grd9}{false}{$o\_CO = Concept\_corresp\_AbstractSet(CO)$}{true}{}
		\Guard{grd10}{false}{$InitialisationAction\_Set \setminus{} InitialisationAction \neq{} \emptyset{}$}{true}{}
		\Guard{grd11}{false}{$o\_ia \in{} InitialisationAction\_Set \setminus{} InitialisationAction$}{true}{}
		\Guard{grd12}{false}{$o\_inds =  Individual\_corresp\_Constant[Individual\_individualOf\_Concept\converse{}[\{CO\}]]$}{true}{}
		\Guard{grd13}{false}{$finite(o\_inds)$}{true}{}
		\Guard{grd14}{false}{$bij\_o\_inds \in{} 1\upto{}card(o\_inds) \tbij{} o\_inds$}{true}{}
	}
	\ACTIONS{true}{
		\Action{act1}{$Variable \bcmeq{}  Variable \bunion{}  \{x\_CO\}$}{true}{}
		\Action{act2}{$Concept\_corresp\_Variable(CO)\bcmeq{}x\_CO$}{true}{}
		\Action{act3}{$Variable\_definedIn\_Component(x\_CO) \bcmeq{} DomainModel\_corresp\_Component(\\Concept\_definedIn\_DomainModel(CO))$}{true}{}
		\Action{act4}{$Invariant \bcmeq{} Invariant \bunion{} \{o\_lg\}$}{true}{}
		\Action{act5}{$LogicFormula \bcmeq{} LogicFormula \bunion{} \{o\_lg\}$}{true}{}
		\Action{act6}{$LogicFormula\_uses\_Operators(o\_lg) \bcmeq{} \{1\mapsto{}Inclusion\_OP\}$}{true}{}
		\Action{act7}{$Invariant\_involves\_Variables(o\_lg) \bcmeq{} \{1\mapsto{}x\_CO\}$}{true}{}
		\Action{act8}{$LogicFormula\_involves\_Sets(o\_lg) \bcmeq{} \{2\mapsto{}o\_CO\}$}{true}{}
		\Action{act9}{$LogicFormula\_definedIn\_Component(o\_lg) \bcmeq{} DomainModel\_corresp\_Component(\\Concept\_definedIn\_DomainModel(CO))$}{true}{}
		\Action{act10}{$InitialisationAction \bcmeq{} InitialisationAction \bunion{} \{o\_ia\}$}{true}{}
		\Action{act11}{$InitialisationAction\_uses\_Operators(o\_ia) \bcmeq{} \{1\mapsto{}BecomeEqual2SetOf\_OP\}$}{true}{}
		\Action{act12}{$Variable\_init\_InitialisationAction(x\_CO)  \bcmeq{}  o\_ia$}{true}{}
		\Action{act13}{$InitialisationAction\_involves\_Constants(o\_ia) \bcmeq{} bij\_o\_inds$}{true}{}
		\Action{act14}{$Variable\_typing\_Invariant(x\_CO) \bcmeq{} o\_lg$}{true}{}
	}
}
\EVT{rule\_9\_2}{false}{ordinary}{}{\\handling the variability of a concept and initializing the associated variable (when the  concept corresponds to a constant)}{
	\ANY{
		\Param{CO}{true}{}
		\Param{x\_CO}{true}{}
		\Param{o\_lg}{true}{}
		\Param{o\_CO}{true}{}
		\Param{o\_ia}{true}{}
		\Param{o\_inds}{true}{}
		\Param{bij\_o\_inds}{true}{}
	}
	\GUARDS{true}{
		\Guard{grd0}{false}{$(dom(Concept\_corresp\_Constant)\binter{}Concept\_isVariable\converse{}[\{TRUE\}]) \setminus{} dom(Concept\_corresp\_Variable) \neq{}\emptyset{}$}{true}{}
		\Guard{grd1}{false}{$CO \in{} (dom(Concept\_corresp\_Constant)\binter{}Concept\_isVariable\converse{}[\{TRUE\}]) \setminus{} dom(Concept\_corresp\_Variable)$}{true}{}
		\Guard{grd2}{false}{$Concept\_definedIn\_DomainModel(CO) \in{} dom(DomainModel\_corresp\_Component)$}{true}{}
		\Guard{grd3}{false}{$Individual\_individualOf\_Concept\converse{}[\{CO\}] \subseteq{}  dom(Individual\_corresp\_Constant)$}{true}{}
		\Guard{grd4}{false}{$Variable\_Set \setminus{} Variable \neq{}\emptyset{}$}{true}{}
		\Guard{grd5}{false}{$x\_CO \in{} Variable\_Set \setminus{} Variable$}{true}{}
		\Guard{grd6}{false}{$LogicFormula\_Set \setminus{} LogicFormula \neq{} \emptyset{}$}{true}{}
		\Guard{grd7}{false}{$o\_lg \in{} LogicFormula\_Set \setminus{} LogicFormula$}{true}{}
		\Guard{grd8}{false}{$o\_CO \in{} Constant$}{true}{}
		\Guard{grd9}{false}{$o\_CO = Concept\_corresp\_Constant(CO)$}{true}{}
		\Guard{grd10}{false}{$InitialisationAction\_Set \setminus{} InitialisationAction \neq{} \emptyset{}$}{true}{}
		\Guard{grd11}{false}{$o\_ia \in{} InitialisationAction\_Set \setminus{} InitialisationAction$}{true}{}
		\Guard{grd12}{false}{$o\_inds =  Individual\_corresp\_Constant[Individual\_individualOf\_Concept\converse{}[\{CO\}]]$}{true}{}
		\Guard{grd13}{false}{$finite(o\_inds)$}{true}{}
		\Guard{grd14}{false}{$bij\_o\_inds \in{} 1\upto{}card(o\_inds) \tbij{} o\_inds$}{true}{}
	}
	\ACTIONS{true}{
		\Action{act1}{$Variable \bcmeq{}  Variable \bunion{}  \{x\_CO\}$}{true}{}
		\Action{act2}{$Concept\_corresp\_Variable(CO)\bcmeq{}x\_CO$}{true}{}
		\Action{act3}{$Variable\_definedIn\_Component(x\_CO) \bcmeq{} DomainModel\_corresp\_Component(\\Concept\_definedIn\_DomainModel(CO))$}{true}{}
		\Action{act4}{$Invariant \bcmeq{} Invariant \bunion{} \{o\_lg\}$}{true}{}
		\Action{act5}{$LogicFormula \bcmeq{} LogicFormula \bunion{} \{o\_lg\}$}{true}{}
		\Action{act6}{$LogicFormula\_uses\_Operators(o\_lg) \bcmeq{} \{1\mapsto{}Inclusion\_OP\}$}{true}{}
		\Action{act7}{$Invariant\_involves\_Variables(o\_lg) \bcmeq{} \{1\mapsto{}x\_CO\}$}{true}{}
		\Action{act8}{$Constant\_isInvolvedIn\_LogicFormulas(o\_CO) \bcmeq{}Constant\_isInvolvedIn\_LogicFormulas(o\_CO)\bunion{}\{2\mapsto{}o\_lg\}$}{true}{}
		\Action{act9}{$LogicFormula\_involves\_Sets(o\_lg) \bcmeq{} \emptyset{}$}{true}{}
		\Action{act10}{$LogicFormula\_definedIn\_Component(o\_lg) \bcmeq{} DomainModel\_corresp\_Component(\\Concept\_definedIn\_DomainModel(CO))$}{true}{}
		\Action{act11}{$InitialisationAction \bcmeq{} InitialisationAction \bunion{} \{o\_ia\}$}{true}{}
		\Action{act12}{$InitialisationAction\_uses\_Operators(o\_ia) \bcmeq{} \{1\mapsto{}BecomeEqual2SetOf\_OP\}$}{true}{}
		\Action{act13}{$Variable\_init\_InitialisationAction(x\_CO)  \bcmeq{}  o\_ia$}{true}{}
		\Action{act14}{$InitialisationAction\_involves\_Constants(o\_ia) \bcmeq{} bij\_o\_inds$}{true}{}
		\Action{act15}{$Variable\_typing\_Invariant(x\_CO) \bcmeq{} o\_lg$}{true}{}
	}
}

\END

\paragraph{Rule 13 : variable relation}
\MACHINE{Ontologies\_BSystem\_specs\_translation\_ref\_1}{Ontologies\_BSystem\_specs\_translation}{BSystem\_Metamodel\_Context,Domain\_Metamodel\_Context}{}

\EVT{rule\_13\_1}{false}{ordinary}{}{\\correspondence of an instance of Relation having its isVariable property set to true (case where domain and range correspond to abstract sets. The others cases will not explicitely included here, since they can easily be obtained based on rules 10$\_$2, 10$\_$3 and 10$\_$4)}{
	\ANY{
		\Param{RE}{true}{}
		\Param{T\_RE}{true}{}
		\Param{o\_RE}{true}{}
		\Param{CO1}{true}{}
		\Param{o\_CO1}{true}{}
		\Param{CO2}{true}{}
		\Param{o\_CO2}{true}{}
		\Param{o\_lg1}{true}{}
		\Param{o\_lg2}{true}{}
		\Param{DM}{true}{}
		\Param{o\_ia}{true}{}
	}
	\GUARDS{true}{
		\Guard{grd0}{false}{$Relation\_isVariable\converse{}[\{TRUE\}] \setminus{} dom(Relation\_Type) \neq{}\emptyset{}$}{true}{}
		\Guard{grd1}{false}{$RE \in{} Relation\_isVariable\converse{}[\{TRUE\}] \setminus{} dom(Relation\_Type)$}{true}{}
		\Guard{grd2}{false}{$dom(Concept\_corresp\_AbstractSet) \neq{} \emptyset{}$}{true}{}
		\Guard{grd3}{false}{$CO1 = Relation\_domain\_Concept(RE)$}{true}{}
		\Guard{grd4}{false}{$CO2 = Relation\_range\_Concept(RE)$}{true}{}
		\Guard{grd5}{false}{$\{CO1, CO2\} \subseteq{} dom(Concept\_corresp\_AbstractSet)$}{true}{}
		\Guard{grd6}{false}{$Relation\_definedIn\_DomainModel(RE) \in{} dom(DomainModel\_corresp\_Component)$}{true}{}
		\Guard{grd7}{false}{$Constant\_Set \setminus{} Constant \neq{}\emptyset{}$}{true}{}
		\Guard{grd8}{false}{$T\_RE \in{} Constant\_Set \setminus{} Constant$}{true}{}
		\Guard{grd9}{false}{$Variable\_Set \setminus{} Variable \neq{}\emptyset{}$}{true}{}
		\Guard{grd10}{false}{$o\_RE \in{} Variable\_Set \setminus{} Variable$}{true}{}
		\Guard{grd11}{false}{$LogicFormula\_Set \setminus{} LogicFormula \neq{} \emptyset{}$}{true}{}
		\Guard{grd12}{false}{$\{o\_lg1,o\_lg2\}  \subseteq{} LogicFormula\_Set \setminus{} LogicFormula$}{true}{}
		\Guard{grd13}{false}{$o\_CO1 = Concept\_corresp\_AbstractSet(CO1)$}{true}{}
		\Guard{grd14}{false}{$o\_CO2 = Concept\_corresp\_AbstractSet(CO2)$}{true}{}
		\Guard{grd15}{false}{$DM = Relation\_definedIn\_DomainModel(RE)$}{true}{}
		\Guard{grd16}{false}{$o\_lg1 \neq{}  o\_lg2$}{true}{}
		\Guard{grd17}{false}{$InitialisationAction\_Set \setminus{} InitialisationAction \neq{} \emptyset{}$}{true}{}
		\Guard{grd18}{false}{$o\_ia \in{} InitialisationAction\_Set \setminus{} InitialisationAction$}{true}{}
	}
	\ACTIONS{true}{
		\Action{act1}{$Constant \bcmeq{}  Constant \bunion{}  \{T\_RE\}$}{true}{}
		\Action{act2}{$Variable \bcmeq{}  Variable \bunion{}  \{o\_RE\}$}{true}{}
		\Action{act3}{$Relation\_Type(RE)\bcmeq{}T\_RE$}{true}{}
		\Action{act4}{$Relation\_corresp\_Variable(RE)\bcmeq{}o\_RE$}{true}{}
		\Action{act5}{$Constant\_definedIn\_Component(T\_RE) \bcmeq{} DomainModel\_corresp\_Component(DM)$}{true}{}
		\Action{act6}{$Variable\_definedIn\_Component(o\_RE) \bcmeq{} DomainModel\_corresp\_Component(DM)$}{true}{}
		\Action{act7}{$Property \bcmeq{} Property \bunion{} \{o\_lg1\}$}{true}{}
		\Action{act8}{$Invariant \bcmeq{} Invariant \bunion{} \{o\_lg2\}$}{true}{}
		\Action{act9}{$LogicFormula \bcmeq{} LogicFormula \bunion{} \{o\_lg1,o\_lg2\}$}{true}{}
		\Action{act10}{$Constant\_typing\_Property(T\_RE)\bcmeq{}  o\_lg1$}{true}{}
		\Action{act11}{$Variable\_typing\_Invariant(o\_RE) \bcmeq{} o\_lg2$}{true}{}
		\Action{act12}{$Constant\_isInvolvedIn\_LogicFormulas(T\_RE)\bcmeq{}  \{1\mapsto{}o\_lg1, 2\mapsto{}o\_lg2\}$}{true}{}
		\Action{act13}{$Invariant\_involves\_Variables(o\_lg2) \bcmeq{} \{1\mapsto{}o\_RE\}$}{true}{}
		\Action{act14}{$LogicFormula\_uses\_Operators\bcmeq{}  LogicFormula\_uses\_Operators \bunion{} \{o\_lg1\mapsto{}\{1\mapsto{}RelationSet\_OP\}, o\_lg2\mapsto{}\{1\mapsto{}Belonging\_OP\}\}$}{true}{}
		\Action{act15}{$LogicFormula\_involves\_Sets\bcmeq{}  LogicFormula\_involves\_Sets \bunion{} \{o\_lg1\mapsto{}\{2\mapsto{}o\_CO1, 3\mapsto{}o\_CO2\}, o\_lg2\mapsto{}\emptyset{}\}$}{true}{}
		\Action{act16}{$LogicFormula\_definedIn\_Component\bcmeq{}LogicFormula\_definedIn\_Component\bunion{}\{\\o\_lg1\mapsto{} DomainModel\_corresp\_Component(DM), o\_lg2\mapsto{} DomainModel\_corresp\_Component(DM)\}$}{true}{}
		\Action{act17}{$InitialisationAction \bcmeq{} InitialisationAction \bunion{} \{o\_ia\}$}{true}{}
		\Action{act18}{$InitialisationAction\_uses\_Operators(o\_ia) \bcmeq{} \{1\mapsto{}BecomeEqual2EmptySet\_OP\}$}{true}{}
		\Action{act19}{$Variable\_init\_InitialisationAction(o\_RE)  \bcmeq{}  o\_ia$}{true}{}
		\Action{act20}{$InitialisationAction\_involves\_Constants(o\_ia) \bcmeq{} \emptyset{}$}{true}{}
	}
}

\END

An instance of \textsf{Relation}, of \textsf{Concept} or of \textsf{Attribute}, having its \textsf{isVariable} property set to \textit{true} gives rise to a variable (Fig. \ref{OurBusinesDomainModel_min_version_2_EventB_variable}). For a concept, the variable represents the set of \textit{B System} elements having this concept as type.  For a relation or an attribute, it represents the set of links between individuals (in case of relation) or between individuals and data values (in case of attribute) defined through it.\textbf{\underline{Example : }}  in Figure \ref{lgsystem_event_b_model_refinment_1}, variables named \textit{landingSetState} and \textit{handleState} appear because of  \textsf{Attribute} instances \textbf{\textit{landingSetState}} and \textbf{\textit{handleState}} for which the \textsf{isVariable} property is set to \textit{true} (Fig. \ref{lgsystem_refinment_1_ontology}).

\subsubsection{Invariants and Properties}

In this section, we are interested in the correspondences between the domain model and the \textit{B System} model that are likely to give rise to \textit{invariants}, \textit{properties} or \textit{initialization} clauses.

\paragraph{Rule 12 : closure property or action raised by relation maplets}
\MACHINE{Ontologies\_BSystem\_specs\_translation\_ref\_1}{Ontologies\_BSystem\_specs\_translation}{BSystem\_Metamodel\_Context,Domain\_Metamodel\_Context}{}

\EVT{rule\_12\_1}{false}{ordinary}{}{\\closure property for constant relations}{
	\ANY{
		\Param{RE}{true}{}
		\Param{o\_lg}{true}{}
		\Param{o\_RE}{true}{}
		\Param{maplets}{true}{}
		\Param{o\_maplets}{true}{}
	}
	\GUARDS{true}{
		\Guard{grd0}{false}{$dom(Relation\_corresp\_Constant) \neq{}\emptyset{}$}{true}{}
		\Guard{grd1}{false}{$RE \in{} dom(Relation\_corresp\_Constant)$}{true}{}
		\Guard{grd2}{false}{$o\_RE = Relation\_corresp\_Constant(RE)$}{true}{}
		\Guard{grd3}{false}{$LogicFormula\_uses\_Operators\converse{}[\{\{1\mapsto{}Equal2SetOf\_OP\}\}]\binter{}\\ran(Constant\_isInvolvedIn\_LogicFormulas(o\_RE))=\emptyset{}$}{true}{}
		\Guard{grd4}{false}{$RelationMaplet\_mapletOf\_Relation\converse{}[\{RE\}] = maplets$}{true}{}
		\Guard{grd5}{false}{$maplets\subseteq{}dom(RelationMaplet\_corresp\_Constant)$}{true}{}
		\Guard{grd6}{false}{$o\_maplets = RelationMaplet\_corresp\_Constant[maplets]$}{true}{}
		\Guard{grd7}{false}{$Relation\_definedIn\_DomainModel(RE) \in{} dom(DomainModel\_corresp\_Component)$}{true}{}
		\Guard{grd8}{false}{$LogicFormula\_Set \setminus{} LogicFormula \neq{} \emptyset{}$}{true}{}
		\Guard{grd9}{false}{$o\_lg \in{} LogicFormula\_Set \setminus{} LogicFormula$}{true}{}
		\Guard{grd10}{false}{$o\_RE \notin{}  o\_maplets$}{true}{}
	}
	\ACTIONS{true}{
		\Action{act1}{$Property \bcmeq{} Property \bunion{} \{o\_lg\}$}{true}{}
		\Action{act2}{$LogicFormula \bcmeq{} LogicFormula \bunion{} \{o\_lg\}$}{true}{}
		\Action{act3}{$LogicFormula\_uses\_Operators(o\_lg) \bcmeq{} \{1\mapsto{}Equal2SetOf\_OP\}$}{true}{}
		\Action{act4}{$Constant\_isInvolvedIn\_LogicFormulas\bcmeq{}  Constant\_isInvolvedIn\_LogicFormulas \ovl{}(\{o\_RE\mapsto{}\{1\mapsto{}o\_lg\}\bunion{}Constant\_isInvolvedIn\_LogicFormulas(o\_RE)\} \bunion{} \{co\mapsto{}lgs | co\in{} o\_maplets \land{} lgs = \{2\mapsto{}o\_lg\}\bunion{}Constant\_isInvolvedIn\_LogicFormulas(co)\})$}{true}{\\appearence order does not matter}
		\Action{act5}{$LogicFormula\_involves\_Sets(o\_lg) \bcmeq{} \emptyset{}$}{true}{}
		\Action{act6}{$LogicFormula\_definedIn\_Component(o\_lg) \bcmeq{} DomainModel\_corresp\_Component(\\Relation\_definedIn\_DomainModel(RE))$}{true}{}
	}
}
\EVT{rule\_12\_2}{false}{ordinary}{}{\\closure action for variable relations}{
	\ANY{
		\Param{RE}{true}{}
		\Param{o\_ia}{true}{}
		\Param{o\_RE}{true}{}
		\Param{maplets}{true}{}
		\Param{o\_maplets}{true}{}
		\Param{ex\_o\_ia}{true}{}
		\Param{bij\_o\_maplets}{true}{}
	}
	\GUARDS{true}{
		\Guard{grd0}{false}{$dom(Relation\_corresp\_Variable) \neq{}\emptyset{}$}{true}{}
		\Guard{grd1}{false}{$RE \in{} dom(Relation\_corresp\_Variable)$}{true}{}
		\Guard{grd2}{false}{$o\_RE = Relation\_corresp\_Variable(RE)$}{true}{}
		\Guard{grd3}{false}{$Variable\_init\_InitialisationAction(o\_RE)\notin{} InitialisationAction\_uses\_Operators\converse{}[\\\{\{1\mapsto{}BecomeEqual2SetOf\_OP\}\}]$}{true}{}
		\Guard{grd4}{false}{$RelationMaplet\_mapletOf\_Relation\converse{}[\{RE\}] = maplets$}{true}{}
		\Guard{grd5}{false}{$maplets\subseteq{}dom(RelationMaplet\_corresp\_Constant)$}{true}{}
		\Guard{grd6}{false}{$o\_maplets = RelationMaplet\_corresp\_Constant[maplets]$}{true}{}
		\Guard{grd7}{false}{$Relation\_definedIn\_DomainModel(RE) \in{} dom(DomainModel\_corresp\_Component)$}{true}{}
		\Guard{grd8}{false}{$InitialisationAction\_Set \setminus{} InitialisationAction \neq{} \emptyset{}$}{true}{}
		\Guard{grd9}{false}{$o\_ia \in{} InitialisationAction\_Set \setminus{} InitialisationAction$}{true}{}
		\Guard{grd10}{false}{$ex\_o\_ia = Variable\_init\_InitialisationAction(o\_RE)$}{true}{}
		\Guard{grd11}{false}{$Variable\_init\_InitialisationAction\converse{}[\{ex\_o\_ia\}]= \{o\_RE\}$}{true}{}
		\Guard{grd12}{false}{$finite(o\_maplets)$}{true}{}
		\Guard{grd13}{false}{$bij\_o\_maplets \in{} 1\upto{}card(o\_maplets) \tbij{} o\_maplets$}{true}{}
	}
	\ACTIONS{true}{
		\Action{act1}{$InitialisationAction \bcmeq{} (InitialisationAction\setminus{}\{ex\_o\_ia\}) \bunion{} \{o\_ia\}$}{true}{}
		\Action{act2}{$InitialisationAction\_uses\_Operators\bcmeq{}(InitialisationAction\_uses\_Operators\setminus{}\{\\ex\_o\_ia\mapsto{}InitialisationAction\_uses\_Operators(ex\_o\_ia)\})\ovl{} \{o\_ia\mapsto{}\{1\mapsto{}BecomeEqual2SetOf\_OP\}\}$}{true}{}
		\Action{act3}{$Variable\_init\_InitialisationAction(o\_RE)  \bcmeq{}  o\_ia$}{true}{}
		\Action{act4}{$InitialisationAction\_involves\_Constants\bcmeq{}(InitialisationAction\_involves\_Constants\setminus{}\{ex\_o\_ia\mapsto{}InitialisationAction\_involves\_Constants(ex\_o\_ia)\})\ovl{} \{o\_ia\mapsto{}bij\_o\_maplets\}$}{true}{}
	}
}

\END

\paragraph{Rule 15 : closure property or action raised by relation maplets}
\MACHINE{Ontologies\_BSystem\_specs\_translation\_ref\_1}{Ontologies\_BSystem\_specs\_translation}{BSystem\_Metamodel\_Context,Domain\_Metamodel\_Context}{}

\EVT{rule\_15\_1}{false}{ordinary}{}{\\closure property for constant attribute}{
	\ANY{
		\Param{AT}{true}{}
		\Param{o\_lg}{true}{}
		\Param{o\_AT}{true}{}
		\Param{maplets}{true}{}
		\Param{o\_maplets}{true}{}
	}
	\GUARDS{true}{
		\Guard{grd0}{false}{$dom(Attribute\_corresp\_Constant) \neq{}\emptyset{}$}{true}{}
		\Guard{grd1}{false}{$AT \in{} dom(Attribute\_corresp\_Constant)$}{true}{}
		\Guard{grd2}{false}{$o\_AT = Attribute\_corresp\_Constant(AT)$}{true}{}
		\Guard{grd3}{false}{$LogicFormula\_uses\_Operators\converse{}[\{\{1\mapsto{}Equal2SetOf\_OP\}\}]\binter{}\\ran(Constant\_isInvolvedIn\_LogicFormulas(o\_AT))=\emptyset{}$}{true}{}
		\Guard{grd4}{false}{$AttributeMaplet\_mapletOf\_Attribute\converse{}[\{AT\}] = maplets$}{true}{}
		\Guard{grd5}{false}{$maplets\subseteq{}dom(AttributeMaplet\_corresp\_Constant)$}{true}{}
		\Guard{grd6}{false}{$o\_maplets = AttributeMaplet\_corresp\_Constant[maplets]$}{true}{}
		\Guard{grd7}{false}{$Attribute\_definedIn\_DomainModel(AT) \in{} dom(DomainModel\_corresp\_Component)$}{true}{}
		\Guard{grd8}{false}{$LogicFormula\_Set \setminus{} LogicFormula \neq{} \emptyset{}$}{true}{}
		\Guard{grd9}{false}{$o\_lg \in{} LogicFormula\_Set \setminus{} LogicFormula$}{true}{}
		\Guard{grd10}{false}{$o\_AT \notin{}  o\_maplets$}{true}{}
	}
	\ACTIONS{true}{
		\Action{act1}{$Property \bcmeq{} Property \bunion{} \{o\_lg\}$}{true}{}
		\Action{act2}{$LogicFormula \bcmeq{} LogicFormula \bunion{} \{o\_lg\}$}{true}{}
		\Action{act3}{$LogicFormula\_uses\_Operators(o\_lg) \bcmeq{} \{1\mapsto{}Equal2SetOf\_OP\}$}{true}{}
		\Action{act4}{$Constant\_isInvolvedIn\_LogicFormulas\bcmeq{}  Constant\_isInvolvedIn\_LogicFormulas \ovl{} (\{o\_AT\mapsto{}(\{1\mapsto{}o\_lg\}\bunion{}Constant\_isInvolvedIn\_LogicFormulas(o\_AT))\}\bunion{}\{co\mapsto{}lgs | co\in{} o\_maplets \land{} lgs = \{2\mapsto{}o\_lg\}\bunion{}Constant\_isInvolvedIn\_LogicFormulas(co)\})$}{true}{\\appearence order does not matter}
		\Action{act5}{$LogicFormula\_involves\_Sets(o\_lg) \bcmeq{} \emptyset{}$}{true}{}
		\Action{act6}{$LogicFormula\_definedIn\_Component(o\_lg) \bcmeq{} DomainModel\_corresp\_Component(\\Attribute\_definedIn\_DomainModel(AT))$}{true}{}
	}
}

\END

\paragraph{Rule 16 : optional characteristics of relations}

\MACHINE{Ontologies\_BSystem\_specs\_translation\_ref\_1}{Ontologies\_BSystem\_specs\_translation}{BSystem\_Metamodel\_Context,Domain\_Metamodel\_Context}{}

\EVT{rule\_16\_1}{false}{ordinary}{}{\\handling the transitivity of a constant relation}{
	\ANY{
		\Param{RE}{true}{}
		\Param{o\_lg1}{true}{}
		\Param{o\_lg2}{true}{}
		\Param{o\_RE}{true}{}
		\Param{composition}{true}{}
	}
	\GUARDS{true}{
		\Guard{grd0}{false}{$(dom(Relation\_corresp\_Constant)\binter{}Relation\_isTransitive\converse{}[\{TRUE\}])  \neq{}\emptyset{}$}{true}{}
		\Guard{grd1}{false}{$RE \in{} (dom(Relation\_corresp\_Constant)\binter{}Relation\_isTransitive\converse{}[\{TRUE\}])$}{true}{}
		\Guard{grd2}{false}{$(\{RE\mapsto{}isTransitive\}) \notin{} dom(RelationCharacteristic\_corresp\_LogicFormula)$}{true}{}
		\Guard{grd3}{false}{$o\_RE = Relation\_corresp\_Constant(RE)$}{true}{}
		\Guard{grd4}{false}{$Relation\_definedIn\_DomainModel(RE) \in{} dom(DomainModel\_corresp\_Component)$}{true}{}
		\Guard{grd5}{false}{$LogicFormula\_Set \setminus{} LogicFormula \neq{} \emptyset{}$}{true}{}
		\Guard{grd6}{false}{$\{o\_lg1, o\_lg2\} \subseteq{} LogicFormula\_Set \setminus{} LogicFormula$}{true}{}
		\Guard{grd7}{false}{$partition(\{o\_lg1, o\_lg2\}, \{o\_lg1\}, \{o\_lg2\})$}{true}{}
		\Guard{grd8}{false}{$Constant\_Set \setminus{} Constant \neq{}\emptyset{}$}{true}{}
		\Guard{grd9}{false}{$composition \in{} Constant\_Set \setminus{} Constant$}{true}{}
	}
	\ACTIONS{true}{
		\Action{act0}{$Constant \bcmeq{} Constant \bunion{} \{composition\}$}{true}{}
		\Action{act1}{$Property \bcmeq{} Property \bunion{} \{o\_lg1, o\_lg2\}$}{true}{}
		\Action{act2}{$LogicFormula \bcmeq{} LogicFormula \bunion{} \{o\_lg1, o\_lg2\}$}{true}{}
		\Action{act3}{$Constant\_typing\_Property(composition)\bcmeq{}  o\_lg1$}{true}{}
		\Action{act4}{$RelationCharacteristic\_corresp\_LogicFormula(\{RE\mapsto{}isTransitive\}) \bcmeq{} o\_lg2$}{true}{}
		\Action{act5}{$Constant\_isInvolvedIn\_LogicFormulas\bcmeq{}  Constant\_isInvolvedIn\_LogicFormulas \ovl{} \{composition\mapsto{}\{1\mapsto{}o\_lg1, 1\mapsto{}o\_lg2\}, o\_RE\mapsto{}\{2\mapsto{}o\_lg1, 3\mapsto{}o\_lg1, 2\mapsto{}o\_lg2\}\bunion{}Constant\_isInvolvedIn\_LogicFormulas(o\_RE)\}$}{true}{}
		\Action{act6}{$LogicFormula\_uses\_Operators\bcmeq{}  LogicFormula\_uses\_Operators \bunion{}\\ \{o\_lg1\mapsto{}\{1\mapsto{}RelationComposition\_OP\}, o\_lg2\mapsto{}\{1\mapsto{}Inclusion\_OP\}\}$}{true}{}
		\Action{act7}{$LogicFormula\_involves\_Sets\bcmeq{}  LogicFormula\_involves\_Sets \bunion{} \{o\_lg1\mapsto{}\emptyset{}, o\_lg2\mapsto{}\emptyset{}\}$}{true}{}
		\Action{act8}{$LogicFormula\_definedIn\_Component\bcmeq{}LogicFormula\_definedIn\_Component\bunion{}\{\\o\_lg1\mapsto{} DomainModel\_corresp\_Component(Relation\_definedIn\_DomainModel(RE)), \\o\_lg2\mapsto{} DomainModel\_corresp\_Component(Relation\_definedIn\_DomainModel(RE))\}$}{true}{}
		\Action{act9}{$Constant\_definedIn\_Component(composition) \bcmeq{}  DomainModel\_corresp\_Component(\\Relation\_definedIn\_DomainModel(RE))$}{true}{}
	}
}
\EVT{rule\_16\_2}{false}{ordinary}{}{\\handling the symmetrie of a constant relation}{
	\ANY{
		\Param{RE}{true}{}
		\Param{o\_lg1}{true}{}
		\Param{o\_lg2}{true}{}
		\Param{o\_RE}{true}{}
		\Param{inverse}{true}{}
	}
	\GUARDS{true}{
		\Guard{grd0}{false}{$(dom(Relation\_corresp\_Constant)\binter{}Relation\_isSymmetric\converse{}[\{TRUE\}])  \neq{}\emptyset{}$}{true}{}
		\Guard{grd1}{false}{$RE \in{} (dom(Relation\_corresp\_Constant)\binter{}Relation\_isSymmetric\converse{}[\{TRUE\}])$}{true}{}
		\Guard{grd2}{false}{$(\{RE\mapsto{}isSymmetric\}) \notin{} dom(RelationCharacteristic\_corresp\_LogicFormula)$}{true}{}
		\Guard{grd3}{false}{$o\_RE = Relation\_corresp\_Constant(RE)$}{true}{}
		\Guard{grd4}{false}{$Relation\_definedIn\_DomainModel(RE) \in{} dom(DomainModel\_corresp\_Component)$}{true}{}
		\Guard{grd5}{false}{$LogicFormula\_Set \setminus{} LogicFormula \neq{} \emptyset{}$}{true}{}
		\Guard{grd6}{false}{$\{o\_lg1, o\_lg2\} \subseteq{} LogicFormula\_Set \setminus{} LogicFormula$}{true}{}
		\Guard{grd7}{false}{$partition(\{o\_lg1, o\_lg2\}, \{o\_lg1\}, \{o\_lg2\})$}{true}{}
		\Guard{grd8}{false}{$Constant\_Set \setminus{} Constant \neq{}\emptyset{}$}{true}{}
		\Guard{grd9}{false}{$inverse \in{} Constant\_Set \setminus{} Constant$}{true}{}
	}
	\ACTIONS{true}{
		\Action{act0}{$Constant \bcmeq{} Constant \bunion{} \{inverse\}$}{true}{}
		\Action{act1}{$Property \bcmeq{} Property \bunion{} \{o\_lg1, o\_lg2\}$}{true}{}
		\Action{act2}{$LogicFormula \bcmeq{} LogicFormula \bunion{} \{o\_lg1, o\_lg2\}$}{true}{}
		\Action{act3}{$Constant\_typing\_Property(inverse)\bcmeq{}  o\_lg1$}{true}{}
		\Action{act4}{$RelationCharacteristic\_corresp\_LogicFormula(\{RE\mapsto{}isSymmetric\}) \bcmeq{} o\_lg2$}{true}{}
		\Action{act5}{$Constant\_isInvolvedIn\_LogicFormulas\bcmeq{}  Constant\_isInvolvedIn\_LogicFormulas \ovl{} \{inverse\mapsto{}\{1\mapsto{}o\_lg1, 1\mapsto{}o\_lg2\}, o\_RE\mapsto{}\{2\mapsto{}o\_lg1, 2\mapsto{}o\_lg2\}\bunion{}Constant\_isInvolvedIn\_LogicFormulas(o\_RE)\}$}{true}{}
		\Action{act6}{$LogicFormula\_uses\_Operators\bcmeq{}  LogicFormula\_uses\_Operators \bunion{} \{o\_lg1\mapsto{}\{1\mapsto{}Inversion\_OP\}, o\_lg2\mapsto{}\{1\mapsto{}Equality\_OP\}\}$}{true}{}
		\Action{act7}{$LogicFormula\_involves\_Sets\bcmeq{}  LogicFormula\_involves\_Sets \bunion{} \{o\_lg1\mapsto{}\emptyset{}, o\_lg2\mapsto{}\emptyset{}\}$}{true}{}
		\Action{act8}{$LogicFormula\_definedIn\_Component\bcmeq{}LogicFormula\_definedIn\_Component\bunion{}\{\\o\_lg1\mapsto{} DomainModel\_corresp\_Component(Relation\_definedIn\_DomainModel(RE)), \\o\_lg2\mapsto{} DomainModel\_corresp\_Component(Relation\_definedIn\_DomainModel(RE))\}$}{true}{}
		\Action{act9}{$Constant\_definedIn\_Component(inverse) \bcmeq{}  DomainModel\_corresp\_Component(\\Relation\_definedIn\_DomainModel(RE))$}{true}{}
	}
}

\END

 \subsubsection{Predicates}
We were not interested in validating the transformation rules of predicates expressed using the SysML/KAOS Domain Modeling formalism to B System logical formulas because either uses first-order logic for predicate expression. As a result, the predicates expressed in one of the formalisms are integrally replicated, without additional transformations in the other.
 When the predicate is an instance of \textsf{GluingInvariant}, the assertion raised is an Event-B gluing invariant.
For example, in Figure \ref{lgsystem_event_b_model_refinment_1},   assertion \textit{\textsf{(1.21)}} is a gluing invariant.

\subsection{Handling Updates on B System Specifications within SysML/KAOS Domain Models}
Here, we are interested in handling modifications on B System specifications within \textit{SysML/KAOS} domain models.
We choose to support only the most repetitive operations that can be performed within the formal specification,   the domain model remaining the one to be updated in case of any major changes. Currently supported operations include : addition of sets and of items in existing sets, addition of subsets of existing sets, addition of individuals  and of data values, addition of relations and of attributes and finally addition of relation and of attribute maplets.

\subsubsection{Addition of Non-Existing Sets}
 \paragraph{Rules 101-102 : addition of a new abstract set}

\MACHINE{Ontologies\_BSystem\_specs\_translation\_ref\_1}{Ontologies\_BSystem\_specs\_translation}{BSystem\_Metamodel\_Context,Domain\_Metamodel\_Context}{}
\EVT{rule\_101}{false}{ordinary}{}{\\handling the addition of a new abstract set (correspondence to a concept)}{
	\ANY{
		\Param{CO}{true}{}
		\Param{o\_CO}{true}{}
	}
	\GUARDS{true}{
		\Guard{grd0}{false}{$AbstractSet \setminus{} (ran(Concept\_corresp\_AbstractSet) \bunion{} ran(DataSet\_corresp\_Set)) \neq{}\emptyset{}$}{true}{}
		\Guard{grd1}{false}{$o\_CO \in{} AbstractSet \setminus{} (ran(Concept\_corresp\_AbstractSet) \bunion{} ran(DataSet\_corresp\_Set))$}{true}{}
		\Guard{grd2}{false}{$Set\_definedIn\_Component(o\_CO) \in{} ran(DomainModel\_corresp\_Component)$}{true}{}
		\Guard{grd3}{false}{$Concept\_Set \setminus{} Concept \neq{}\emptyset{}$}{true}{}
		\Guard{grd4}{false}{$CO \in{} Concept\_Set \setminus{} Concept$}{true}{}
	}
	\ACTIONS{true}{
		\Action{act1}{$Concept \bcmeq{}  Concept \bunion{}  \{CO\}$}{true}{}
		\Action{act2}{$Concept\_corresp\_AbstractSet(CO)\bcmeq{}o\_CO$}{true}{}
		\Action{act3}{$Concept\_definedIn\_DomainModel(CO) \bcmeq{} DomainModel\_corresp\_Component\converse{}(\\Set\_definedIn\_Component(o\_CO))$}{true}{}
		\Action{act4}{$Concept\_isVariable(CO) \bcmeq{}  FALSE$}{true}{}
	}
}
\EVT{rule\_102}{false}{ordinary}{}{\\handling the addition of a new abstract set (correspondence to a custom data set)}{
	\ANY{
		\Param{DS}{true}{}
		\Param{o\_DS}{true}{}
	}
	\GUARDS{true}{
		\Guard{grd0}{false}{$AbstractSet \setminus{} (ran(Concept\_corresp\_AbstractSet) \bunion{} ran(DataSet\_corresp\_Set)) \neq{}\emptyset{}$}{true}{}
		\Guard{grd1}{false}{$o\_DS \in{} AbstractSet \setminus{} (ran(Concept\_corresp\_AbstractSet) \bunion{} ran(DataSet\_corresp\_Set))$}{true}{}
		\Guard{grd2}{false}{$Set\_definedIn\_Component(o\_DS) \in{} ran(DomainModel\_corresp\_Component)$}{true}{}
		\Guard{grd3}{false}{$DataSet\_Set \setminus{} DataSet \neq{}\emptyset{}$}{true}{}
		\Guard{grd4}{false}{$DS \in{} DataSet\_Set \setminus{} DataSet$}{true}{}
		\Guard{grd5}{false}{$DS \notin{} \{\_NATURAL,\_INTEGER,\_FLOAT,\_BOOL,\_STRING\}$}{true}{}
	}
	\ACTIONS{true}{
		\Action{act1}{$CustomDataSet \bcmeq{}  CustomDataSet \bunion{}  \{DS\}$}{true}{}
		\Action{act2}{$DataSet \bcmeq{}  DataSet \bunion{}  \{DS\}$}{true}{}
		\Action{act3}{$CustomDataSet\_corresp\_AbstractSet(DS)\bcmeq{}o\_DS$}{true}{}
		\Action{act4}{$DataSet\_definedIn\_DomainModel(DS) \bcmeq{} DomainModel\_corresp\_Component\converse{}(\\Set\_definedIn\_Component(o\_DS))$}{true}{}
		\Action{act5}{$DataSet\_corresp\_Set(DS) \bcmeq{}  o\_DS$}{true}{}
	}
}
\END

\paragraph{Rule 103 : addition of an enumerated set}

\MACHINE{Ontologies\_BSystem\_specs\_translation\_ref\_1}{Ontologies\_BSystem\_specs\_translation}{BSystem\_Metamodel\_Context,Domain\_Metamodel\_Context}{}
\EVT{rule\_103}{false}{ordinary}{}{\\handling  the addition of an enumerated set}{
	\ANY{
		\Param{EDS}{true}{}
		\Param{o\_EDS}{true}{}
		\Param{elements}{true}{}
		\Param{o\_elements}{true}{}
		\Param{mapping\_elements\_o\_elements}{true}{}
	}
	\GUARDS{true}{
		\Guard{grd0}{false}{$EnumeratedSet \setminus{} ran(DataSet\_corresp\_Set) \neq{}\emptyset{}$}{true}{}
		\Guard{grd1}{false}{$o\_EDS \in{} EnumeratedSet \setminus{} ran(DataSet\_corresp\_Set)$}{true}{}
		\Guard{grd2}{false}{$Set\_definedIn\_Component(o\_EDS) \in{} ran(DomainModel\_corresp\_Component)$}{true}{}
		\Guard{grd3}{false}{$DataSet\_Set \setminus{} DataSet \neq{}\emptyset{}$}{true}{}
		\Guard{grd4}{false}{$EDS \in{} DataSet\_Set \setminus{} DataSet$}{true}{}
		\Guard{grd5}{false}{$DataValue\_Set \setminus{} DataValue \neq{}\emptyset{}$}{true}{}
		\Guard{grd6}{false}{$elements \subseteq{}  DataValue\_Set \setminus{} DataValue$}{true}{}
		\Guard{grd7}{false}{$o\_elements = SetItem\_itemOf\_EnumeratedSet\converse{}[\{o\_EDS\}]$}{true}{}
		\Guard{grd8}{false}{$card(o\_elements) = card(elements)$}{true}{}
		\Guard{grd9}{false}{$mapping\_elements\_o\_elements \in{} elements \tbij{} o\_elements$}{true}{}
		\Guard{grd10}{false}{$ran(DataValue\_corresp\_SetItem)\binter{}o\_elements=\emptyset{}$}{true}{}
		\Guard{grd11}{false}{$EDS \notin{} \{\_NATURAL,\_INTEGER,\_FLOAT,\_BOOL,\_STRING\}$}{true}{}
	}
	\ACTIONS{true}{
		\Action{act1}{$EnumeratedDataSet \bcmeq{}  EnumeratedDataSet \bunion{}  \{EDS\}$}{true}{}
		\Action{act2}{$DataSet \bcmeq{}  DataSet \bunion{}  \{EDS\}$}{true}{}
		\Action{act3}{$EnumeratedDataSet\_corresp\_EnumeratedSet(EDS)\bcmeq{}o\_EDS$}{true}{}
		\Action{act4}{$DataSet\_definedIn\_DomainModel(EDS) \bcmeq{} DomainModel\_corresp\_Component\converse{}(\\Set\_definedIn\_Component(o\_EDS))$}{true}{}
		\Action{act5}{$DataValue \bcmeq{} DataValue \bunion{}  elements$}{true}{}
		\Action{act6}{$DataValue\_elements\_EnumeratedDataSet \bcmeq{}  DataValue\_elements\_EnumeratedDataSet \bunion{} \{(xx\mapsto{}yy) |xx\in{}elements \land{}yy=EDS\}$}{true}{}
		\Action{act7}{$DataValue\_corresp\_SetItem \bcmeq{}  DataValue\_corresp\_SetItem \bunion{} mapping\_elements\_o\_elements$}{true}{}
		\Action{act8}{$DataSet\_corresp\_Set \bcmeq{}  DataSet\_corresp\_Set \ovl{} \{EDS\mapsto{}o\_EDS\}$}{true}{}
		\Action{act9}{$DataValue\_valueOf\_DataSet \bcmeq{}  DataValue\_valueOf\_DataSet \bunion{} \{(xx\mapsto{}yy) |xx\in{}elements \land{}yy=EDS\}$}{true}{}
		\Action{act10}{$CustomDataSet \bcmeq{}  CustomDataSet \bunion{}  \{EDS\}$}{true}{}
	}
}
\END

\subsubsection{Addition of Non-Existing Set Items or Constants}

\paragraph{Rule 104 : addition of a set item}

\MACHINE{Ontologies\_BSystem\_specs\_translation\_ref\_1}{Ontologies\_BSystem\_specs\_translation}{BSystem\_Metamodel\_Context,Domain\_Metamodel\_Context}{}

\EVT{rule\_104}{false}{ordinary}{}{\\handling  the addition of a new element in an existing enumerated set}{
	\ANY{
		\Param{EDS}{true}{}
		\Param{o\_EDS}{true}{}
		\Param{element}{true}{}
		\Param{o\_element}{true}{}
	}
	\GUARDS{true}{
		\Guard{grd0}{false}{$dom(SetItem\_itemOf\_EnumeratedSet) \setminus{} ran(DataValue\_corresp\_SetItem) \neq{}\emptyset{}$}{true}{}
		\Guard{grd1}{false}{$o\_element \in{} dom(SetItem\_itemOf\_EnumeratedSet) \setminus{} ran(DataValue\_corresp\_SetItem)$}{true}{}
		\Guard{grd2}{false}{$o\_EDS = SetItem\_itemOf\_EnumeratedSet(o\_element)$}{true}{}
		\Guard{grd3}{false}{$o\_EDS \in{} ran(EnumeratedDataSet\_corresp\_EnumeratedSet)$}{true}{}
		\Guard{grd4}{false}{$EDS = EnumeratedDataSet\_corresp\_EnumeratedSet\converse{}(o\_EDS)$}{true}{}
		\Guard{grd5}{false}{$DataValue\_Set \setminus{} DataValue \neq{}\emptyset{}$}{true}{}
		\Guard{grd6}{false}{$element \in{} DataValue\_Set \setminus{} DataValue$}{true}{}
	}
	\ACTIONS{true}{
		\Action{act1}{$DataValue \bcmeq{}  DataValue \bunion{}  \{element\}$}{true}{}
		\Action{act2}{$DataValue\_elements\_EnumeratedDataSet(element) \bcmeq{}  EDS$}{true}{}
		\Action{act3}{$DataValue\_corresp\_SetItem(element) \bcmeq{}  o\_element$}{true}{}
		\Action{act4}{$DataValue\_valueOf\_DataSet(element) \bcmeq{}  EDS$}{true}{}
	}
}

\END

 \paragraph{Rule 105 : addition of a constant, sub set of an instance of Concept}

\MACHINE{Ontologies\_BSystem\_specs\_translation\_ref\_1}{Ontologies\_BSystem\_specs\_translation}{BSystem\_Metamodel\_Context,Domain\_Metamodel\_Context}{}

\EVT{rule\_105\_1}{false}{ordinary}{}{\\handling  the addition of a constant, sub set of an instance of Concept (case where the concept corresponds to an abstract set)}{
	\ANY{
		\Param{CO}{true}{}
		\Param{o\_CO}{true}{}
		\Param{PCO}{true}{}
		\Param{o\_lg}{true}{}
		\Param{o\_PCO}{true}{}
	}
	\GUARDS{true}{
		\Guard{grd0}{false}{$dom(Constant\_typing\_Property) \setminus{} ran(Concept\_corresp\_Constant) \neq{}\emptyset{}$}{true}{}
		\Guard{grd1}{false}{$o\_CO \in{} dom(Constant\_typing\_Property) \setminus{} ran(Concept\_corresp\_Constant)$}{true}{}
		\Guard{grd2}{false}{$o\_lg = Constant\_typing\_Property(o\_CO)$}{true}{}
		\Guard{grd3}{false}{$LogicFormula\_uses\_Operators(o\_lg) = \{1\mapsto{}Inclusion\_OP\}$}{true}{}
		\Guard{grd4}{false}{$LogicFormula\_involves\_Sets(o\_lg) \neq{} \emptyset{}$}{true}{}
		\Guard{grd5}{false}{$(2\mapsto{}o\_PCO)\in{}LogicFormula\_involves\_Sets(o\_lg)$}{true}{}
		\Guard{grd6}{false}{$o\_PCO \in{} ran(Concept\_corresp\_AbstractSet)$}{true}{}
		\Guard{grd7}{false}{$PCO = Concept\_corresp\_AbstractSet\converse{}(o\_PCO)$}{true}{}
		\Guard{grd8}{false}{$Concept\_Set \setminus{} Concept \neq{}\emptyset{}$}{true}{}
		\Guard{grd9}{false}{$CO \in{} Concept\_Set \setminus{} Concept$}{true}{}
		\Guard{grd10}{false}{$Constant\_definedIn\_Component(o\_CO) \in{} ran(DomainModel\_corresp\_Component)$}{true}{}
	}
	\ACTIONS{true}{
		\Action{act1}{$Concept \bcmeq{}  Concept \bunion{}  \{CO\}$}{true}{}
		\Action{act2}{$Concept\_corresp\_Constant(CO)\bcmeq{}o\_CO$}{true}{}
		\Action{act3}{$Concept\_definedIn\_DomainModel(CO) \bcmeq{} DomainModel\_corresp\_Component\converse{}(\\Constant\_definedIn\_Component(o\_CO))$}{true}{}
		\Action{act4}{$Concept\_parentConcept\_Concept(CO) \bcmeq{} PCO$}{true}{}
		\Action{act5}{$Concept\_isVariable(CO) \bcmeq{}  FALSE$}{true}{}
	}
}
\EVT{rule\_105\_2}{false}{ordinary}{}{\\handling  the addition of a constant, sub set of an instance of Concept (case where the concept corresponds to a constant)}{
	\ANY{
		\Param{CO}{true}{}
		\Param{o\_CO}{true}{}
		\Param{PCO}{true}{}
		\Param{o\_lg}{true}{}
		\Param{o\_PCO}{true}{}
	}
	\GUARDS{true}{
		\Guard{grd0}{false}{$dom(Constant\_typing\_Property) \setminus{} ran(Concept\_corresp\_Constant) \neq{}\emptyset{}$}{true}{}
		\Guard{grd1}{false}{$o\_CO \in{} dom(Constant\_typing\_Property) \setminus{} ran(Concept\_corresp\_Constant)$}{true}{}
		\Guard{grd2}{false}{$o\_lg = Constant\_typing\_Property(o\_CO)$}{true}{}
		\Guard{grd3}{false}{$LogicFormula\_uses\_Operators(o\_lg) = \{1\mapsto{}Inclusion\_OP\}$}{true}{}
		\Guard{grd4}{false}{$LogicFormula\_involves\_Sets(o\_lg) = \emptyset{}$}{true}{}
		\Guard{grd5}{false}{$o\_PCO \in{} dom(Constant\_isInvolvedIn\_LogicFormulas)$}{true}{}
		\Guard{grd6}{false}{$(2\mapsto{}o\_lg) \in{} Constant\_isInvolvedIn\_LogicFormulas(o\_PCO)$}{true}{}
		\Guard{grd7}{false}{$o\_PCO \in{} ran(Concept\_corresp\_Constant)$}{true}{}
		\Guard{grd8}{false}{$PCO = Concept\_corresp\_Constant\converse{}(o\_PCO)$}{true}{}
		\Guard{grd9}{false}{$Concept\_Set \setminus{} Concept \neq{}\emptyset{}$}{true}{}
		\Guard{grd10}{false}{$CO \in{} Concept\_Set \setminus{} Concept$}{true}{}
		\Guard{grd11}{false}{$Constant\_definedIn\_Component(o\_CO) \in{} ran(DomainModel\_corresp\_Component)$}{true}{}
	}
	\ACTIONS{true}{
		\Action{act1}{$Concept \bcmeq{}  Concept \bunion{}  \{CO\}$}{true}{}
		\Action{act2}{$Concept\_corresp\_Constant(CO)\bcmeq{}o\_CO$}{true}{}
		\Action{act3}{$Concept\_definedIn\_DomainModel(CO) \bcmeq{} DomainModel\_corresp\_Component\converse{}(\\Constant\_definedIn\_Component(o\_CO))$}{true}{}
		\Action{act4}{$Concept\_parentConcept\_Concept(CO) \bcmeq{} PCO$}{true}{}
		\Action{act5}{$Concept\_isVariable(CO) \bcmeq{}  FALSE$}{true}{}
	}
}

\END

\paragraph{Rule 106 : addition of an individual }

\MACHINE{Ontologies\_BSystem\_specs\_translation\_ref\_1}{Ontologies\_BSystem\_specs\_translation}{BSystem\_Metamodel\_Context,Domain\_Metamodel\_Context}{}

\EVT{rule\_106\_1}{false}{ordinary}{}{\\handling  the addition of an individual (case where the concept corresponds to an abstract set)}{
	\ANY{
		\Param{ind}{true}{}
		\Param{o\_ind}{true}{}
		\Param{CO}{true}{}
		\Param{o\_lg}{true}{}
		\Param{o\_CO}{true}{}
	}
	\GUARDS{true}{
		\Guard{grd0}{false}{$dom(Constant\_typing\_Property) \setminus{} ran(Individual\_corresp\_Constant) \neq{}\emptyset{}$}{true}{}
		\Guard{grd1}{false}{$o\_ind \in{} dom(Constant\_typing\_Property) \setminus{} ran(Individual\_corresp\_Constant)$}{true}{}
		\Guard{grd2}{false}{$o\_lg = Constant\_typing\_Property(o\_ind)$}{true}{}
		\Guard{grd3}{false}{$LogicFormula\_uses\_Operators(o\_lg) = \{1\mapsto{}Belonging\_OP\}$}{true}{}
		\Guard{grd4}{false}{$LogicFormula\_involves\_Sets(o\_lg) \neq{} \emptyset{}$}{true}{}
		\Guard{grd5}{false}{$(2\mapsto{}o\_CO)\in{}LogicFormula\_involves\_Sets(o\_lg)$}{true}{}
		\Guard{grd6}{false}{$o\_CO \in{} ran(Concept\_corresp\_AbstractSet)$}{true}{}
		\Guard{grd7}{false}{$CO = Concept\_corresp\_AbstractSet\converse{}(o\_CO)$}{true}{}
		\Guard{grd8}{false}{$Individual\_Set \setminus{} Individual \neq{}\emptyset{}$}{true}{}
		\Guard{grd9}{false}{$ind \in{} Individual\_Set \setminus{} Individual$}{true}{}
	}
	\ACTIONS{true}{
		\Action{act1}{$Individual \bcmeq{}  Individual \bunion{}  \{ind\}$}{true}{}
		\Action{act2}{$Individual\_individualOf\_Concept(ind) \bcmeq{}  CO$}{true}{}
		\Action{act3}{$Individual\_corresp\_Constant(ind) \bcmeq{}  o\_ind$}{true}{}
	}
}
\EVT{rule\_106\_2}{false}{ordinary}{}{\\handling  the addition of an individual (case where the concept corresponds to a constant)}{
	\ANY{
		\Param{ind}{true}{}
		\Param{o\_ind}{true}{}
		\Param{CO}{true}{}
		\Param{o\_lg}{true}{}
		\Param{o\_CO}{true}{}
	}
	\GUARDS{true}{
		\Guard{grd0}{false}{$dom(Constant\_typing\_Property) \setminus{} ran(Individual\_corresp\_Constant) \neq{}\emptyset{}$}{true}{}
		\Guard{grd1}{false}{$o\_ind \in{} dom(Constant\_typing\_Property) \setminus{} ran(Individual\_corresp\_Constant)$}{true}{}
		\Guard{grd2}{false}{$o\_lg = Constant\_typing\_Property(o\_ind)$}{true}{}
		\Guard{grd3}{false}{$LogicFormula\_uses\_Operators(o\_lg) = \{1\mapsto{}Belonging\_OP\}$}{true}{}
		\Guard{grd4}{false}{$LogicFormula\_involves\_Sets(o\_lg) = \emptyset{}$}{true}{}
		\Guard{grd5}{false}{$o\_CO\in{}dom(Constant\_isInvolvedIn\_LogicFormulas)$}{true}{}
		\Guard{grd6}{false}{$(2\mapsto{}o\_lg) \in{} Constant\_isInvolvedIn\_LogicFormulas(o\_CO)$}{true}{}
		\Guard{grd7}{false}{$o\_CO \in{} ran(Concept\_corresp\_Constant)$}{true}{}
		\Guard{grd8}{false}{$CO = Concept\_corresp\_Constant\converse{}(o\_CO)$}{true}{}
		\Guard{grd9}{false}{$Individual\_Set \setminus{} Individual \neq{}\emptyset{}$}{true}{}
		\Guard{grd10}{false}{$ind \in{} Individual\_Set \setminus{} Individual$}{true}{}
	}
	\ACTIONS{true}{
		\Action{act1}{$Individual \bcmeq{}  Individual \bunion{}  \{ind\}$}{true}{}
		\Action{act2}{$Individual\_individualOf\_Concept(ind) \bcmeq{}  CO$}{true}{}
		\Action{act3}{$Individual\_corresp\_Constant(ind) \bcmeq{}  o\_ind$}{true}{}
	}
}

\END

\paragraph{Rule 107 : addition of a data value}

\MACHINE{Ontologies\_BSystem\_specs\_translation\_ref\_1}{Ontologies\_BSystem\_specs\_translation}{BSystem\_Metamodel\_Context,Domain\_Metamodel\_Context}{}

\EVT{rule\_107}{false}{ordinary}{}{\\handling  the addition of a data value}{
	\ANY{
		\Param{dva}{true}{}
		\Param{o\_dva}{true}{}
		\Param{DS}{true}{}
		\Param{o\_lg}{true}{}
		\Param{o\_DS}{true}{}
	}
	\GUARDS{true}{
		\Guard{grd0}{false}{$dom(Constant\_typing\_Property) \setminus{} ran(DataValue\_corresp\_Constant) \neq{}\emptyset{}$}{true}{}
		\Guard{grd1}{false}{$o\_dva \in{} dom(Constant\_typing\_Property) \setminus{} ran(DataValue\_corresp\_Constant)$}{true}{}
		\Guard{grd2}{false}{$o\_lg = Constant\_typing\_Property(o\_dva)$}{true}{}
		\Guard{grd3}{false}{$LogicFormula\_uses\_Operators(o\_lg) = \{1\mapsto{}Belonging\_OP\}$}{true}{}
		\Guard{grd4}{false}{$LogicFormula\_involves\_Sets(o\_lg) \neq{} \emptyset{}$}{true}{}
		\Guard{grd5}{false}{$(2\mapsto{}o\_DS)\in{}LogicFormula\_involves\_Sets(o\_lg)$}{true}{}
		\Guard{grd6}{false}{$o\_DS \in{} ran(DataSet\_corresp\_Set)$}{true}{}
		\Guard{grd7}{false}{$DS = DataSet\_corresp\_Set\converse{}(o\_DS)$}{true}{}
		\Guard{grd8}{false}{$DataValue\_Set \setminus{} DataValue \neq{}\emptyset{}$}{true}{}
		\Guard{grd9}{false}{$dva \in{} DataValue\_Set \setminus{} DataValue$}{true}{}
	}
	\ACTIONS{true}{
		\Action{act1}{$DataValue \bcmeq{}  DataValue \bunion{}  \{dva\}$}{true}{}
		\Action{act2}{$DataValue\_valueOf\_DataSet(dva) \bcmeq{}  DS$}{true}{}
		\Action{act3}{$DataValue\_corresp\_Constant(dva) \bcmeq{}  o\_dva$}{true}{}
	}
}

\END

\paragraph{Rule 109 : addition of a constant, defined as a maplet}

\MACHINE{Ontologies\_BSystem\_specs\_translation\_ref\_1}{Ontologies\_BSystem\_specs\_translation}{BSystem\_Metamodel\_Context,Domain\_Metamodel\_Context}{}

\EVT{rule\_109\_1}{false}{ordinary}{}{\\handling  the addition of a constant, defined as a maplet, element of a relation (case where the relation corresponds to a constant relation)}{
	\ANY{
		\Param{o\_maplet}{true}{}
		\Param{maplet}{true}{}
		\Param{o\_RE}{true}{}
		\Param{RE}{true}{}
		\Param{o\_lg1}{true}{}
		\Param{o\_lg2}{true}{}
		\Param{antecedent}{true}{}
		\Param{image}{true}{}
		\Param{o\_antecedent}{true}{}
		\Param{o\_image}{true}{}
	}
	\GUARDS{true}{
		\Guard{grd0}{false}{$dom(Constant\_typing\_Property) \setminus{} ran(RelationMaplet\_corresp\_Constant) \neq{}\emptyset{}$}{true}{}
		\Guard{grd1}{false}{$o\_maplet \in{} dom(Constant\_typing\_Property) \setminus{} ran(RelationMaplet\_corresp\_Constant)$}{true}{}
		\Guard{grd2}{false}{$o\_lg1 = Constant\_typing\_Property(o\_maplet)$}{true}{}
		\Guard{grd3}{false}{$LogicFormula\_uses\_Operators(o\_lg1) = \{1\mapsto{}Maplet\_OP\}$}{true}{}
		\Guard{grd4}{false}{$\{o\_antecedent, o\_image\}\subseteq{}(dom(Constant\_isInvolvedIn\_LogicFormulas)\binter{}ran(Individual\_corresp\_Constant))$}{true}{}
		\Guard{grd5}{false}{$(2\mapsto{}o\_lg1) \in{} Constant\_isInvolvedIn\_LogicFormulas(o\_antecedent)$}{true}{}
		\Guard{grd6}{false}{$(3\mapsto{}o\_lg1) \in{} Constant\_isInvolvedIn\_LogicFormulas(o\_image)$}{true}{}
		\Guard{grd7}{false}{$antecedent = Individual\_corresp\_Constant\converse{}(o\_antecedent)$}{true}{}
		\Guard{grd8}{false}{$image = Individual\_corresp\_Constant\converse{}(o\_image)$}{true}{}
		\Guard{grd9}{false}{$o\_lg2 \in{} LogicFormula$}{true}{}
		\Guard{grd10}{false}{$LogicFormula\_uses\_Operators(o\_lg2) = \{1\mapsto{}Equal2SetOf\_OP\}$}{true}{}
		\Guard{grd11}{false}{$(2\mapsto{}o\_lg2) \in{} Constant\_isInvolvedIn\_LogicFormulas(o\_maplet)$}{true}{}
		\Guard{grd12}{false}{$o\_RE \in{} ran(Relation\_corresp\_Constant)$}{true}{}
		\Guard{grd13}{false}{$(1\mapsto{}o\_lg2) \in{} Constant\_isInvolvedIn\_LogicFormulas(o\_RE)$}{true}{}
		\Guard{grd14}{false}{$RE = Relation\_corresp\_Constant\converse{}(o\_RE)$}{true}{}
		\Guard{grd15}{false}{$Relation\_Maplet\_Set \setminus{} RelationMaplet \neq{}\emptyset{}$}{true}{}
		\Guard{grd16}{false}{$maplet \in{} Relation\_Maplet\_Set \setminus{} RelationMaplet$}{true}{}
		\Guard{grd17}{false}{$Individual\_individualOf\_Concept(antecedent)=Relation\_domain\_Concept(RE)$}{true}{}
		\Guard{grd18}{false}{$Individual\_individualOf\_Concept(image)=Relation\_range\_Concept(RE)$}{true}{}
	}
	\ACTIONS{true}{
		\Action{act1}{$RelationMaplet \bcmeq{}  RelationMaplet \bunion{}  \{maplet\}$}{true}{}
		\Action{act2}{$RelationMaplet\_corresp\_Constant(maplet) \bcmeq{}  o\_maplet$}{true}{}
		\Action{act3}{$RelationMaplet\_mapletOf\_Relation(maplet) \bcmeq{}RE$}{true}{}
		\Action{act4}{$RelationMaplet\_antecedent\_Individual(maplet) \bcmeq{} antecedent$}{true}{}
		\Action{act5}{$RelationMaplet\_image\_Individual(maplet) \bcmeq{} image$}{true}{}
	}
}

\END

\subsubsection{Addition of Non-Existing Variables}

\paragraph{Rule 108 : addition of a variable, sub set of an instance of Concept}

\MACHINE{Ontologies\_BSystem\_specs\_translation\_ref\_1}{Ontologies\_BSystem\_specs\_translation}{BSystem\_Metamodel\_Context,Domain\_Metamodel\_Context}{}

\EVT{rule\_108\_1}{false}{ordinary}{}{\\handling  the addition of a variable, sub set of an instance of Concept (case where the concept corresponds to an abstract set)}{
	\ANY{
		\Param{x\_CO}{true}{}
		\Param{CO}{true}{}
		\Param{o\_lg}{true}{}
		\Param{o\_CO}{true}{}
	}
	\GUARDS{true}{
		\Guard{grd0}{false}{$dom(Variable\_typing\_Invariant) \setminus{} ran(Concept\_corresp\_Variable) \neq{}\emptyset{}$}{true}{}
		\Guard{grd1}{false}{$x\_CO \in{} dom(Variable\_typing\_Invariant) \setminus{} ran(Concept\_corresp\_Variable)$}{true}{}
		\Guard{grd2}{false}{$o\_lg = Variable\_typing\_Invariant(x\_CO)$}{true}{}
		\Guard{grd3}{false}{$LogicFormula\_uses\_Operators(o\_lg) = \{1\mapsto{}Inclusion\_OP\}$}{true}{}
		\Guard{grd4}{false}{$LogicFormula\_involves\_Sets(o\_lg) \neq{} \emptyset{}$}{true}{}
		\Guard{grd5}{false}{$(2\mapsto{}o\_CO)\in{}LogicFormula\_involves\_Sets(o\_lg)$}{true}{}
		\Guard{grd6}{false}{$o\_CO \in{} ran(Concept\_corresp\_AbstractSet)$}{true}{}
		\Guard{grd7}{false}{$CO = Concept\_corresp\_AbstractSet\converse{}(o\_CO)$}{true}{}
	}
	\ACTIONS{true}{
		\Action{act1}{$Concept\_isVariable(CO) \bcmeq{}  TRUE$}{true}{}
		\Action{act2}{$Concept\_corresp\_Variable(CO) \bcmeq{} x\_CO$}{true}{}
	}
}
\EVT{rule\_108\_2}{false}{ordinary}{}{\\handling  the addition of a variable, sub set of an instance of Concept (case where the concept corresponds to a constant)}{
	\ANY{
		\Param{x\_CO}{true}{}
		\Param{CO}{true}{}
		\Param{o\_lg}{true}{}
		\Param{o\_CO}{true}{}
	}
	\GUARDS{true}{
		\Guard{grd0}{false}{$dom(Variable\_typing\_Invariant) \setminus{} ran(Concept\_corresp\_Variable) \neq{}\emptyset{}$}{true}{}
		\Guard{grd1}{false}{$x\_CO \in{} dom(Variable\_typing\_Invariant) \setminus{} ran(Concept\_corresp\_Variable)$}{true}{}
		\Guard{grd2}{false}{$o\_lg = Variable\_typing\_Invariant(x\_CO)$}{true}{}
		\Guard{grd3}{false}{$LogicFormula\_uses\_Operators(o\_lg) = \{1\mapsto{}Inclusion\_OP\}$}{true}{}
		\Guard{grd4}{false}{$LogicFormula\_involves\_Sets(o\_lg) = \emptyset{}$}{true}{}
		\Guard{grd5}{false}{$o\_CO\in{}dom(Constant\_isInvolvedIn\_LogicFormulas)$}{true}{}
		\Guard{grd6}{false}{$(2\mapsto{}o\_lg) \in{} Constant\_isInvolvedIn\_LogicFormulas(o\_CO)$}{true}{}
		\Guard{grd7}{false}{$o\_CO \in{} ran(Concept\_corresp\_Constant)$}{true}{}
		\Guard{grd8}{false}{$CO = Concept\_corresp\_Constant\converse{}(o\_CO)$}{true}{}
	}
	\ACTIONS{true}{
		\Action{act1}{$Concept\_isVariable(CO) \bcmeq{}  TRUE$}{true}{}
		\Action{act2}{$Concept\_corresp\_Variable(CO) \bcmeq{} x\_CO$}{true}{}
	}
}

\END

\subsection{The SysML/KAOS Domain Model Parser Tool}\label{the_tool_description_section}
\begin{figure}[!h]
\begin{center}
\includegraphics[width=1.1\textwidth]{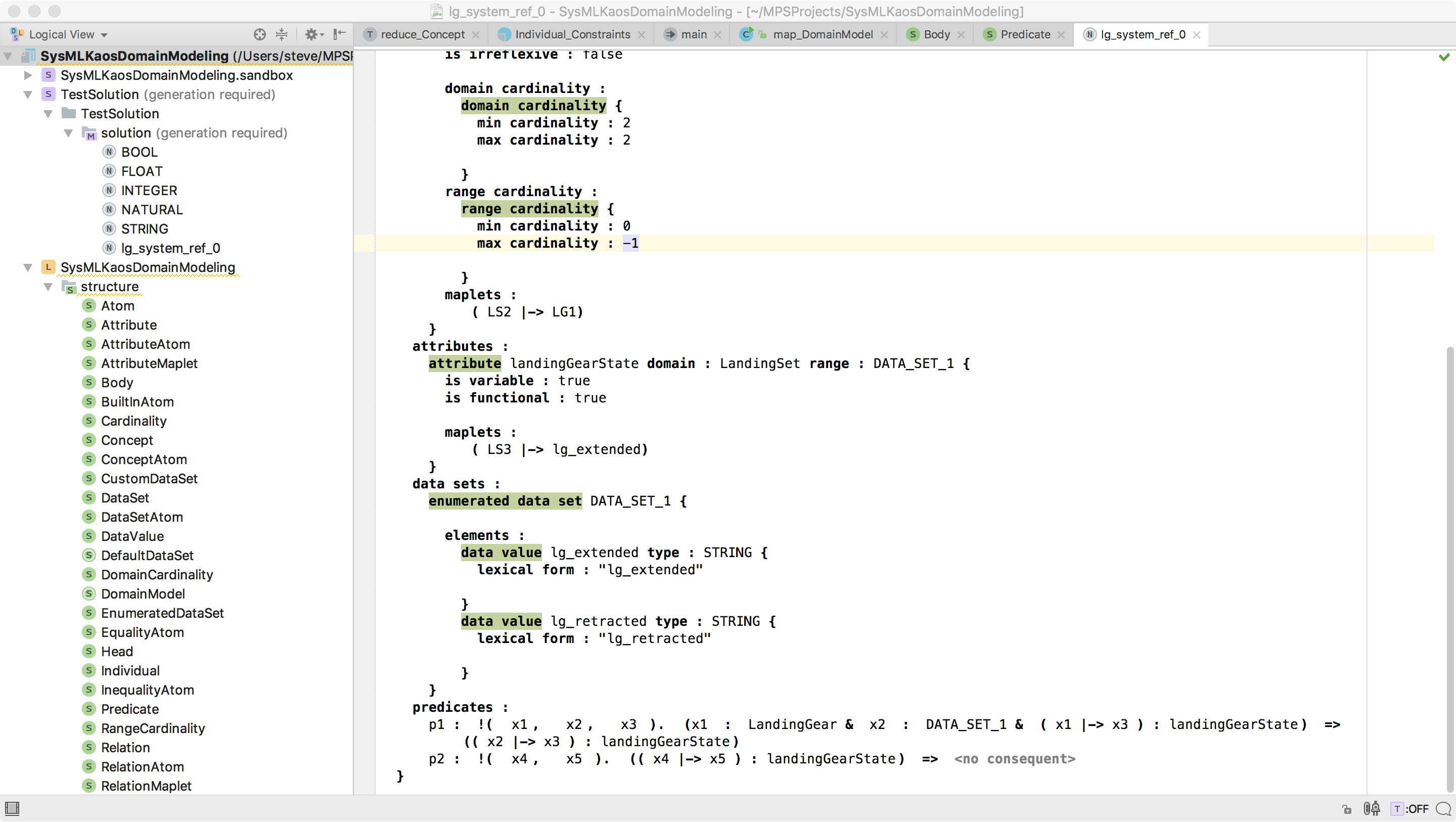}
\end{center}
\caption{\label{symlkaos_domain_modeling_tool_capture1} Preview of the SysML/KAOS Domain Model Parser Tool}
\end{figure}

\begin{figure}[!h]
\begin{center}
\includegraphics[width=1.1\textwidth]{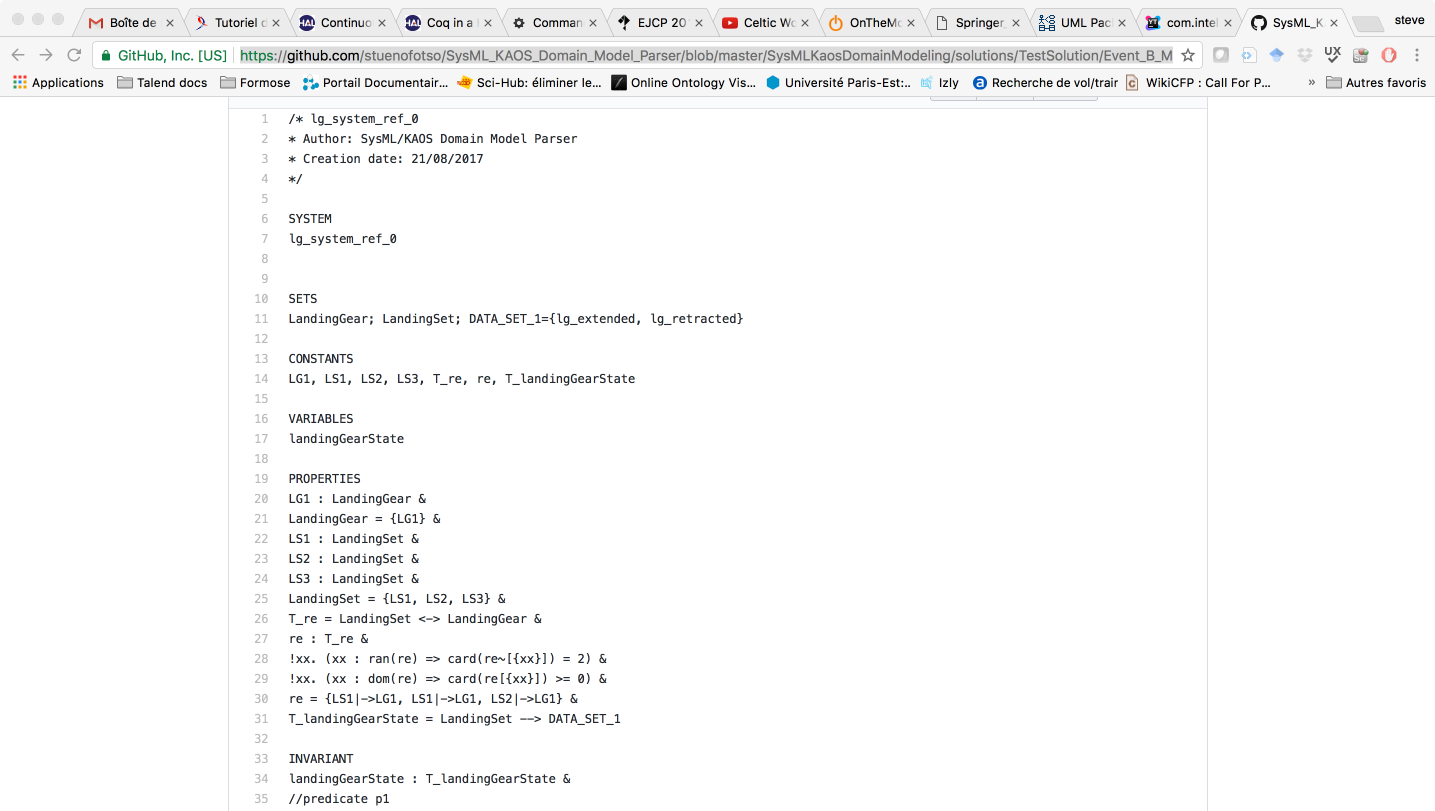}
\end{center}
\caption{\label{symlkaos_domain_modeling_tool_capture2} Preview of \textit{B System} Specifications Generated by the SysML/KAOS Domain Model Parser Tool for the Landing Gear System Case Study}
\end{figure}

The correspondence rules outlined here have been implemented within an open source tool called \textit{SysML/KAOS Domain Model Parser} \cite{SysML_KAOS_Domain_Model_Parser_link}. It allows the construction of domain models \textit{(Fig. \ref{symlkaos_domain_modeling_tool_capture1})}  and generates  the corresponding \textit{B System} specifications \textit{(Fig. \ref{symlkaos_domain_modeling_tool_capture2})}. It is build through \textit{Jetbrains Meta Programming System} \cite{jetbrains_mps}, a tool to design domain specific languages using language-oriented programming.

\section{Conclusion and Future Works}
This paper was focused on a presentation of  mapping rules between SysML/KAOS domain models and B System specifications illustrated through a case study dealing with a landing gear system. The specifications thus obtained can also be seen as a formal semantics for SysML/KAOS domain models. They complement  the formalization of the SysML/KAOS goal model by providing a description of the state of the system.

Work in progress is aimed  at 
integrating our approach, implemented through the \textit{SysML/KAOS Domain Model Parser } tool,  within the open-source platform  \textit{Openflexo}  \cite{openflexo_link} and at evaluating the impact of updates  on domain models on B System specifications.

\section*{Acknowledgment}
This work is carried out within the framework of the  \textit{FORMOSE} project \cite{anr_FORMOSE_reference_link} funded by the French National Research Agency (ANR).

%
%
\bibliographystyle{splncs03}
\bibliography{references}

\begin{thebibliography}{10}
\providecommand{\url}[1]{\texttt{#1}}
\providecommand{\urlprefix}{URL }

\bibitem{DBLP:books/daglib/0024570}
Abrial, J.: Modeling in {Event-B} - System and Software Engineering. Cambridge
  University Press (2010)

\bibitem{DBLP:journals/scp/AlkhammashBFC15}
Alkhammash, E., Butler, M.J., Fathabadi, A.S., C{\^{\i}}rstea, C.: Building
  traceable {Event-B} models from requirements. Sci. Comput. Program.  111,
  318--338 (2015)

\bibitem{h.Alkhammash}
{Alkhammash, Eman H.}: {Derivation of Event-B Models from OWL Ontologies}.
  MATEC Web Conf.  76,  04008 (2016)

\bibitem{anr_FORMOSE_reference_link}
{ANR-14-CE28-0009}: {Formose ANR} project (2017)

\bibitem{DBLP:conf/birthday/BjornerE10}
Bj{\o}rner, D., Eir, A.: Compositionality: Ontology and mereology of domains.
  Essays in Honor of Willem-Paul de Roever, LNCS, vol. 5930, pp. 22--59.
  Springer (2010)

\bibitem{Boniol2014}
Boniol, F., Wiels, V.: The landing gear system case study. pp. 1--18. ABZ,
  Springer (2014)

\bibitem{clearsy_b_system_link}
ClearSy: {Atelier B: B System} (2014), \url{http://clearsy.com/}

\bibitem{DBLP:journals/stt/Doberkat01}
Doberkat, E.: The {Object-Z} specification language. Softwaretechnik-Trends
  21(1) (2001)

\bibitem{DBLP:conf/icfem/DongSW02}
Dong, J.S., Sun, J., Wang, H.H.: Z approach to semantic web. In: Formal Methods
  and Software Engineering - {ICFEM}, {LNCS}. vol. 2495, pp. 156--167. Springer
  (2002)

\bibitem{DBLP:conf/inforsid/GnahoS10}
Gnaho, C., Semmak, F.: Une extension {SysML} pour l'ing{\'{e}}nierie des
  exigences dirig{\'{e}}e par les buts. In: 28e Congr{\`{e}}s INFORSID, France.
  pp. 277--292 (2010)

\bibitem{van2001reference}
van Harmelen, F., Patel-Schneider, P.F., Horrocks, I.: Reference description of
  the {DAML+ OIL} ontology markup language  (2001)

\bibitem{jetbrains_mps}
Jetbrains: Jetbrains mps (2017), \url{https://www.jetbrains.com/mps/}

\bibitem{DBLP:conf/kbse/LaleauM00}
Laleau, R., Mammar, A.: An overview of a method and its support tool for
  generating {B} specifications from {UML} notations. In: The Fifteenth {IEEE}
  International Conference on Automated Software Engineering, {ASE} 2000,
  Grenoble, France, September 11-15, 2000. pp. 269--272. {IEEE} Computer
  Society (2000), \url{https://doi.org/10.1109/ASE.2000.873675}

\bibitem{DBLP:books/daglib/0025377}
van Lamsweerde, A.: Requirements Engineering - From System Goals to {UML}
  Models to Software Specifications. Wiley (2009)

\bibitem{DBLP:conf/isola/MammarL16}
Mammar, A., Laleau, R.: On the use of domain and system knowledge modeling in
  goal-based {Event-B} specifications. In: ISoLA 2016, LNCS. pp. 325--339.
  Springer (2016)

\bibitem{DBLP:conf/iceccs/MatoussiGL11}
Matoussi, A., Gervais, F., Laleau, R.: A goal-based approach to guide the
  design of an abstract {Event-B} specification. In: {ICECCS} 2011. pp.
  139--148. {IEEE} Computer Society (2011)

\bibitem{openflexo_link}
Openflexo: Openflexo project (2015), \url{http://www.openflexo.org}

\bibitem{DBLP:conf/ifip2/PoernomoU09}
Poernomo, I., Umarov, T.: A mapping from normative requirements to {Event-B} to
  facilitate verified data-centric business process management. CEE-SET LNCS,
  vol. 7054, pp. 136--149. Springer (2009)

\bibitem{DBLP:reference/snam/SenguptaH14}
Sengupta, K., Hitzler, P.: Web ontology language {(OWL)}. In: Encyclopedia of
  Social Network Analysis and Mining, pp. 2374--2378 (2014)

\bibitem{Snook:2006:UFM:1125808.1125811}
Snook, C., Butler, M.: {UML-B: Formal Modeling and Design Aided by UML}. ACM
  Trans. Softw. Eng. Methodol.  15(1),  92--122 (Jan 2006)

\bibitem{SysML_KAOS_Domain_Model_Parser_link}
Tueno, S.: {SysML/KAOS Domain Model Parser} (2017),
  \url{https://github.com/stuenofotso/SysML_KAOS_Domain_Model_Parser}

\bibitem{sysml_kaos_domain_modeling}
Tueno, S., Laleau, R., Mammar, A., Frappier, M.: {Towards Using Ontologies for
  Domain Modeling within the SysML/KAOS Approach}. IEEE proceedings of MoDRE
  workshop, 25th IEEE International Requirements Engineering Conference  (2017)

\bibitem{owlgred_reference_link}
UL, I.: Owlgred home (2017), \url{http://owlgred.lumii.lv/}

\bibitem{DBLP:conf/icfem/WangDS10}
Wang, H.H., Damljanovic, D., Sun, J.: Enhanced semantic access to formal
  software models. In: Formal Methods and Software Engineering - {ICFEM},
  {LNCS}. vol. 6447, pp. 237--252. Springer (2010)

\end{thebibliography}

\end{document}